\documentclass[aps,prd,amssymb,amsmath,eqsecnum,showpacs,nofootinbib,twocolumn] {revtex4}
\usepackage{color}
\usepackage{footnote}
\usepackage{graphicx}
\usepackage{ulem}
\begin{document}
\author{Gast\' on Briozzo$^{1}$, Emanuel Gallo$^1$, Thomas M\"adler$^2$ }

\newcommand{\red}[1]{\textcolor{red}{#1}}
\newcommand{\blue}[1]{\textcolor{blue}{#1}}

\affiliation{$^1$FaMAF, UNC; Instituto de Física Enrique Gaviola (IFEG), CONICET, \\
Ciudad Universitaria, (5000) C\'ordoba, Argentina. }

\affiliation{$^2$Escuela de Obras Civiles and N\'ucleo de Astronom\'ia, Facultad de Ingenier\'{i}a y Ciencias, Universidad Diego Portales, Avenida Ej\'{e}rcito
Libertador 441, Casilla 298-V, Santiago, Chile. }

%\affiliation{$^3$ }

\title{Shadows of rotating black holes in plasma environments with aberration effects}

\begin{abstract}

The shadows of black holes encode significant information about the properties of black holes and the spacetime surrounding them. So far, the effects of dispersive media, such as plasma, and relativistic aberration on the propagation of light around compact objects have been treated separately in the literature. In this paper, we will employ the Konoplya, Stuchlík, and Zhidenko family of stationary, axially symmetric, and asymptotically flat metrics to describe the spacetime around rotating black holes. We will study how the parameters of the black hole, the chromatic effects resulting from the presence of a non-magnetized, pressureless plasma environment, and the effects of relativistic aberration of a moving observer modify the morphology of the shadow.

%\begin{keywords} 
 %   General Relativity -- Gravitational Lensing -- Astrophysical Plasma -- Relativistic Aberration -- Black Holes -– Black Hole Shadow
%\end{keywords}

\end{abstract}

%\pacs{ 04.20.-q, 04.20.Cv, 04.20.Ex, 04.25.D- }

\maketitle
\section{Introduction} 
 %\cite{1967ApJ...150.1005H,1968ApJ...153..807H}
 
For most observations in the field of general relativity, the influence of optical media along the photon trajectory is negligible. In fact, the theory of gravitational lensing is generally applied to the propagation of light rays in vacuum, where light follows geodesic curves and the deflection angles and trajectories are independent of the photon frequency, resulting in achromatic phenomena. However, the effects of the medium are not negligible in the radio frequency range. At all scales, we are motivated to consider galaxies, galaxy clusters, black holes, or other compact objects to be immersed in a dispersive plasma medium or surrounded by dense plasma-rich magnetospheres that can form accretion disks \cite{GJ_1969, petri2016theory, hitomi2018temperature}.

In such optical media, photon propagation becomes frequency dependent, giving rise to chromatic phenomena. While these effects are negligible in the visible spectrum, they generate strong modifications of the usual gravitational lensing behavior at radio frequencies. Therefore, it is worth investigating how plasma influences gravitational lensing phenomena.

Moreover, there is a well-known correspondence between the dynamics of photons in a homogeneous plasma and massive test particles under the influence of gravitational fields \cite{kulsrud1992dynamics}. They follow the same geodesics, making this topic of interest for studying the behavior of massive neutrinos (see, for example, \cite{Caballero:2011dw} where the influence of the bending of neutrino trajectories and the energy shifts on the nucleosynthesis resulting from the interaction of the emitted neutrinos and hot outflowing material ejected from the disk is studied).

The effect of plasma on light propagation has been studied since the 1960s. In 1966, Muhleman and Johnston studied the influence of the electronic plasma in the solar corona on the time delay of radio frequencies under the Sun's gravitational field, using the gravitational and plasma refractive indices to derive a weak-field approximation \cite{MJ_1966}. In 1970, Muhleman, Ekers, and Fomalont measured the deflection of the quasar 3C273 when it was occulted by the Sun and obtained an estimate of the electron density in the solar corona, calculating for the first time the deflection of light in the presence of a plasma in the weak-field approximation \cite{MDE_1970}. A variety of studies focusing on the solar wind and the electron density profile in its outer corona were also performed, using light propagation analysis on different space missions such as Viking, Mariner 6 and 7, and the Cassini mission. In 1980, Breuer and Ehlers performed a rigorous derivation of a Hamiltonian for light rays including a magnetized plasma and a curved spacetime \cite{BE_1980}. Recently, plasma lensing effects have been used to resolve the emission regions of a pulsar \cite{2018Main}. For a recent (but incomplete) list of works studying the influence of plasma on the propagation of light rays in different astrophysical situations, we refer to \cite{Er:2013efa, Rogers_2015, Tsupko_2017, CG_2018, CGV_2019, CGR_2019, Crisnejo:2019ril, Crisnejo:2022qlv, BG_2023,Tsupko_2019, Turimov_2019, Kimpson:2019mji, Er:2022lad, Tsupko:2019axo, Sun:2022ujt}. 

In recent years, there has been increasing interest in studying the shadows of different rotating black holes with plasma environments, following the pioneering work of Perlick and Tsupko \cite{PyT_2017} where light propagation in a Kerr spacetime filled by a cold plasma was analytically studied. In particular, in \cite{BPB_2022}, general necessary and sufficient conditions for a stationary and axially-symmetric metric to achieve separability of the Hamilton-Jacobi (HJ) equation in non-magnetized cold plasma media were obtained.  

    The first problem encountered when investigating black hole shadows in plasma environments is that photons in general do not follow null geodesics \cite{CG_2018}.
Additionally, when considering lensing effects in strong gravitational fields, we must treat both gravitational and plasma lensing effects simultaneously since the deflection is too significant to be linearized as is done in the weak gravitational field regime.
The magnitude of the plasma influence is determined by the ratio between the plasma frequency and the photon frequency. Lower photon frequency (longer wavelength) or higher plasma density leads to higher refractive deflection \cite{Tsupko_2019}.
When a photon travels through the plasma atmosphere of a star or through the accretion disk around a black hole, the influence of the plasma on the light deflection can be significant due to its high concentration \cite{tsupko2013gravitational, MJ_1966}. If the plasma deflection is large enough, it can compensate or completely overcome the gravitational deflection.
The shadow of a plasma-free black hole was first calculated by Synge \cite{Synge_1966} for the Schwarzschild spacetime and by Bardeen \cite{bardeen1973timelike} for the Kerr spacetime. Recently, Grenzebach et al. \cite{GPL_2014, Grenzebach_2015, Grenzebach_2016} generalized the calculations for the shadow of a Kerr black hole in the Plebański-Demiański class limit, incorporating relativistic aberration effects arising from a moving observer.
Studying the dependence of the shadow on the parameters of alternative metrics to Kerr can aid in testing Einstein's General Theory of Relativity in the strong field regime. Additionally, taking into account the presence of plasma generally requires numerical methods to obtain a detailed understanding of the shadow. However, there exists a large family of metrics (which includes the Kerr metric as a particular case) where these shadows can be obtained analytically. This allows for an understanding of how different metric parameters (and the plasma environment) can influence the shape of the shadow.

    So far, the effects of dispersive media, such as plasma, and relativistic aberration on the propagation of light around compact objects have been treated separately in the literature. In this article, we  consider these two effects simultaneously to study the shadow of a family of alternative metrics that contain the Kerr black hole as a particular case.
{
Although there exists a plethora of stationary and axially symmetric metrics in the literature that generalize the Kerr metric from alternative theories to general relativity, we employ the family of metrics described by Konoplya, Stuchlík, and Zhidenko (KSZ) in \cite{KSZ_2018} to represent the spacetime around rotating black holes immersed in plasma environments. The peculiarity of this metric family lies in its ability to disentangle variables in the Klein-Gordon and Hamilton-Jacobi (HJ) equations. However, it should be noted that this family is a subset of a larger family described in \cite{Konoplya:2016jvv} that includes the KSZ as a specific instance. Nonetheless, the latter family does not generally exhibit the desirable property of Hamilton-Jacobi equation separability, which allows for the analytical study of shadows. Consequently, in our study, we focus exclusively on the KSZ metric family. We will follow the approach seen in \cite{PyT_2017} to obtain the equations that parameterize the black hole shadow, and generalizing the Grenzebach equations \cite{Grenzebach_2015}, we will present a model that considers both the effects of the plasma environment and the effects of relativistic aberration.}

{
The organization of the paper is as follows. In Sect. \ref{c3s2}, we will summarize the metrics described in \cite{KSZ_2018} and present the specific spacetime models on which we will work. In Sec. \ref{c3s3}, we will develop the general formulas needed to parameterize the contour curve of the black hole shadow in cold non-magnetic plasma environments adapted to the KSZ metric family, and in Sec. \ref{c3s3ss2}, we will set the values of parameters used in the rest of this work. In Sec. \ref{sec:sinplasma}, we start with general considerations of shadows in plasma-free environments, and we follow in Sec. \ref{sec:plasma}, showing how the shadows and photon regions are affected by the different parameters of the black hole and plasma profiles. In Sec. \ref{c3s4}, we will modify the previously obtained procedure to consider now the effects of relativistic aberration. In Sec. \ref{c3s4ss2}, we show the modifications of the shadows for some particular observers moving with respect to the standard observers (as defined in Sec. \ref{c3s4}). We end with final remarks in Sec. \ref{c3s5}.
}    
\section{Generalized rotating black hole metrics}
\label{c3s2}

\subsection{The Metrics of Konoplya, Stuchlík and Zhidenko}
\label{c3s2ss1}

    A family of stationary, axially symmetric and asymptotically flat metrics with an event horizon was presented in 
    \cite{KSZ_2018} having the line element  
    \begin{equation}
    \begin{split}
        ds^2&=-\frac{N^2-W^2\sin^2\vartheta}{K^2}dt^2-2Wr\sin^2\vartheta dtd\varphi\\&
        +K^2r^2\sin^2\vartheta d\varphi^2+\Sigma\frac{B^2}{N^2}dr^2+\Sigma r^2d\vartheta^2,
        \label{eq:KSZ_ds1}
        \end{split}
    \end{equation}
    where $N$, $W$, $K$, $\Sigma$ and $B$ are arbitrary functions of $r$  and $\vartheta$. The main motivation in the work of Konoplya, Stuchlík and Zhidenko was to find a general structure for metrics  allowing  separability of the Klein-Gordon and HJ equations.  
    The form of the metric components found by KSZ was 
        \begin{eqnarray}
        \label{eq:KSZ_C2}
        %\begin{split}
        B(r,y)     &=&R_B(r), \\
        \Sigma(r,y)&=&R_\Sigma(r)+\frac{a^2y^2}{r^2}, \\
        W(r,y)     &=&\frac{aR_M(r)}{r^2\Sigma(r,y)}, \\
        N^2(r,y)   &=&R_\Sigma(r)-\frac{R_M(r)}{r}+\frac{a^2}{r^2}, \\
        K^2(r,y)   &=&\frac{1}{\Sigma(r,y)}\left[R_\Sigma^2(r)+\frac{a^2R_\Sigma(r)}{r^2}+\frac{a^2R_M(r)}{r^2}\right.\nonumber\\&&\left.+\frac{a^2y^2}{r^2}N^2(r,y)\right].
        %\end{split}
    \end{eqnarray}
    where $y=\cos\vartheta$. As the metric is asymptotically flat, the functions $R_B(r)$, $R_\Sigma(r)$ and $R_M(r)$ must satisfy
    \begin{equation}
        \label{eq:KSZ_lim}
        \lim_{r\rightarrow\infty} R_B(r)      =1, ~~~~ 
        \lim_{r\rightarrow\infty} R_\Sigma(r) =1, ~~~~
        \lim_{r\rightarrow\infty} \frac{R_M(r)}{r} =0.
    \end{equation}
    
    The co and contravariant metric components can be expressed as
    \begin{widetext}
    \begin{equation}
        [g_{\mu\nu}]=
        \left(
        \begin{matrix}
        -\left(1-\frac{rR_M}{\rho^2}\right) & 0                          & 0      & -\frac{arR_M\sin^2\vartheta}{\rho^2}  \\
        0                                   & \frac{\rho^2R_B^2}{\Delta} & 0      & 0 \\
        0                                   & 0                          & \rho^2 & 0 \\
        -\frac{arR_M\sin^2\vartheta}{\rho^2}   & 0                          & 0      & \sin^2\vartheta\left(r^2R_\Sigma+a^2+\frac{a^2rR_M\sin^2\vartheta}{\rho^2}\right)\\
        \end{matrix}
        \right),
        \label{eq:KSZ_g}
    \end{equation}
    \begin{equation}
        [g^{\mu\nu}]=\frac{1}{\rho^2}
        \left(
        \begin{matrix}
        a^2\sin^2\vartheta-\frac{(r^2R_\Sigma+a^2)^2}{\Delta} & 0                    & 0 & -\frac{arR_M}{\Delta} \\
        0                                                  & \frac{\Delta}{R_B^2} & 0 & 0 \\
        0                                                  & 0                    & 1 & 0 \\
        -\frac{arR_M}{\Delta}                              & 0                    & 0 & \frac{1}{\sin^2\vartheta}-\frac{a^2}{\Delta} \\
        \end{matrix}
        \right),
        \label{eq:KSZ_ginv}
    \end{equation}
    \end{widetext}
    where we have introduced the auxiliary functions
    \begin{equation}
        \rho^2(r,\vartheta)=r^2\Sigma(r,\vartheta)=r^2R_\Sigma(r)+a^2\cos^2\vartheta,
        \label{eq:rho2}
    \end{equation}
    \begin{equation}
        \Delta(r)=r^2N^2(r)=r^2R_\Sigma(r)-rR_M(r)+a^2.
        \label{eq:Delta}
    \end{equation}

    We remark that the KSZ metrics are not the most generic ones  allowing a separation in the HJ equation. An alternative method to obtain stationary and axially symmetric spacetimes employs  the Newman-Janis algorithm (see e.g. \cite{Shaikh:2019fpu,Chen_2019}).
    
\subsection{Spacetime models}
\label{c3s2ss4}

    In this subsection we will introduce some of the most common spacetime models that belong to the KSZ family of metrics. Table \ref{tab:Metrics} expresses the metric components of the main models mentioned here, as well as the location of the event horizon. The latter is obtained from the condition $g^{rr}=0$, which occurs for the bigger root of $\Delta(r_h)=0$.

    The Kerr black hole \cite{Kerr_1963, wei2019curvature} is the simplest model of a rotating black hole, considering only its mass $m$ and spin parameter $a$ associated to the angular momentum $J$ by $a=J/m$.   
    In order to avoid a naked singularity, the spin value is constrained between $0$ and $m$.

    The Kerr-Newman black hole  \cite{Newman_1965} (KN) incorporates an electric charge $Q$ into the Kerr model. To avoid a naked singularity, we must require $a^2+Q^2<m^2$, so that for a fixed value of the mass $m$, both the angular momentum and the charge of the black hole are constrained, with the maximum value of one depending on the magnitude of the other.
    
    The modified Kerr model \cite{KZ_2016} takes the Kerr black hole and introduces a small static deformation $\eta$ that alters the relation between the mass $m$ and the radius $r_h$ of the event horizon, preserving the asymptotic properties of Kerr spacetime.
   
    The Kerr-Sen model is a solution in heterotic string theory \cite{Sen_1992}  for a rotating black hole with electric charge $Q$, which is qualitatively similar to the Kerr model. Here, the parameter $b=Q^2/2m$ is introduced to facilitate the treatment of the model.
    Again, by requiring the radius of the event horizon to be a positive real number, we obtain the conditions $a^2<(m-b)^2$ and $\blue{0<}b<m$, so that both parameters are constrained. 
    
    The Brane-world theory \cite{BW_2005} consider a black hole with (nonelectrical) tidal charge $\beta$. This metric is identical in appearance to the Kerr-Newman model with electric charge $Q$ when $\beta>0$, after doing the identification$\beta=Q^2$. However, the tidal charge can take negative values, allowing the angular momentum of the black hole to considerably exceed the limit allowed in Kerr-Newman. To distinguish from Kerr-Newman, we will only consider a negative tidal charge in the Braneworld scenario, taking $\beta=-Q^{*2}$. By calculating the radius of the event horizon, the relation $a^2-Q^{*2}<m^2$ must be satisfied. So while the value of $|a|$ is constrained between $0$ and $\sqrt{m^2+Q^{*2}}$, there is not an upper bound constraint on the value of $Q^*$. In any case, $Q^*$ is expected to be small.
    
    Finally, we will consider the modified dilaton black hole \cite{Shaikh:2019fpu}. This spacetime contemplates the existence of both electric $Q_E$ and magnetic $Q_M$ charges, being qualitatively similar to the Kerr-Newman model. We introduce the Dilaton charge $r_0=(Q_M^2-Q_E^2)/2m$ and the parameter $q=Q_E^2+Q_M^2-r_0^2$. To avoid a naked singularity, the condition $a^2+q<m^2$ must be satisfied. Thus, the absolute value of $a$ is constrained between $0$ and $\sqrt{m^2-q}$. In turn, $q$ is upper bounded by $m^2$.

    \begin{table*}[htbp]
        \centering
        \begin{tabular}{|l|l|l|l|l|l}
        \hline 
        Metric & $R_M(r)$               & $R_\Sigma(r)$         & $\rho^2(r,\vartheta)$            & Horizon                            \\ \hline     
        Kerr    & $2m$                   & $1$                   & $r^2+a^2\cos^2\vartheta$       & $r_h=m\pm\sqrt{m^2-a^2}$             \\ \hline 
        Kerr-Newman     & $2m-\frac{Q^2}{r}$     & $1$                   & $r^2+a^2\cos^2\vartheta$       & $r_h=m+\sqrt{m^2-a^2-Q^2}$         \\ \hline 
        Modified Kerr     & $2m+\frac{\eta}{r^2}$  & $1$                   & $r^2+a^2\cos^2\vartheta$       & $0=r_h^2-2mr_h-\frac{\eta}{r_h}+a^2$ \\ \hline 
        Kerr-Sen     & $2m$                   & $1+\frac{2b}{r}$      & $r^2+a^2\cos^2\vartheta+2br$   & $r_h=m-b+\sqrt{(m-b)^2-a^2}$       \\ \hline 
        Brane-World      & $2m+\frac{Q^{*2}}{r}$     & $1$                   & $r^2+a^2\cos^2\vartheta$       & $r_h=m+\sqrt{m^2-a^2+Q^{*2}}$         \\ \hline 
        Dilaton      & $2m-\frac{q+r_0^2}{r}$ & $1-\frac{r_0^2}{r^2}$ & $r^2+a^2\cos^2\vartheta-r_0^2$ & $r_h=m+\sqrt{m^2-a^2-q}$           \\ \hline 
        \end{tabular}
        \caption{Spacetime models. All metrics take $R_B=1$.}
        \label{tab:Metrics}
    \end{table*}    
    Note that all the aformentionded models turn out to be identical to the Kerr spacetime if their respective characteristic charge parameters become zero. For this reason, from now on we will not consider the Kerr model as a metric independent of the rest, but as a particular case of them.

\section{Shadow of a black hole in a plasma environment}
\subsection{The general framework}\label{c3s3}
    {In the following,  we present the conditions on the plasma profile necessary to separate the Hamilton-Jacobi equations associated with a cold non-magnetic plasma in a KSZ background. To do so, we will closely follow the pioneering work of \cite{PyT_2017}, which was originally developed to separate the HJ equation in the Kerr metric.}
    
    For an observer pointing a telescope in the direction of a black hole, there will be a region in the sky that will remain dark as long as there are no light sources in between. To determine the shape of the limiting surface between brightness and darkness, it is technically convenient to consider light rays sent backward from the observer's position in the Boyer-Lindquist like coordinates $(r_O,\vartheta_O)$. The position of the observer is arbitrary but  fixed in the domain of external communication.
    The time component of the $4$-momentum $p_\mu$ of the photon is set to be positive, so that a photon's path is traced backward from the observer into the photon region, i.e. the region in spacetime where light rays follow spherical orbits determined by $r=const$ around the black hole\cite{PyT_2017}.

    Assuming a continuous and uniform distribution of light sources filling the Universe while excluding the region between the observer and the black hole, we can distinguish between two types of trajectories.
    On the one hand, those trajectories where the radial coordinate increases to infinity reaching a light source after being deflected by the black hole. Here we assign brightness to the initial direction of such trajectories.
    On the other hand, trajectories for along which the radial coordinate decreases until reaching the horizon at $r=r_h$. We will assign darkness to the initial directions of these trajectories, which will determine the shadow of the black hole.
    The shadow boundary corresponds to the light rays between the two types. These decay in a spiral asymptotically approaching unstable spherical orbits in the photon region, where the essential information for determining the shape of the black hole shadow is found. Consequently, the shadow is an image of the photon region and not of the event horizon.  
    
    The dynamics of a light ray traveling through a dispersive medium 
    is given by the  Hamilton-Jacobi Equation (see \cite{Perlick_2021} and reference therein),
    \begin{widetext}
            \begin{equation}
        H\left(x^\mu, \frac{\partial{S}}{\partial{x^\mu}}\right) = \frac{1}{2}\left[g^{\mu\nu}(x^\alpha)\frac{ \partial S}{\partial x^\mu} \frac{\partial S}{\partial x^\nu} + [n^2(x^\alpha) -1](p_\mu V^\mu)^2\right]=0,
        \label{eq:SH_HJ}
    \end{equation}
    \end{widetext}
    where  $V^\mu$ is the $4$-velocity of the plasma medium and $n$ its refractive index.

    As we are considering axially symmetric and stationary spacetimes , we assume a solution of the HJ equation of the form
    \begin{equation}       S(x^\mu)=p_tt+p_\varphi\varphi+S_r(r)+S_\vartheta(\vartheta).
        \label{eq:SH_S}
    \end{equation}    
    
    For a cold non-magnetic and pressuresless plasma, the refractive index of the plasma take the general form
    \begin{equation}
        n^2(x)=1-\frac{\omega_p^2(x)}{\omega^2(x)},
        \label{eq:SH_n}
    \end{equation}
    where $x\equiv \{x^\mu\}$, $\omega(x)$ is the photon frequency and $\omega_p(x)$ the plasma frequency of the medium, given by 
    \begin{equation}
        \omega_p(x)=\frac{4\pi e^2}{m_e}N(x),
        \label{eq:SH_wp1}
    \end{equation}
    where $e$ and $m_e$ are the electron charge and mass respectively, and $N(x)$ is the plasma number density.    
    
    Since plasma is a dispersive medium, the propagation of light will depend on the  frequency of the light $\omega(x)$, which in turn will depend on the $4$-velocity of the plasma $V^\mu(x)$, resulting in
    \begin{equation}
        \label{eq:SH_wk}
        \begin{split}
        \hbar\omega(x) &= p_\mu V^\mu(x), \\
        k^\mu(x)  &= p^\mu+p_\nu V^\nu(x) V^\mu (x).
        \end{split}
    \end{equation}
    where $k^\mu$ is a spacelike vector representing the wave number propagation vector.
    Its dispersion relation is 
    \begin{equation}        \hbar^2\omega^2(x)=k_\mu(x)k^\mu(x)+\hbar^2\omega_p^2(x).
        \label{eq:SH_dis}
    \end{equation}
    As $k^\mu$ is spacelike, we have the following propagation condition
    \begin{equation}
        \omega^2(x)\geq\omega_p^2(x).
        \label{eq:SH_pc}
    \end{equation}
    
    Using \eqref{eq:SH_n} and \eqref{eq:SH_wk}, the general HJ equation can be rewritten as
    \begin{equation}
        H\left(x, \frac{\partial{S}}{\partial{x^\mu}}\right) = \frac{1}{2}\left[g^{\mu\nu}({x})\frac{ \partial S}{\partial x^\mu} \frac{\partial S}{\partial x^\nu} + \hbar^2\omega_p^2(x)\right]=0.
        \label{eq:SH_H}
    \end{equation}
    Using \eqref{eq:KSZ_g} and  \eqref{eq:KSZ_ginv} in \eqref{eq:SH_H}, we find
\begin{widetext}    
    \begin{equation}
        H=\frac{1}{2}\frac{1}{\rho^2}\left[ 
               -\frac{\left[(r^2R_\Sigma+a^2)p_t+ap_\varphi\right]^2}{\Delta}
               +\frac{\Delta}{R_B^2}\left(\frac{\partial S}{\partial r}\right)^2
               +\left(a\sin\vartheta p_t+\frac{1}{\sin\vartheta}p_\varphi\right)^2
               +\left(\frac{\partial S}{\partial \vartheta}\right)^2
               +\rho^2\hbar^2\omega_p^2
               \right]=0.
        \label{eq:SH_H2}
    \end{equation}
    \end{widetext} 
    where we have introduced the constants of motion $p_t=\frac{ \partial S}{\partial t}$ and $p_\varphi=\frac{ \partial S}{\partial \varphi}$, in correspondence with the conserved energy and angular momentum of the photon respectively. The HJ equation, $H=0$, will be separable in the variables $r$ and $\vartheta$ only if the plasma distribution has the form
    \begin{equation}        \omega_p^2(r,\vartheta)=\frac{f_r(r)+f_\vartheta(\vartheta)}{\rho^2(r,\vartheta)}.
        \label{eq:SH_wp}
    \end{equation}
    Hence, with \eqref{eq:SH_wp}, we can guarantee that the equations of motion are completely integrable. We redefine the generalized Carter constant,
    \begin{equation}
    \begin{split}
        K:
        &=\frac{\left((r^2R_\Sigma(r)+a^2)p_t+ap_\varphi\right)^2}{\Delta(r)}-\frac{\Delta(r)}{R_B^2(r)}\left(\frac{\partial S}{\partial r}\right)^2-f_r(r) \\
        &=\left(a\sin\vartheta p_t+\frac{1}{\sin\vartheta}p_\varphi\right)^2+\left(\frac{\partial S}{\partial \vartheta}\right)^2+f_\vartheta(\vartheta),
    \end{split}
        \label{eq:SH_K}
    \end{equation}
    with $\partial S/\partial r=p_r$ being a function only of $r$ and $\partial S/\partial \vartheta=p_\vartheta$ being a function only of $\vartheta$, as can be seen in Eq. \eqref{eq:SH_S}.
    
    We will assume that the plasma is a static, inhomogeneous medium such that $V^t=\sqrt{-g^{tt}}, V^\alpha=0$, resulting in $\hbar\omega=p_t\sqrt{-g^{tt}}$. Since the metric is asymptotically flat, the frequency $\omega_\infty$ (the frequency of the photon observed by a static observer at infinity) satisfy $\hbar\omega_\infty = p_t$. From now on, we take $\hbar=1$.
    The photon impact parameter $b$ is the distance between the line of sight connecting the black hole tothe observer and the photon trajectory, measured at infinity perpendicular to the line of sight. 
    The asymptotic group velocity $n_0$ is the velocity with which the overall envelope shape of the wave's amplitudes propagates through space at infinity \cite{CG_2018}.
    
    Using the recently incorporated constants, we can express the components of the $4$-momentum of the photon as 
    \begin{equation}
        \label{eq:SH_p}
        \begin{split}
        p_t        &=\omega_\infty,  \\
        p_r^2      &={\frac{R_B^2(r)}{\Delta(r)}\left[-K+\frac{\left((r^2R_\Sigma(r)+a^2)p_t+ap_\varphi\right)^2}{\Delta(r)}-f_r(r)\right]},  \\
        p_\vartheta^2 &={K-\left(a\sin\vartheta p_t+\frac{1}{\sin\vartheta}p_\varphi\right)^2-f_\vartheta(\vartheta)},  \\
        p_\varphi     &=\omega_\infty b n_0,
        \end{split}
    \end{equation}
    where the choice of $p_t>0$ and $p_r<0$ is determined by the consideration of light rays coming from the observer toward the black hole (e.g. they are directed toward the observer's past).
    Moreover, the sign of $p_\vartheta$ will be taken positive to determine the upper edge of the shadow and negative for the determination of the lower edge.     
    
    As the observer is located at finite position $(r_O,\vartheta_O)$, the observed frequency $\omega_{obs}$ will not be equal to $\omega_\infty$, but they are related by
    \begin{equation}
        \omega_{obs}=\frac{(r^2R_\Sigma+a^2)\omega_\infty+ap_\varphi}{\rho\sqrt{\Delta}},
        \label{eq:SH_wo}
    \end{equation}
    where the right hand side is evaluated at $(r_O,\vartheta_O)$.
    Insertion of \eqref{eq:SH_p} into the HJ equation \eqref{eq:SH_HJ} yields the following first-order equations of motion for the photon,
    \begin{equation}
        \label{eq:SH_dot}
        \begin{split}
        \dot{t}        &=\frac{1}{\rho^2}\left[\left(a^2(1-y^2)-\frac{(r^2R_\Sigma+a^2)^2}{\Delta}\right)p_t-\frac{arR_M}{\Delta}p_\varphi\right],  \\
        \dot{r}^2      &=\frac{1}{\rho^4}\frac{{-K\Delta+((r^2R_\Sigma+a^2)p_t+ap_\varphi)^2-\Delta f_r}}{R_B^2},  \\
        \dot{\vartheta}^2 &=\frac{1}{\rho^4}\left[K-\left(a\sqrt{1-y^2}p_t+\frac{1}{\sqrt{1-y^2}}p_\varphi\right)^2-f_\vartheta\right],  \\
        \dot{\varphi}     &=\frac{1}{\rho^2}\left[\left(\frac{1}{1-y^2}-\frac{a^2}{\Delta}\right)p_\varphi-\frac{arR_M}{\Delta}p_t\right],
        \end{split}
    \end{equation}
    where we introduced the dot notation $dx^\mu/ds = \dot x^a$ for the directional tangent vectors along the curves ${x^\mu(s) = \left(t(s),r(s),\vartheta(s),\varphi(s)\right)}$ parameterised by $s$.
    To find the photon region, we must solve $\dot{r}=0$ and $\ddot{r}=0$. These conditions lead to the following pair of equations
    \begin{equation}
        \label{eq:SH_rf}
        \begin{split}
        &0=-(K+f_r)\Delta+\left((r^2R_\Sigma+a^2)p_t+ap_\varphi\right)^2,  \\
        &0=-(K+f_r)\Delta'-f_r'\Delta\\&+2\left(2rR_\Sigma+r^2R_\Sigma'\right)p_t\left((r^2R_\Sigma+a^2)p_t+ap_\varphi\right),
        \end{split}
    \end{equation}
    from which we can deduce the following expressions for the constants of motion
    \begin{widetext} 
    \begin{equation}
        K=\Delta\left(\frac{\left(2rR_\Sigma+r^2R_\Sigma'\right)p_t}{\Delta'}\right)^2\left[1\pm\sqrt{1-\left(\frac{\Delta'}{\left(2rR_\Sigma+r^2R_\Sigma'\right)p_t}\right)^2\frac{f_r'}{\Delta'}}\right]^2-f_r,
        \label{eq:SH_Kr}
    \end{equation}
    \begin{equation}
        p_\varphi=\frac{\Delta}{a}\left(\frac{\left(2rR_\Sigma+r^2R_\Sigma'\right)p_t}{\Delta'}\right)\left[1\pm\sqrt{1-\left(\frac{\Delta'}{\left(2rR_\Sigma+r^2R_\Sigma'\right)p_t}\right)^2\frac{f_r'}{\Delta'}}\right]-\left(\frac{r^2R_\Sigma}{a}+a\right)p_t,
        \label{eq:SH_pfr}
    \end{equation}
    \end{widetext} 
    where the choice of sign will depend on the distribution and density of the plasma. For a wide variety of distributions, including those to be discussed in this paper, these equations have a solution only if the $+$ sign is taken \cite{PyT_2017}.
   
    The boundary of the photon region can be deduced by requiring $\dot{\vartheta}$ to be a real non-negative number in \eqref{eq:SH_dot}, which gives
    \begin{equation}
        {(1-y^2)(K-f_\vartheta)}=\left(p_\varphi+a(1-y^2)p_t\right)^2. 
        \label{eq:SH_rlim}
    \end{equation}    
    To construct the shadow, we propose the following orthogonal tetrad
    \begin{equation}
        \label{eq:SH_e}
        \begin{split}
        e_0 &=\frac{r^2R_\Sigma+a^2}{\rho\sqrt{\Delta}}\partial_t+\frac{a}{\rho\sqrt{\Delta}}\partial_\varphi, \\
        e_1 &=\frac{1}{\rho}\partial_\vartheta, \\
        e_2 &=-\frac{1}{\rho\sin\vartheta}\partial_\varphi-\frac{a\sin\vartheta}{\rho}\partial_t, \\
        e_3 &=-\frac{\sqrt{\Delta}}{\rho R_B}\partial_r.
        \end{split}
    \end{equation}
    {This tetrad has the property of reducing to the tetrad used by Perlick and Tsupko in \cite{PyT_2017} when considering the Kerr spacetime (also known as the Carter tetrad\cite{PhysRev.174.1559}).} Specifically, in this spacetime, the vectors $e_0\pm e_3$ reduce to the principal null directions of the Kerr spacetime. An observer $\mathcal{O}$ with associated  tetrad \eqref{eq:SH_e} will be called a standard observer.
    Another natural choice would be to choose ZAMO observers (zero angular momentum observers).
    It is clear that these vectors must be evaluated at the observer's position $(r_O,\vartheta_O)$. 
    Being $\lambda(r(s),\vartheta(s),\varphi(s),t(s))$ the photon trajectory, the corresponding tangent vector $\dot\lambda(s)$ can be found at the observer's position from
    \begin{equation}
         \dot{\lambda}\equiv
         \frac{d}{ds} =
         \dot{r}\partial_r+\dot{\vartheta }\partial_\vartheta+\dot{\varphi}\partial_\varphi+\dot{t}\partial_t,
        \label{eq:SH_l1}
    \end{equation}
    Moreover, the tangent vector can be written as
    \begin{equation}
        \dot{\lambda}=-\alpha e_0 + \beta\left(\sin\Theta\cos\Phi e_1+\sin\Theta\sin\Phi e_2+\cos\Theta e_3\right),
        \label{eq:SH_l2}
    \end{equation}
    where $\alpha$ and $\beta$ are positive factors, $\Theta$ is the colatitude and $\Phi$ is the azimuth angle in the observer's sky. The direction given by $\Theta=0$ will then be the direction toward the black hole, while $\Theta=\pi$ will be the opposite direction. 

    From \eqref{eq:SH_HJ} we have  $g(\dot{\lambda},\dot{\lambda})=-\omega_p^2$, implying 
    \begin{equation}
        \alpha^2-\beta^2=\omega_p^2(r_O,\vartheta_O).
        \label{eq:SH_abw}
    \end{equation}
    There also is $\alpha=g(\dot{\lambda},e_0)$, so that
    \begin{equation}
        \label{eq:alpha}
        \begin{split}
        \alpha &=\frac{(r^2R_\Sigma+a^2)p_t+ap_\varphi}{\rho\sqrt{\Delta}}, \\
        \beta &=\sqrt{\frac{((r^2R_\Sigma+a^2)p_t+ap_\varphi)^2}{\rho^2\Delta}-\omega_p^2},
        \end{split}
    \end{equation}
    where $\alpha$ and  $\beta$ should be evaluated using at the observer cordinates $(r_O,\vartheta_O)$. 
    
    Comparing the tangent vectors in \eqref{eq:SH_l1} and \eqref{eq:SH_l2}, we can parameterize the shadow edge in the coordinates seen by the observer as follows
    \begin{widetext} 
    \begin{equation}
        \label{eq:SH_STF}
        \begin{split}
        \sin\Theta &=\sqrt{\frac{\Delta(r_O)(K(r_p)-f_\vartheta(\vartheta_O))}{((r_O^2R_\Sigma(r_O)+a^2)p_t+ap_\varphi(r_p))^2-\Delta(r_O)(f_r(r_O)+f_\vartheta(\vartheta_O))}}, \\
        \sin\Phi   &=-\frac{p_\varphi(r_p)+ap_t\sin^2\vartheta_O}{\sin\vartheta_O\sqrt{K(r_p)-f_\vartheta(\vartheta_O)}}, 
        \end{split}
    \end{equation}
    \end{widetext} 
    where $r_p$ is a parameter that will allow us to describe the shadow contour curve.  Its upper and lower bounds are the solutions of \eqref{eq:SH_rlim} that are evaluated at $\vartheta_O$.
    
    Trajectories corresponding to the shadow boundary will asymptotically approximate circular orbits of the photon region. These trajectories are in fact in-falling spirals, for which  their constants of motion $K$ and $p_\varphi$  have the same values as  the trajectories inhabiting at the boundary spheres in the photon region. 
    If $a=0$, a metric with spherical symmetry is obtain and there is no need to parameterize the shadow.
    For $a\neq0$
    we can evaluate the constants $K(r_p)$ and $p_{\varphi}(r_p)$ according to  \eqref{eq:SH_Kr} and \eqref{eq:SH_pfr} and employ them in \eqref{eq:SH_STF} to obtain the shadow boundary. The parameter $r_p$ ranges from a minimum value $r_{p,min}$ to a maximum value $r_{p,max}$, these being determined by the equation $\sin^2\Phi=1$, which corresponds to Eq. \eqref{eq:SH_rlim}. 
    
    By taking the values of $r_p$ in the interval $[r_{p,min},r_{p,max}]$, $\Theta$ and $\Phi$ will describe with \eqref{eq:SH_STF} the shape and angular size of the black hole shadow for an observer with a $4$-velocity $e_0$ at $(r_O,\vartheta_O)$. If the observer moves at a different $4$-velocity, the image will be distorted by aberration. Following \cite{GPL_2014}, we will use stereographic projections on a plane tangent to the celestial sphere at the pole $\Theta=0$ while charting this plane with the dimensionless Cartesian coordinates
    \begin{equation}
        \label{eq:SH_XY}
        \begin{split}
        X(r_p) &=-2\tan\left[\frac{\Theta(r_p)}{2}\right]\sin\left[\Phi(r_p)\right], \\
        Y(r_p) &=-2\tan\left[\frac{\Theta(r_p)}{2}\right]\cos\left[\Phi(r_p)\right].
        \end{split}
    \end{equation}
    These coordinates are directly related to angular measurements in the observer's sky.

\subsection{Setting of the parameters of the metrics}
\label{c3s3ss2}

   {In this subsection we present the notation for the parameters characterizing the different metrics and the observer, setting their values for the rest of this work. The shadow is influenced by the spin parameter $a$, the distance to the observer $r_O$, the tilt angle between the observer and the rotation axis $\vartheta_O$} and  characteristic charges $\{Q,\eta,b, Q^*,r_0\}$ associated to each one of the metrics. Note that some of these charges not have the same physical origin nor the same units. However, for each of the proposed metrics with a given mass $m$, we can associate a single{dimensionless} ``charge" parameter $Q_{(p)}$, whose value encodes their influence on the shape and size of the shadow.  These are related to the electrical charge $Q$ in Kerr-Newman metric, the deviation $\eta$ in modified Kerr metric, the ratio $b$ in Kerr-Sen metric, the tidal charge $Q^*$ in braneworld case and the ratio $r_0$ in Dilaton case (in this last metric, in order to have only one parameter, we will take for simplicity the $q$ parameter as defined in Sec.\eqref{c3s2ss4} as  $q=-2mr_0$, note also that for $Q^2_M<Q^2_E$, $r_0<0$, more precisely, for our choice we are taking $Q^2_M=\frac{-r_0}{4m-r_0}Q^2_E$). From now on, we will refer to all of them just by $Q_{(p)}$ corresponding to the dimensionless parameters associated to each of the metrics defined as follows 
   {
    \begin{equation}
        Q_{(p)} \equiv
        \begin{cases}
        \frac{Q}{m},      & \text{for Kerr-Newmann}, \\
        \frac{\eta}{m^3}, & \text{for Modified Kerr}, \\
        \frac{b}{m},      & \text{for Kerr-Sen}, \\
        \frac{Q^*}{m},    & \text{for Braneworld}, \\
        -\frac{r_0}{m},   & \text{for Dilaton}, \\
        \end{cases}
        \label{eq:SH_Qp}
    \end{equation} 
    }

     Here, we use different values of $Q_{(p)}$ for  an exploration of the metric models and their qualities.

    {In most cases there is a value $Q_{(p)max}$ which is an upper bound for the allowed values for $Q_{(p)}$ in each metric. For metrics with no such a bound (i.e. the modified Kerr and Braneworld case), as the associated $Q_{(p)}$ parameter is expected to be relatively small, we propose an artificial bound. Table \ref{tab:Qamax} shows the values for $Q_{(p)max}$ that we will use for each metric. 
    In turn, from the expression in Table \ref{tab:Metrics}, we can find the upper limit of the angular momentum for each metric{(for a given $Q_{(p)}$)}, also expressed in Table \ref{tab:Qamax} . Again, not all models impose an upper bound for the angular momentum, hence for those cases the choice of $a_{max}$ is free. To standardize the results, the magnitudes $Q_{(p)}$ and $a$ will be expressed in terms of $Q_{(p)max}$ and $a_{max}$ respectively.}
    
    \begin{table*}[htbp]
        \centering
        \begin{tabular}{|c|c|c|c|c|c|}
        \hline 
        Metric      & Kerr-Newman        & Modified Kerr   & Kerr-Sen   & Braneworld         & Dilaton        \\ \hline     
        $Q_{(p)max}$ & $1$                & $0.2$           & $1$        & $1$                & $0.5$          \\ \hline 
        $a_{max}/m$ & $\sqrt{1-Q_{(p)}^2}$ & $1$             & $1-Q_{(p)}$    & $\sqrt{1+Q_{(p)}^2}$ & $\sqrt{1-2Q_{(p)}}$  \\ \hline 
        \end{tabular}
        \caption{Maximum values for the characteristic parameter $Q_{(p)max}$ and the angular momentum $a_{max}$ according to the metric.}
        \label{tab:Qamax}
    \end{table*}

\section{The Shadow in Plasma-filled Environments (without aberration effects)}
\label{c3s3ss3}
\subsection{General considerations of shadows in plasma free environments}\label{sec:sinplasma}
As a warm-up, and only for completeness, we will briefly review how different parameters of various metrics affect the shape of the black hole's shadow when plasma effects are neglected.
    
\begin{figure*}[htbp]
        \centering
        \begin{tabular}{ccccc}
            Kerr-Newman & Modified Kerr & Kerr-Sen & Braneworld & Dilaton \\
            {\includegraphics[scale=0.30,trim=0 0 0 0]{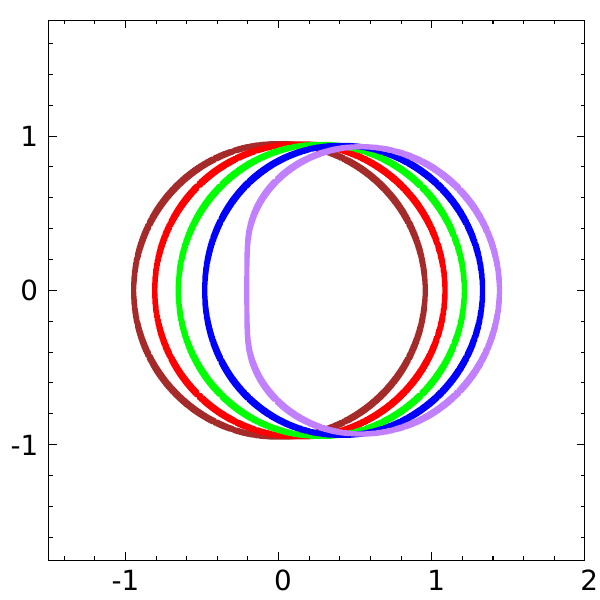}} &
            {\includegraphics[scale=0.30,trim=0 0 0 0]{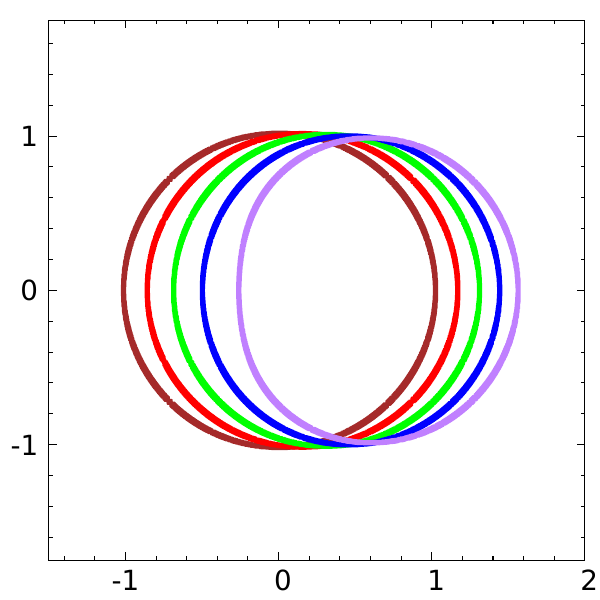}} &
            {\includegraphics[scale=0.30,trim=0 0 0 0]{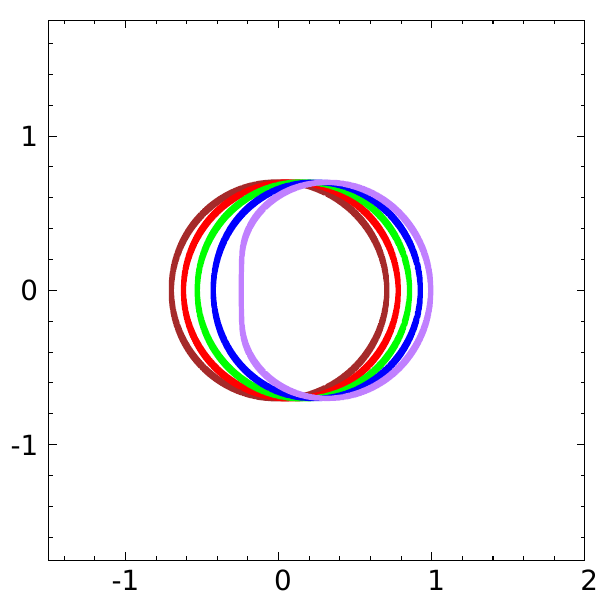}} &
            {\includegraphics[scale=0.30,trim=0 0 0 0]{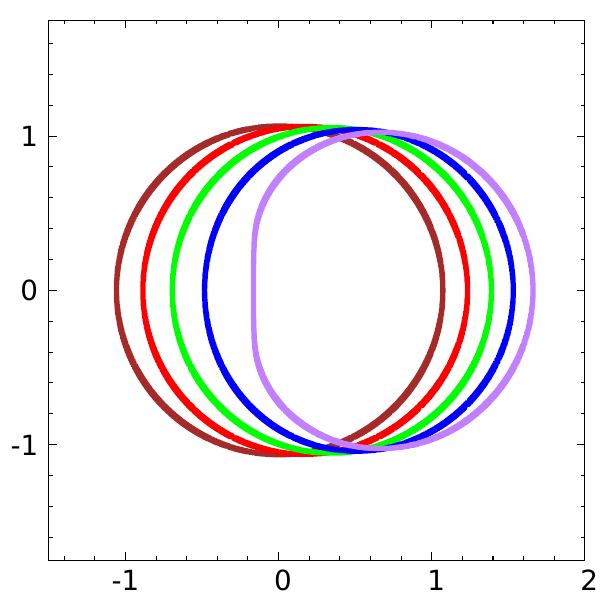}} &
            {\includegraphics[scale=0.30,trim=0 0 0 0]{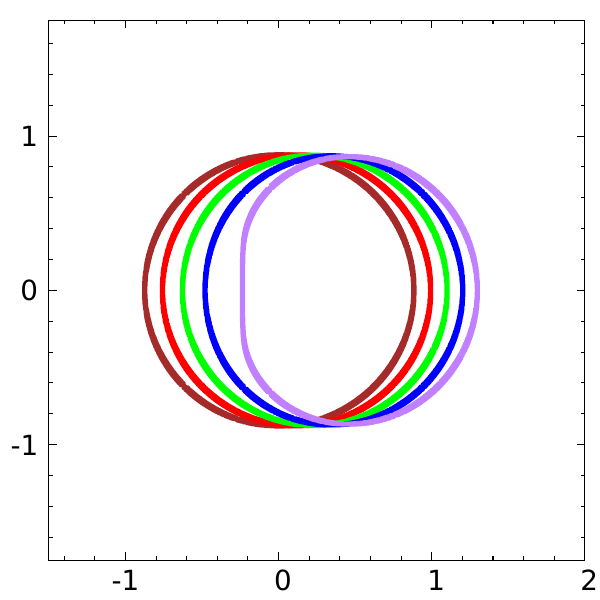}} \\
            {\includegraphics[scale=0.30,trim=0 0 0 0]{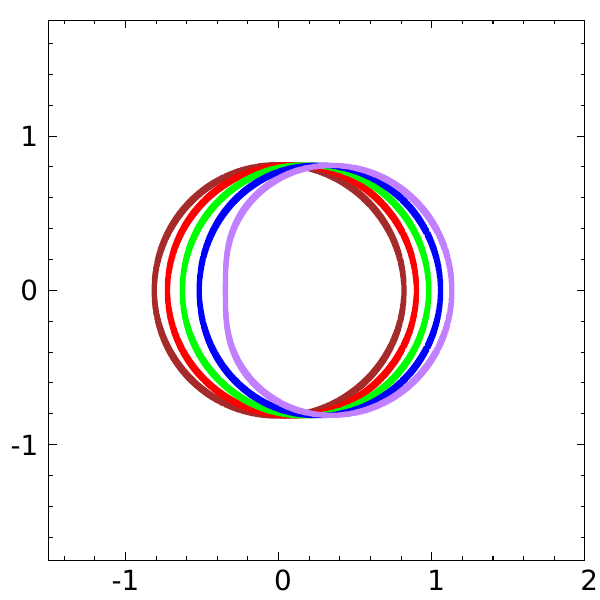}} &
            {\includegraphics[scale=0.30,trim=0 0 0 0]{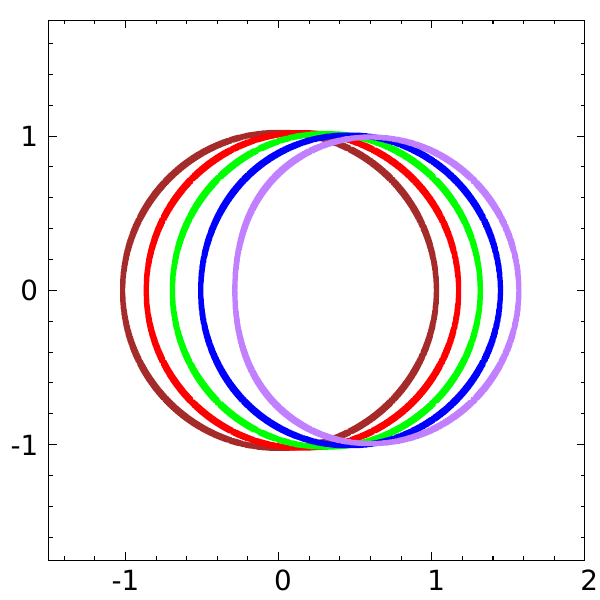}} &
            {\includegraphics[scale=0.30,trim=0 0 0 0]{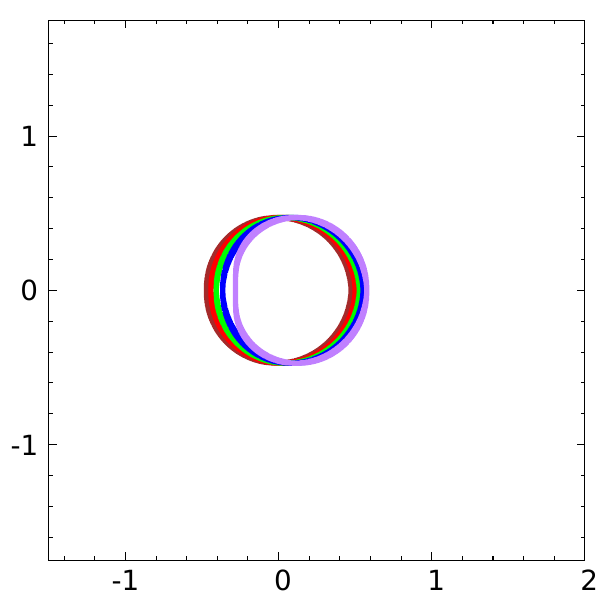}} &
            {\includegraphics[scale=0.30,trim=0 0 0 0]{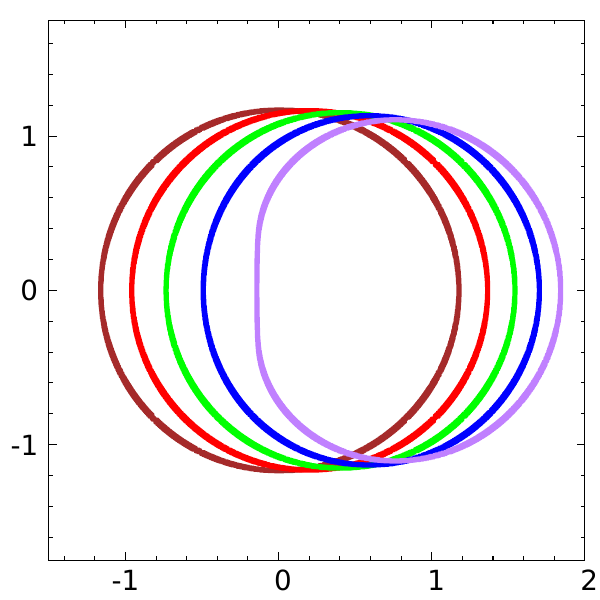}} &
            {\includegraphics[scale=0.30,trim=0 0 0 0]{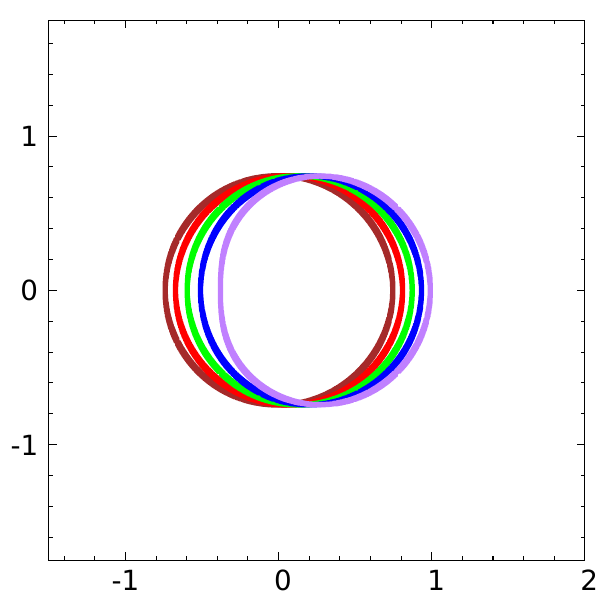}} \\
        \end{tabular}
        \caption{Effect of the angular momentum on the shadows. In each curve $a/a_{max}$ is equal to $0.01$ (Brown),  $0.25$ (red), $0.50$ (green), $0.75$ (blue) and $0.999$ (purple). $r_O=5$, $\vartheta_O=\pi/2$, top $Q_{(p)}/Q_{(p)max}=\sqrt{0.25}$, bottom $Q_{(p)}/Q_{(p)max}=\sqrt{0.75}$.
        }
        \label{fig:SH_a}
    \end{figure*}

\begin{figure*}[htbp]
        \centering
        \begin{tabular}{ccccc}
            Kerr-Newman & Modified Kerr  & Kerr-Sen & Braneworld & Dilaton \\
            {\includegraphics[scale=0.30,trim=0 0 0 0]{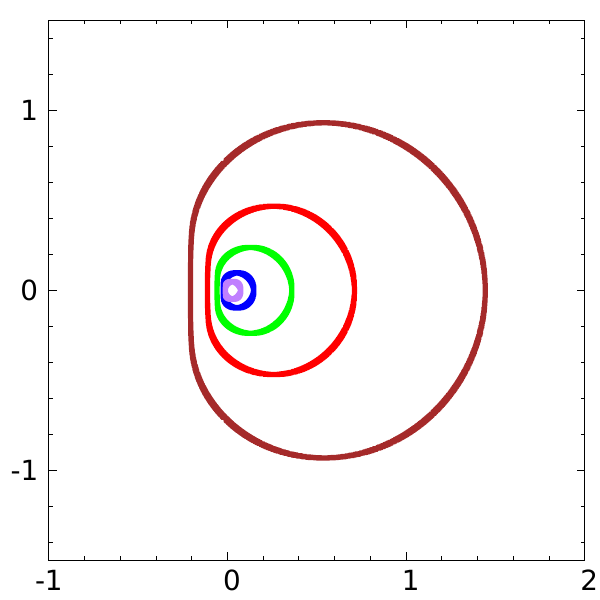}} &
            {\includegraphics[scale=0.30,trim=0 0 0 0]{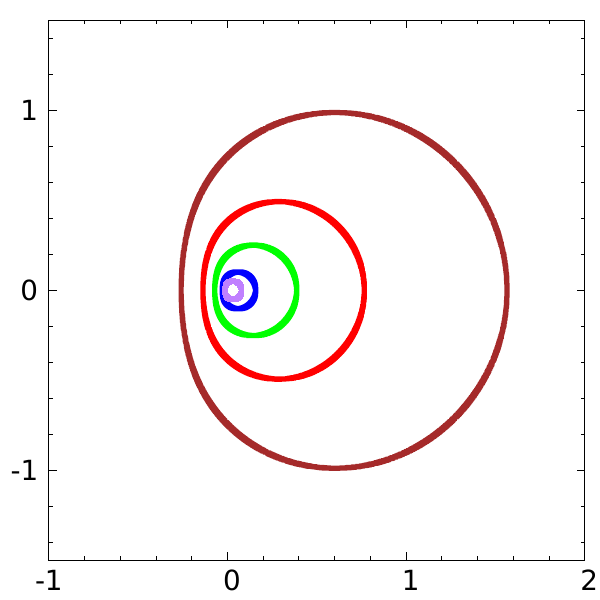}} &
            {\includegraphics[scale=0.30,trim=0 0 0 0]{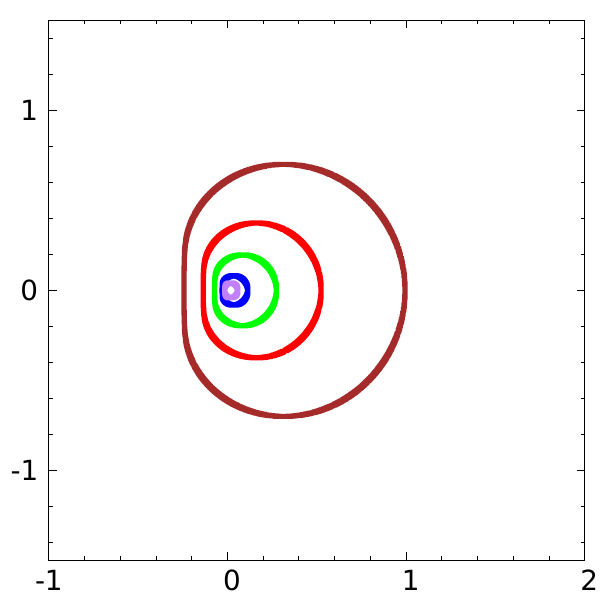}} &
            {\includegraphics[scale=0.30,trim=0 0 0 0]{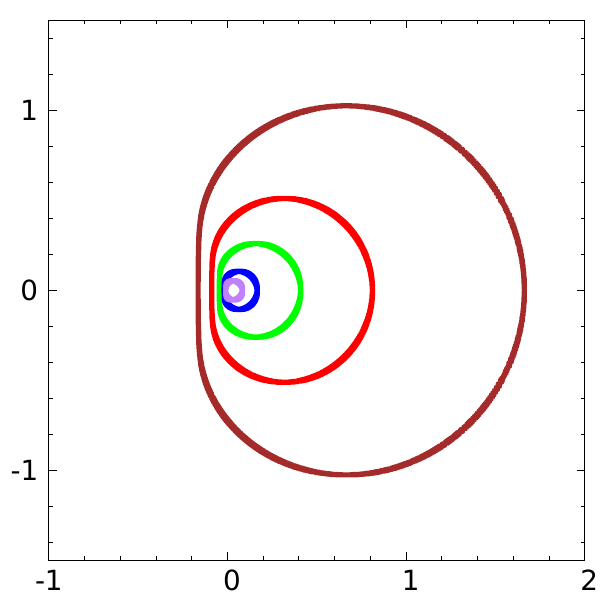}} &
            {\includegraphics[scale=0.30,trim=0 0 0 0]{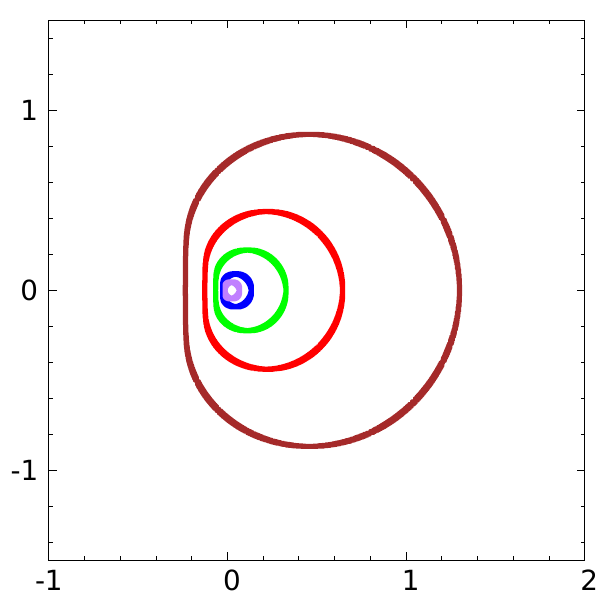}} \\
            {\includegraphics[scale=0.30,trim=0 0 0 0]{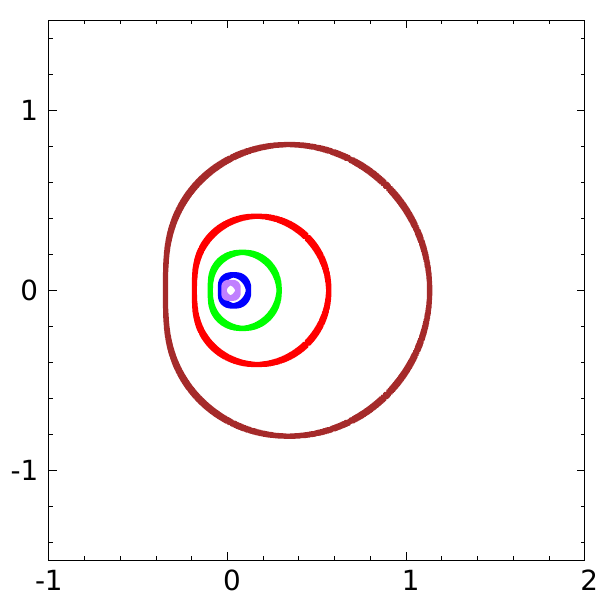}} &
            {\includegraphics[scale=0.30,trim=0 0 0 0]{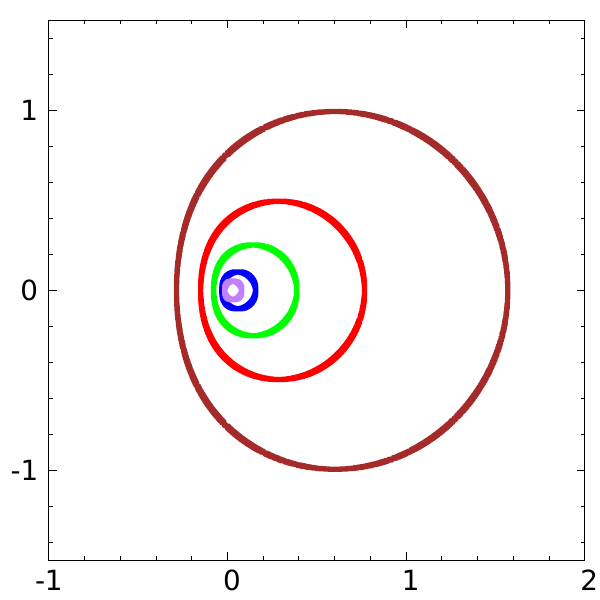}} &
            {\includegraphics[scale=0.30,trim=0 0 0 0]{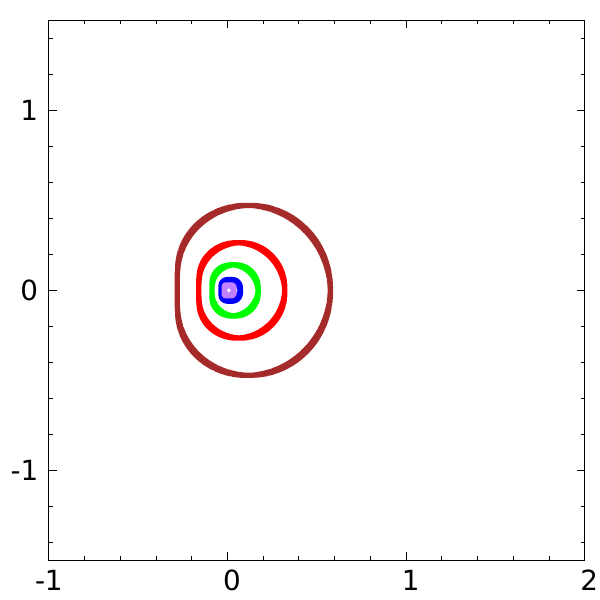}} &
            {\includegraphics[scale=0.30,trim=0 0 0 0]{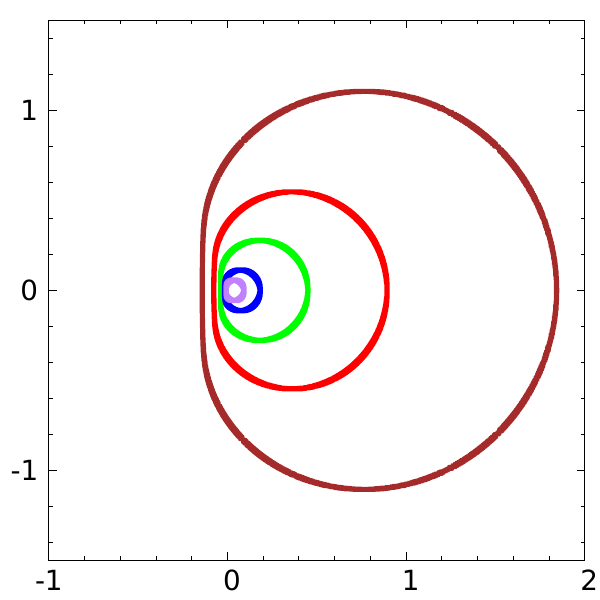}} &
            {\includegraphics[scale=0.30,trim=0 0 0 0]{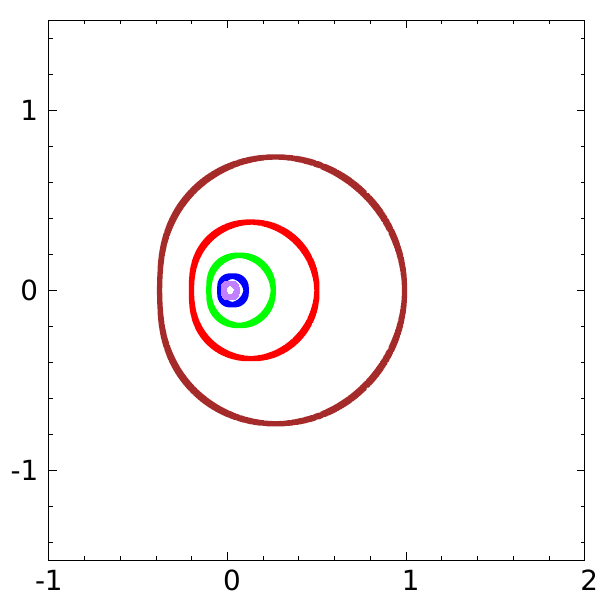}} \\
        \end{tabular}
        \caption{Distance to the observer $r_O$. Brown, red, green, blue and purple curves correspond respectively to $r_O=5m$, $10m$, $20m$, $50m$ and $100m$. $a/a_{max}=0.999$, $\vartheta_O=\pi/2$, top $Q_{(p)}/Q_{(p)max}=\sqrt{0.25}$, bottom $Q_{(p)}/Q_{(p)max}=\sqrt{0.75}$.}
        \label{fig:SH_r}
    \end{figure*}

\begin{figure*}[htbp]
        \centering
        \begin{tabular}{ccccc}
            Kerr-Newman & Modified Kerr & Kerr-Sen & Braneworld & Dilaton \\
            {\includegraphics[scale=0.30,trim=0 0 0 0]{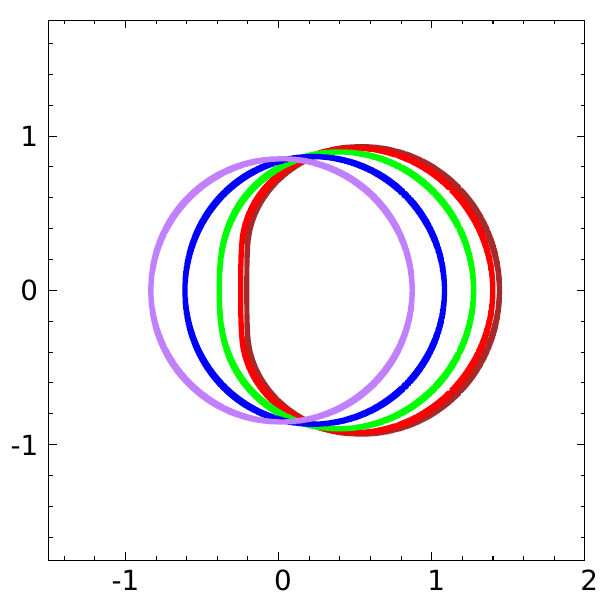}} &
            {\includegraphics[scale=0.30,trim=0 0 0 0]{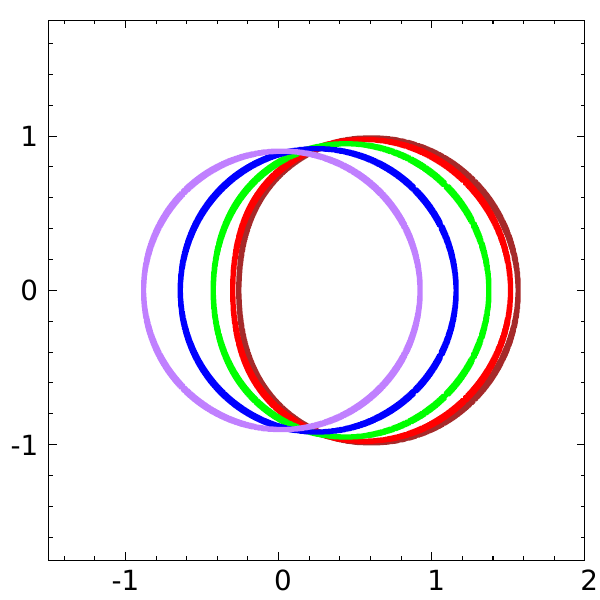}} &
            {\includegraphics[scale=0.30,trim=0 0 0 0]{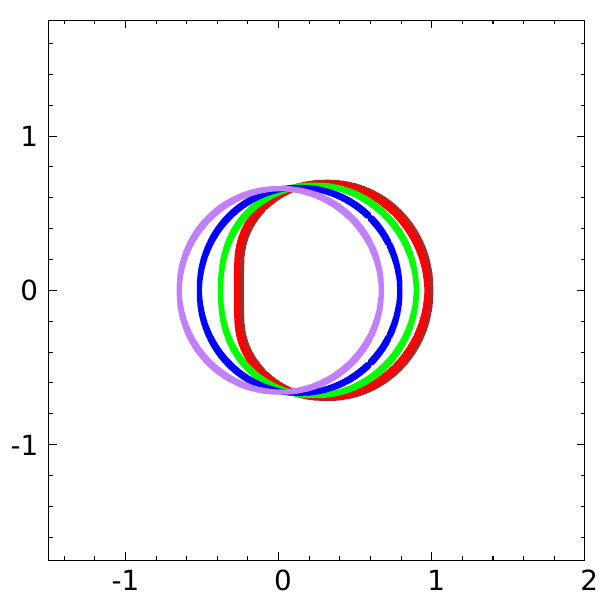}} &
            {\includegraphics[scale=0.30,trim=0 0 0 0]{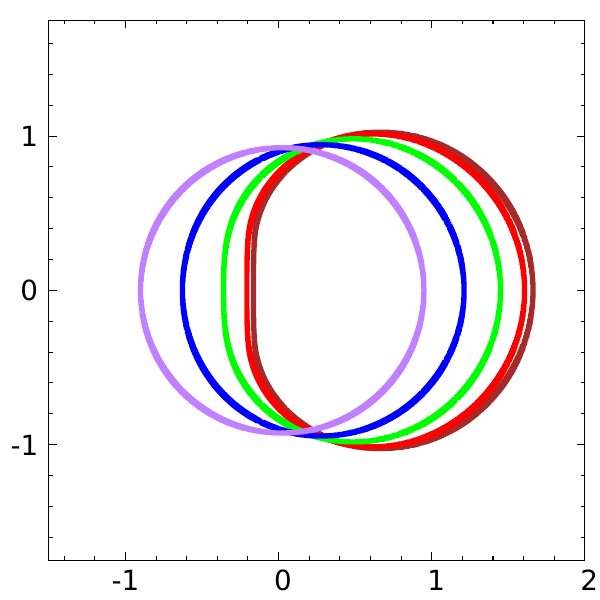}} &
            {\includegraphics[scale=0.30,trim=0 0 0 0]{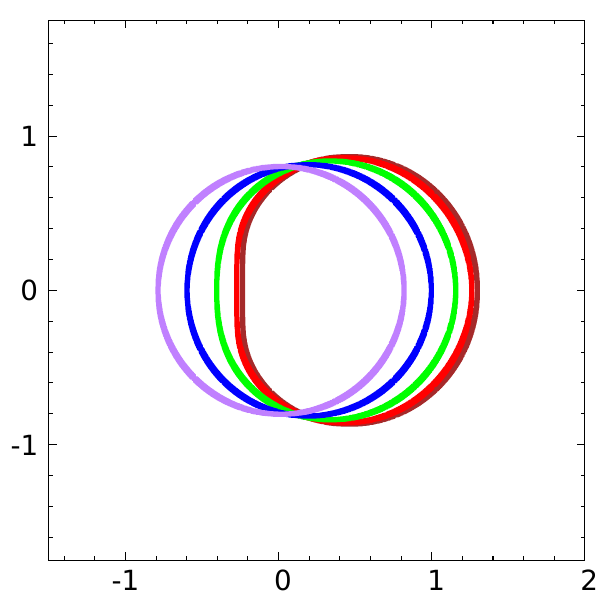}} \\
            {\includegraphics[scale=0.30,trim=0 0 0 0]{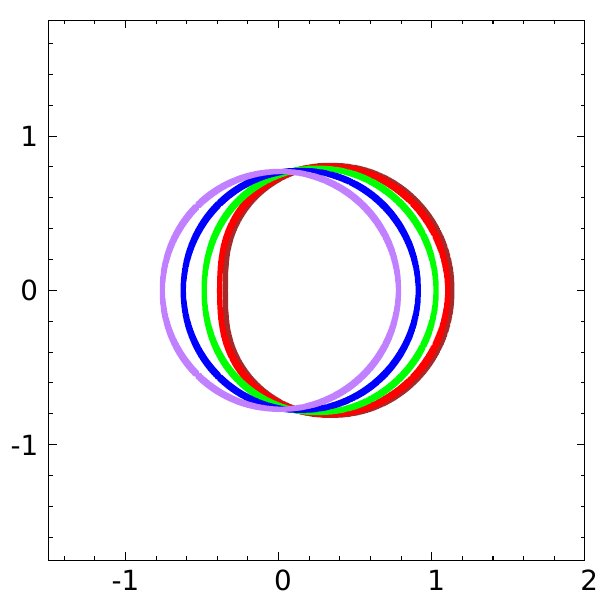}} &
            {\includegraphics[scale=0.30,trim=0 0 0 0]{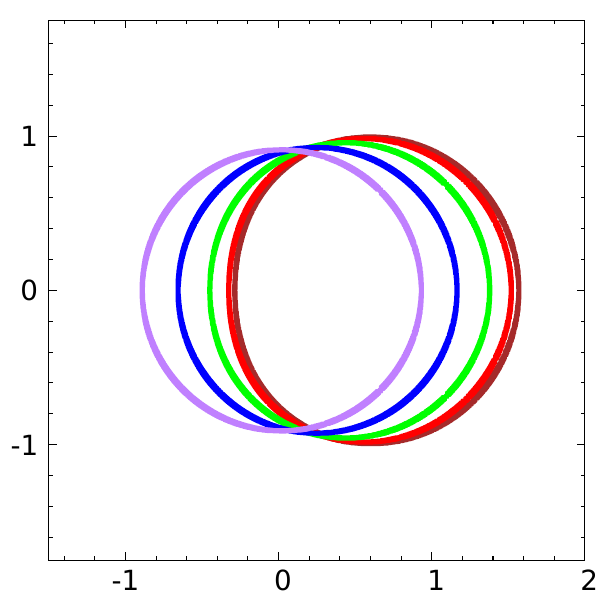}} &
            {\includegraphics[scale=0.30,trim=0 0 0 0]{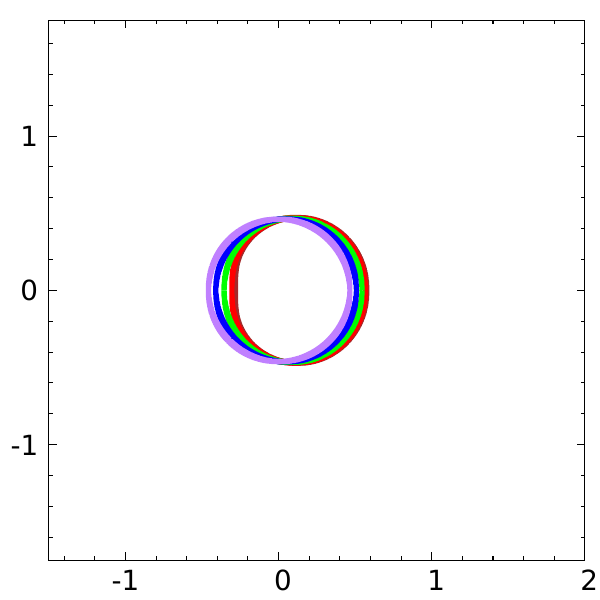}} &
            {\includegraphics[scale=0.30,trim=0 0 0 0]{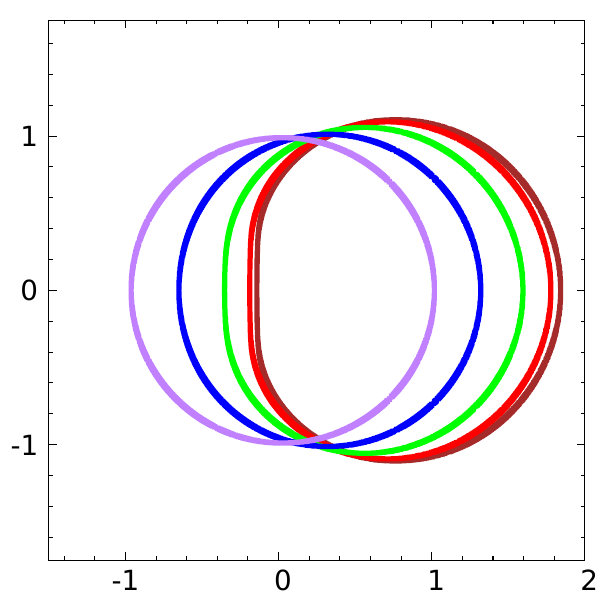}} &
            {\includegraphics[scale=0.30,trim=0 0 0 0]{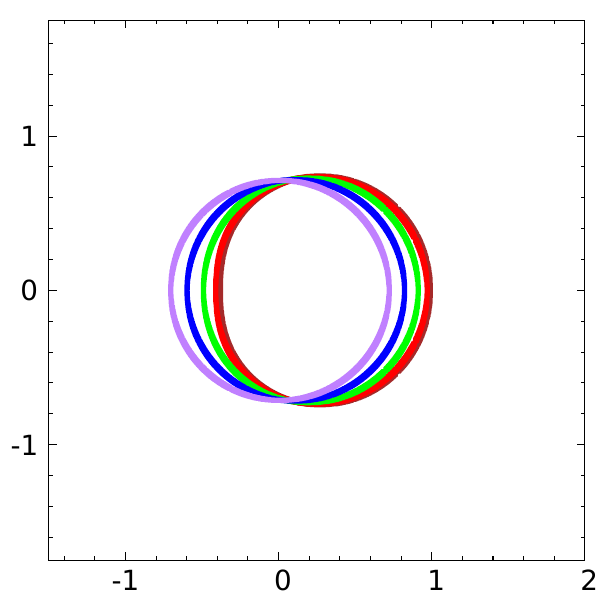}} \\
        \end{tabular}
        \caption{
      {Shadows for different angular positions $\vartheta_O$ of the observer}. Brown, red, green, blue and purple curves correspond  to  $\vartheta_O\in\{\pi/2,3\pi/8,\pi/4,\pi/8,\pi/100\}$.{Here,} $a/a_{max}=0.999$, $r_O=5m$.{In the top panel} $Q_{(p)}/Q_{(p)max}=\sqrt{0.25}$,{ while} in the bottom $Q_{(p)}/Q_{(p)max}=\sqrt{0.75}$.
        }
        \label{fig:SH_t}
    \end{figure*}

    Figs.\eqref{fig:SH_a}, \eqref{fig:SH_r} and \eqref{fig:SH_t} show respectively how the spin parameter $a$ of the black hole, the distance $r_O$ between the observer and the black hole and the angle $\vartheta_O$ between the observer's position and axial symmetry axis of the black hole influence the shape and size of the shadow. Moreover, the graphs are repeated for different values of the $Q_{(p)}$ parameter, allowing us to study their influence in the different spacetime models. In each table, the top row takes $Q_{(p)}/Q_{(p)max}=\sqrt{0.25}$ while the bottom row takes $Q_{(p)}/Q_{(p)max}=\sqrt{0.75}$. 
    Although a physically realistic observer is expected to be far from the event horizon, we will generally take $r_O=5m$ in order to better relate our results with those presented in \cite{PyT_2017}, where $r_O=5m$ is employed.

    Fig.~\eqref{fig:SH_a} shows a comparison of the shadows cast by black holes with different spin values $a$, from near zero spin ($a/a_{max}=0.01$) to almost the maximum spin ($a/a_{max}=0.999$).
    The angular momentum affects the black hole shadow in three characteristic ways. 
    The first is the reduction in size. The larger the angular momentum, the smaller the area the shadow occupies in the observer's sky, and the smaller its horizontal and vertical diameters. 
    The second modification affects the shape of the shadow. While for black holes with low $a$ the shadow is nearly circular, it takes a $D$ shape as $a$ increases, being more noticeable as we approach $a_{max}$. This phenomenon happens in the same way in all metrics, with the flattening side being parallel to the black hole rotation axis (i.e., vertical) and coming, as expected, from photon orbits co-rotating with the black hole (i.e. from the left).
    The third effect is the displacement of the shadow position. The direction of the displacement is opposite to the flattened side (i.e. to the right). Again, this effect is practically negligible for small $a$ values and increases with this. 
    
    In Fig.\eqref{fig:SH_r} we see how the apparent size of the shadow decreases with the distance between the black hole and the observer. However, this distance does not seem to produce noticeable differences in the shape of the shadow.
    
    Fig.\eqref{fig:SH_t} shows how the angle $\vartheta_O$ between the observer's position and the black hole's rotation axis changes the shape, size and position of the shadow.
    Observers near the rotation axis ($\vartheta_O\rightarrow0$) will see a smaller, rounder and centered shadow. Observers near the equatorial plane ($\vartheta_O\rightarrow\pi/2$) will see a displaced shadow with a larger vertical diameter and $D$-shaped. 
    
\begin{figure*}[htbp]
        \centering
        \begin{tabular}{cccc}
            $Q_{(p)}/Q_{(p)max}=\sqrt{0.250}$ & $Q_{(p)}/Q_{(p)max}=\sqrt{0.500}$ & $Q_{(p)}/Q_{(p)max}=\sqrt{0.750}$ & $Q_{(p)}/Q_{(p)max}=\sqrt{0.999}$ \\
            {\includegraphics[scale=0.38,trim=0 0 0 0]{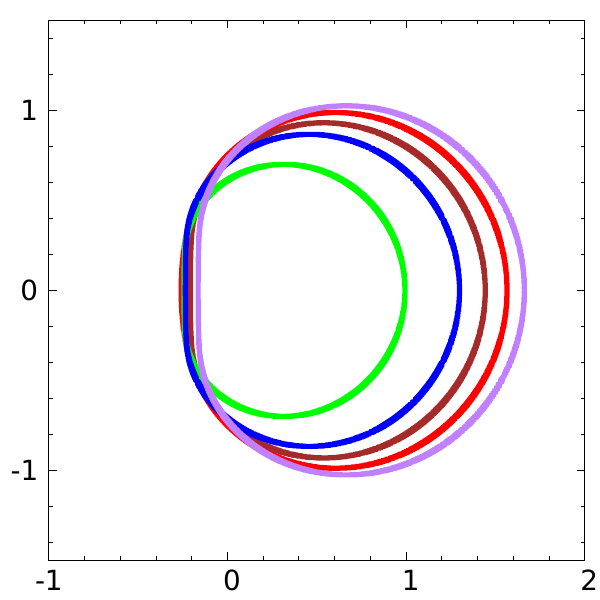}} &
            {\includegraphics[scale=0.38,trim=0 0 0 0]{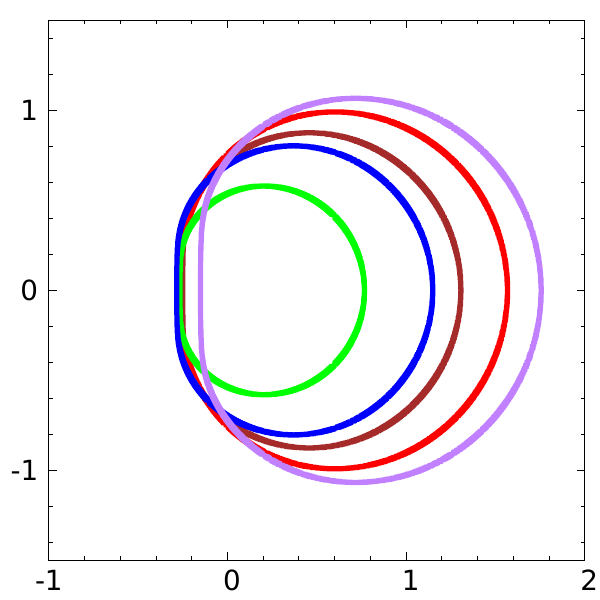}} &
            {\includegraphics[scale=0.38,trim=0 0 0 0]{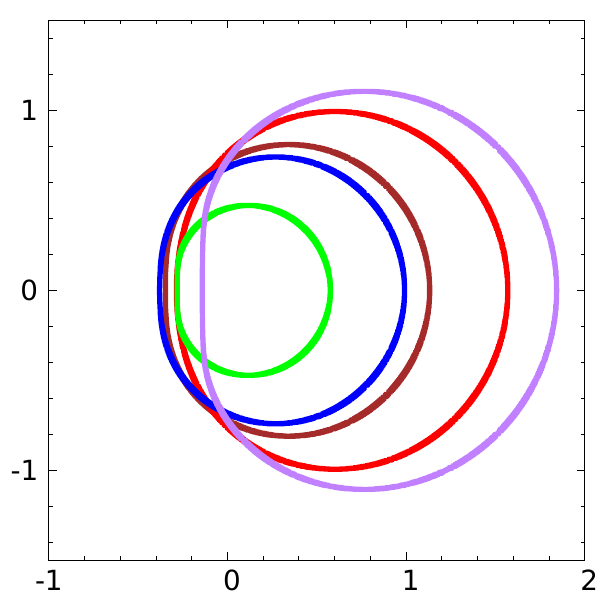}} &
            {\includegraphics[scale=0.38,trim=0 0 0 0]{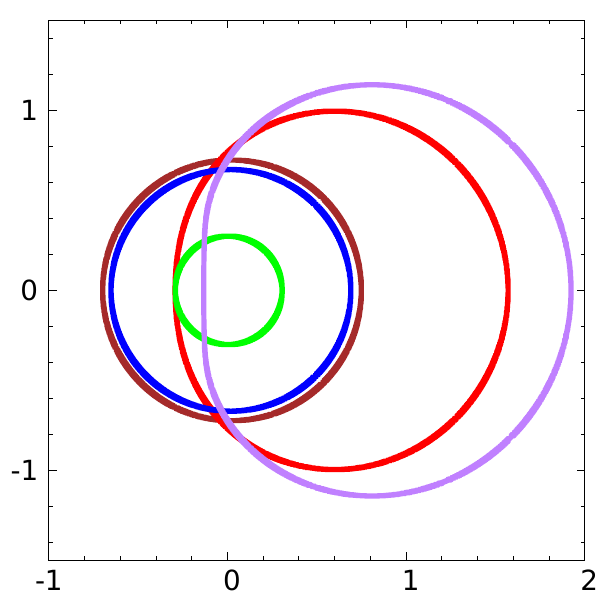}} 
        \end{tabular}
        \caption{Characteristic parameter $Q_{(p)}$. Brown, red, green, blue and purple curves correspond respectively to the Kerr-Newman, modified Kerr, Kerr-Sen, Dilaton and Braneworld metrics. $a/a_{max}=0.999$, $r_O=5m$, and $\vartheta_O=\pi/2$.
       The characteristic parameter reduces the $D$-shape, the size and the displacement of the shadow in Kerr-Newman, Kerr-Sen and Dilaton, and increases them in modified Kerr and Braneworld.
        }
        \label{fig:SH_m}
    \end{figure*}

    Fig.\eqref{fig:SH_m} compares the shadows observed in the different metrics, using the same values for $a$, $r_O$ and $\vartheta_O$, exploring different values for the $Q_{(p)}$ parameter. Here we can see how different spacetime models affect the shape of the shadow, and what role $Q_{(p)}$ plays in each of them. The different metric models converge to Kerr spacetime as $Q_{(p)}\rightarrow0$. As  $Q_{(p)}$ increases, however, the shadows evolve in completely different ways, according to the different metric properties. 
    
    Metrics such as the Kerr-Newman, Kerr-Sen and Dilaton present nearly spherical shadows for $Q_{(p)}\rightarrow Q_{(p)max}$. As can be seen in Table \ref{tab:Qamax}, the angular momentum $a$ vanishes as $Q_{(p)}$ increases, so that the spin effects becomes negligible. In these metrics, the effects on the shadow associated with angular momentum, such as the increase in size, the shape of $D$ and the horizontal displacement will be reduced as the parameter $Q_{(p)}$ approaches $Q_{(p)max}$.     
    Something particular occurs with the Kerr-Sen metric.
    As in the Kerr-Newman and Dilaton cases, the larger the $Q_{(p)}$ the lower the allowed $a$. However, the reduction in size and the loosing of the $D$ shape as $Q_{(p)}$ increases is much more noticeable in the Kerr-Sen spacetime. This is due to the properties of the characteristic parameter and how it affects the metrics.
    On the other hand, in metrics such as modified Kerr and Braneworld where the value of $a$ is not so constrained by $Q_{(p)}$, the effects associated with the angular momentum still present even for high $Q_{(p)}$ values. Moreover, in the Braneworld case such effects can even be increased, since a larger $Q_{(p)}$ allows to increase the maximum spin parameter $a_{max}$, so the effects on the shadow are even more evident. 
    Thus, we see that in general the parameter $Q_{(p)}$ exaggerates or attenuates the effects of angular momentum.

\subsection{Shadow in plasma media}\label{sec:plasma}
    In this subsection we will analyze different plasma distributions and how they affect the formation of the black hole shadow. To guarantee the separability of the HJ equations, these distributions must satisfy Eq. \eqref{eq:SH_wp}, so our work here will be to propose the functions $f_r(r)$ and $f_\vartheta(\vartheta)$. This will be done by mainly considering the plasma distributions presented in \cite{PyT_2017} and \cite{Rogers_2015}, as well as proposing some new ones.

    The first example is a plasma distribution with $f_r'(r)=0$. Since this leaves $f_r(r)$ constant, it can be absorbed by the function $f_\vartheta(\vartheta)$, so we will take $f_r(r)=0$. Additionally, since the electron density $N_e$ is a positive real number, $\omega_p^2\geq 0$, so $f_\vartheta(\vartheta)\geq0$ must be satisfied. The chosen plasma density is 
    \begin{widetext} 
    \begin{equation}
        f_{r,1}(r)=0, ~~~~f_{\vartheta,1}(\vartheta)=\omega_c^2m^2(1+2\sin^2\vartheta), ~~~~
        \omega_{p,1}^2(r,\vartheta)=\frac{\omega_c^2m^2(1+2\sin^2\vartheta)}{r^2R_\Sigma(r)+a^2\cos^2\vartheta},
        \label{eq:SH_pp2}
    \end{equation}
    \end{widetext} 
    where $\omega_c$ is a constant with dimension of frequency, making it easier to relate the plasma frequency $\omega_p$ to the photon frequency $\omega$.
    The shadows obtained for this plasma density are presented in Fig.\eqref{fig:SH_p2}. 
    
    Next, we consider an inhomogeneous plasma with a density asymptotically proportional to $r^{-3/2}$. Considering the separability condition, the plasma distribution turns out to be
    \begin{widetext} 
    \begin{equation}
        f_{r,2}(r)=\omega_c^2\sqrt{m^3r}, ~~~~f_{\vartheta,2}(\vartheta)=0, ~~~~
        \omega_{p,2}^2(r,\vartheta)=\frac{\omega_c^2\sqrt{m^3r}}{r^2R_\Sigma(r)+a^2\cos^2\vartheta}. 
        \label{eq:SH_pp3}
    \end{equation}
    \end{widetext} 
    The shadows modeled with this plasma distribution are shown in Fig.\eqref{fig:SH_p3}. This profile was already discussed for the Kerr-Newman-Dilaton and BW metric in \cite{Badia:2021kpk} and also for the Kerr-Sen metric in \cite{Badia:2022phg}.
    
    A typical example of academic interest is the case of a homogeneous plasma, where 
    \begin{widetext} 
    \begin{equation}
        f_{r,3}(r)=\omega_c^2r^2R_\Sigma(r), ~~~~ f_{\vartheta,3}(\vartheta)=\omega_c^2a^2\cos^2\vartheta, ~~~~
        \omega_{p,3}^2(r,\vartheta)=\omega_c^2,
        \label{eq:SH_pp4}
    \end{equation}
    \end{widetext} 
    which can also be used as a model to study massive test particles trajectories. 
    Black hole shadows formed in such environments are shown in Fig.\eqref{fig:SH_p4}. 
    The characteristic feature of these profile, in comparison with the previous examples, is the existence of stable spherical photon orbits (see Fig. \eqref{fig:SH_pr}). This implies that from some observation positions there are light rays that are sent into the past and go neither to infinity nor to the horizon, but remain within a spatially compact region. Following \cite{PyT_2017}, we will assign darkness only to those light rays going to the horizon, and therefore we assign brightness to the light rays approaching the stable circular orbits. 

    In \cite{Rogers_2015}, Rogers study a plasma profile $\omega_p^2\propto r^{-3}$ to characterize light curves from a slowly rotating neutron star. This work is extended to take into account other metrics and analytical pulse profiles in \cite{Briozzo_2022, BG_2023}. The choice of this plasma profile is based on \cite{GJ_1969}, where a charged pulsar magnetosphere is taken into account. While this is not the environment we are considering, it is an interesting profile to study the differences with other profiles in the shape of the shadow. Nevertheless, while Rogers original profile depended only on $r$, we have to satisfy Eq. \eqref{eq:SH_wp}, resulting in
    \begin{widetext} 
    \begin{equation}
        f_{r,4}(r)=\frac{\omega_c^2m^3}{r}, ~~~~ f_{\vartheta,4}(\vartheta)=0, ~~~~
        \omega_{p,4}^2(r,\vartheta)=\frac{\omega_c^2m^3}{r(r^2R_\Sigma(r)+a^2\cos^2\vartheta)},
        \label{eq:SH_pp5}
    \end{equation}
    \end{widetext} 
    which tends to Rogers' profile when moving away from the black hole. The shadows obtained for this plasma profile are shown in Fig.\eqref{fig:SH_p5}. 

    Finally, we present a profile where the electronic density decays exponentially. This decay is greater than what one could achieve with a simple power law like the ones we have presented above. The plasma density will then be given by 
    \begin{widetext} 
     \begin{equation}
        f_{r,5}(r)=\omega_c^2m^2\exp\left(\frac{r_1-r}{r_0}\right), ~~~~ f_{\vartheta,5}(\vartheta)=0, ~~~~
        \omega_{p,5}^2(r,\vartheta)=\frac{\omega_c^2m^2\exp((r_1-r)/r_0)}{r^2R_\Sigma(r)+a^2\cos^2\vartheta},
        \label{eq:SH_pp6}
    \end{equation} 
    \end{widetext} 
    where $r_0$ and $r_1$ are two arbitrary constants that characterize the profile. For simplicity, we will take $r_0=r_1=m$. Some examples of shadows obtained with this plasma density can be seen in Fig.\eqref{fig:SH_p6}. 
    
\begin{figure*}[htbp]
        \centering
        \begin{tabular}{cc}
            {Profile type 1} & {Profile type 2}  \\
            {\includegraphics[scale=0.53,trim=0 0 0 0]{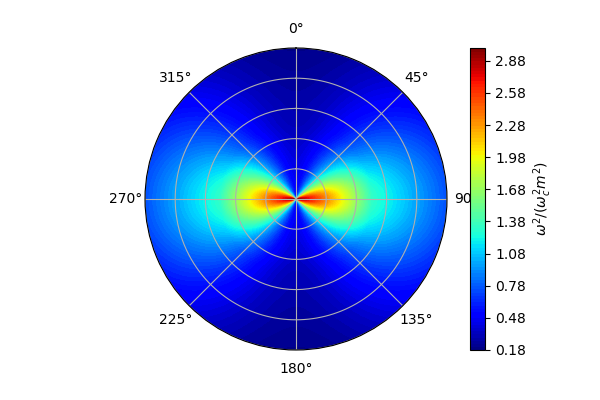}} &
            {\includegraphics[scale=0.53,trim=0 0 0 0]{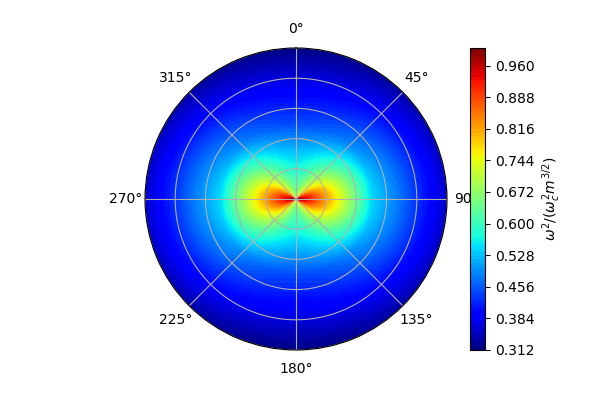}} \\
            {Profile type 4}  & {Profile type 5}  \\
            {\includegraphics[scale=0.53,trim=0 0 0 0]{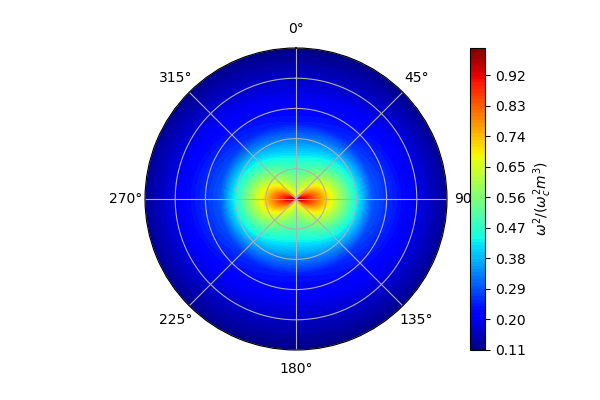}} &
            {\includegraphics[scale=0.53,trim=0 0 0 0]{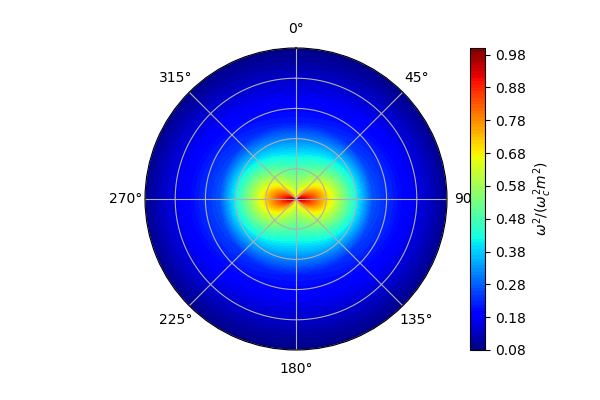}} 
        \end{tabular}
        \caption{Plasma frequency $\omega_p^2(r,\vartheta)$ for profiles of type 1,2,4 and 5.}
        \label{fig:SH_p}
    \end{figure*}

\begin{figure*}[htbp]
        \centering
        \begin{tabular}{ccccc}
            Kerr-Newman, profile 0, $\frac{\omega_c^2}{\omega_\infty^2}=0$ & Kerr-Newman, profile 1, $\frac{\omega_c^2}{\omega_\infty^2}=9$ & Modified Kerr, profile 2, $\frac{\omega_c^2}{\omega_\infty^2}=15$ \\
            {\includegraphics[scale=0.6,trim=0 0 0 0]{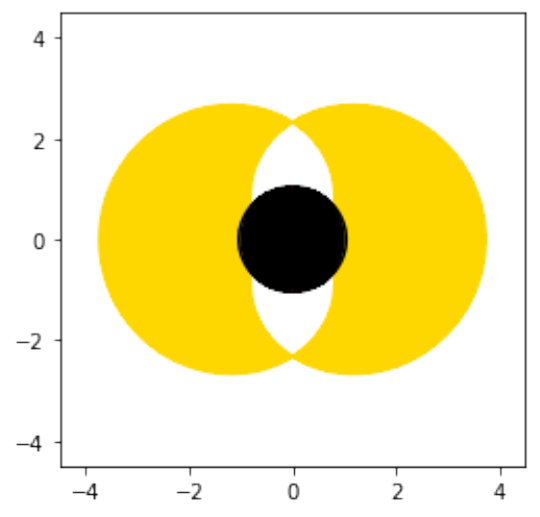}} &
            {\includegraphics[scale=0.6,trim=0 0 0 0]{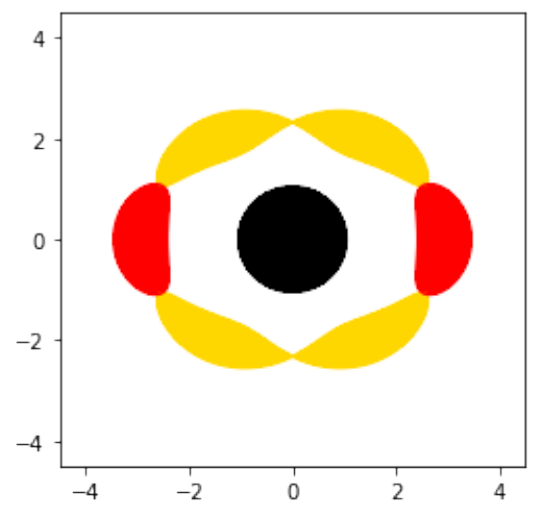}} &
            {\includegraphics[scale=0.6,trim=0 0 0 0]{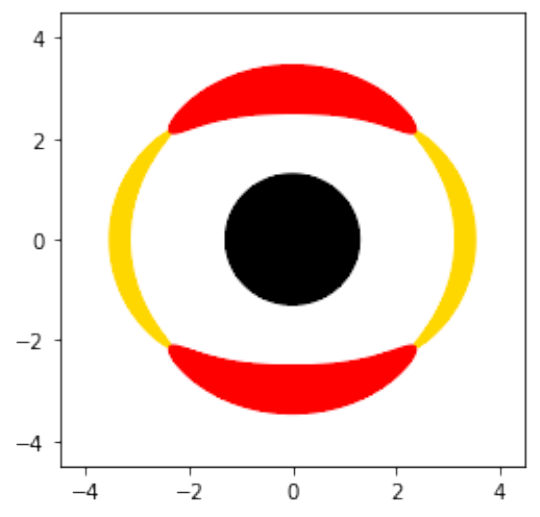}} \\
             Kerr-Sen, profile 3, $\frac{\omega_c^2}{\omega_\infty^2}=1.15$ & Braneworld, profile 4, $\frac{\omega_c^2}{\omega_\infty^2}=70$ & Dilaton, profile 5, $\frac{\omega_c^2}{\omega_\infty^2}=75$  \\
            {\includegraphics[scale=0.6,trim=0 0 0 0]{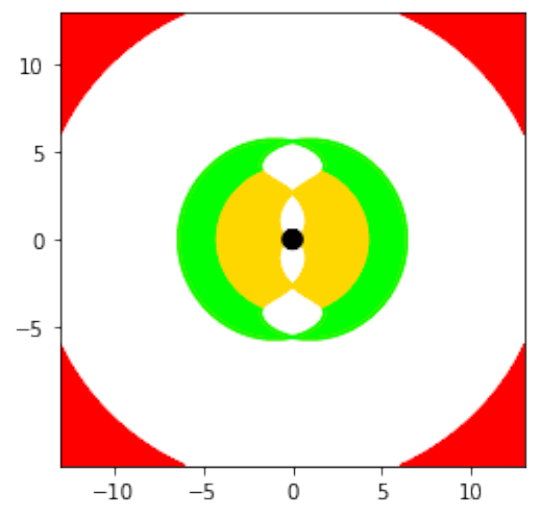}} &
            {\includegraphics[scale=0.6,trim=0 0 0 0]{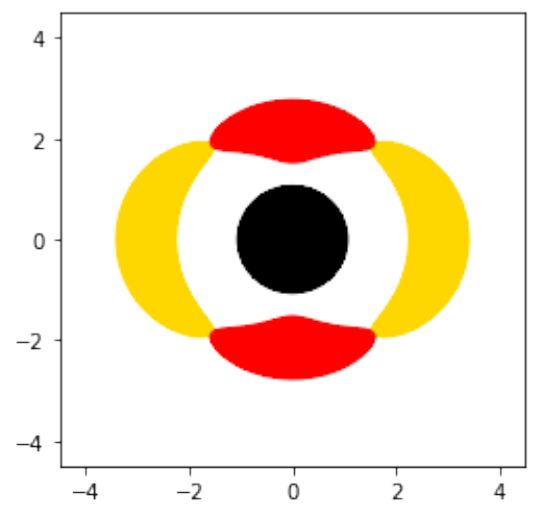}} &
            {\includegraphics[scale=0.6,trim=0 0 0 0]{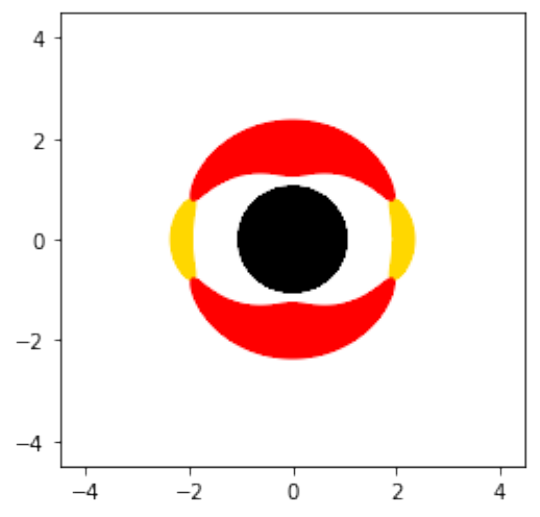}} \\
        \end{tabular}
        \caption{{Photon Regions. Event horizon in black, unstable photon region in yellow, stable photon region in green and forbidden region in red. $Q_{(p)}/Q_{(p)max}=\sqrt{0.25}$. The morphology of the photon region (whether it arises from the poles or from the equator) depends exclusively on the plasma profile. Metrics and other parameters can affect the shape of these regions. For the uniform plasma (profile 3), the forbidden region emerges for $\chi=1$ (as follows from Eq.\eqref{eq:SH_pc}).
        For this profile, when $\omega_c^2/\omega_{\infty}^2\ll1$ the stable and unstable photon regions are separated, and merge as $\omega_c^2/\omega_{\infty}^2$ reaches a critical value.
        }
        }
        \label{fig:SH_pr}
    \end{figure*}

\begin{figure*}[htbp]
        \centering
        \begin{tabular}{ccccc}
            Kerr-Newman & Modified Kerr & Kerr-Sen & Braneworld & Dilaton \\
            {\includegraphics[scale=0.30,trim=0 0 0 0]{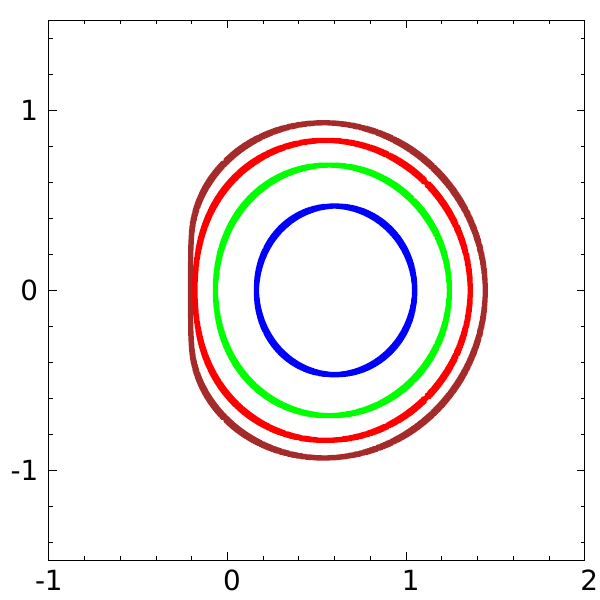}} &
            {\includegraphics[scale=0.30,trim=0 0 0 0]{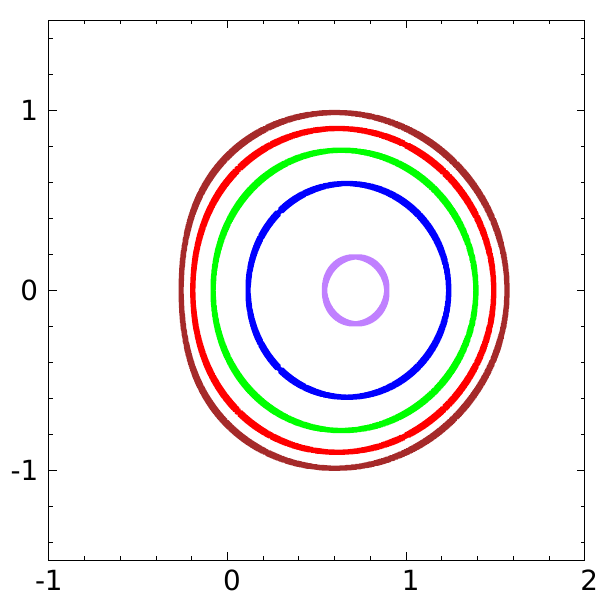}} &
            {\includegraphics[scale=0.30,trim=0 0 0 0]{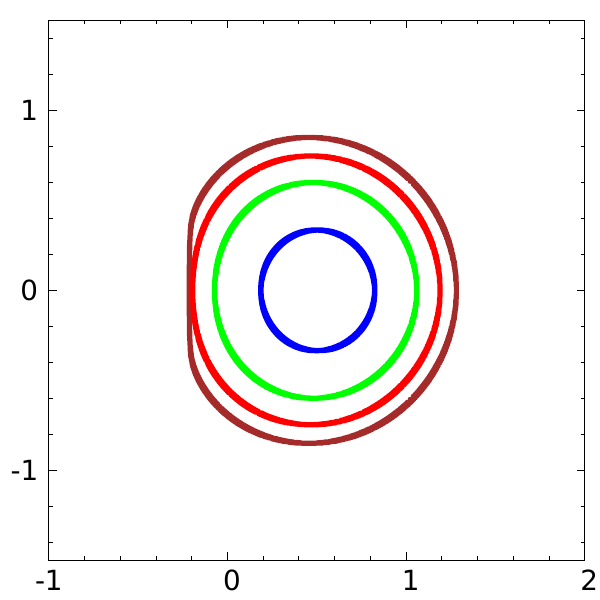}} &
            {\includegraphics[scale=0.30,trim=0 0 0 0]{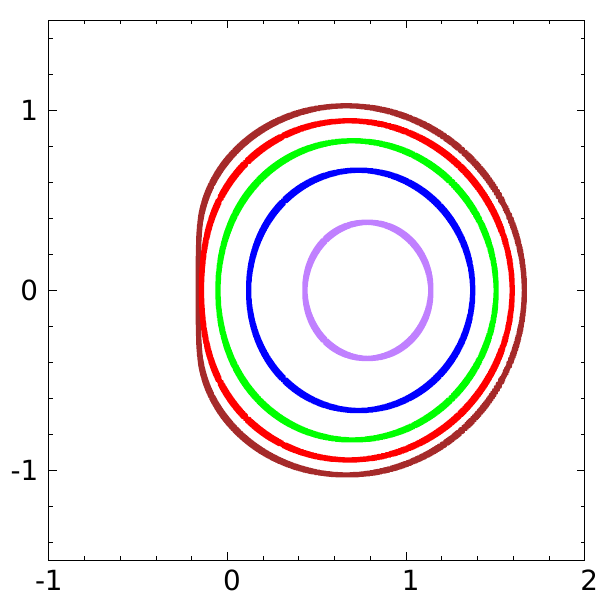}} &
            {\includegraphics[scale=0.30,trim=0 0 0 0]{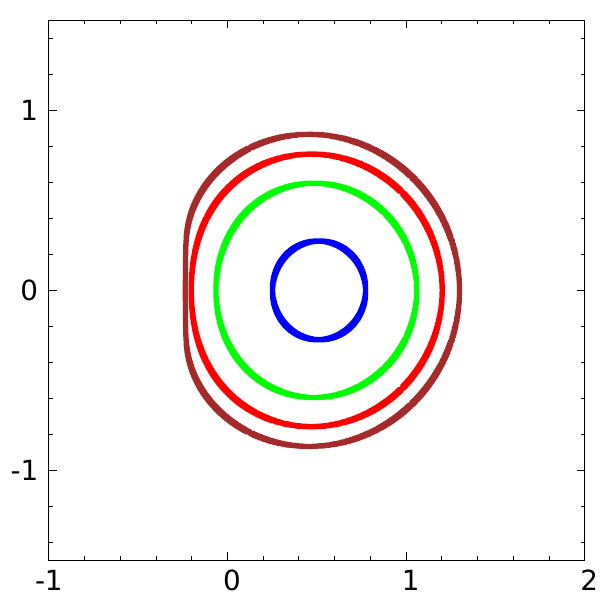}} \\
            {\includegraphics[scale=0.30,trim=0 0 0 0]{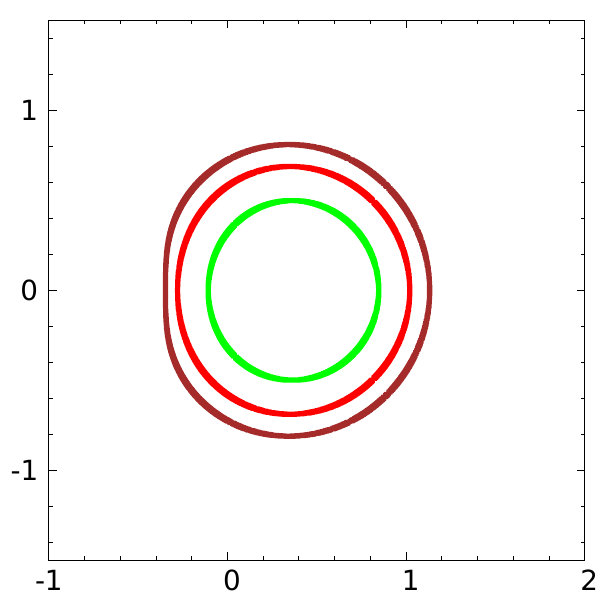}} &
            {\includegraphics[scale=0.30,trim=0 0 0 0]{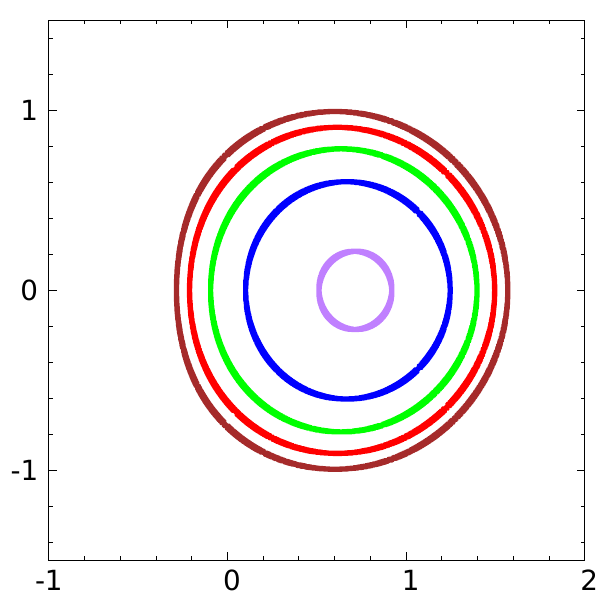}} &
            {\includegraphics[scale=0.30,trim=0 0 0 0]{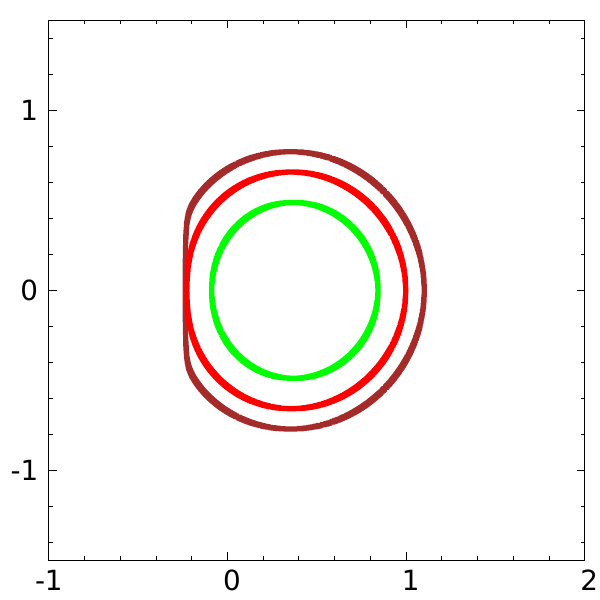}} &
            {\includegraphics[scale=0.30,trim=0 0 0 0]{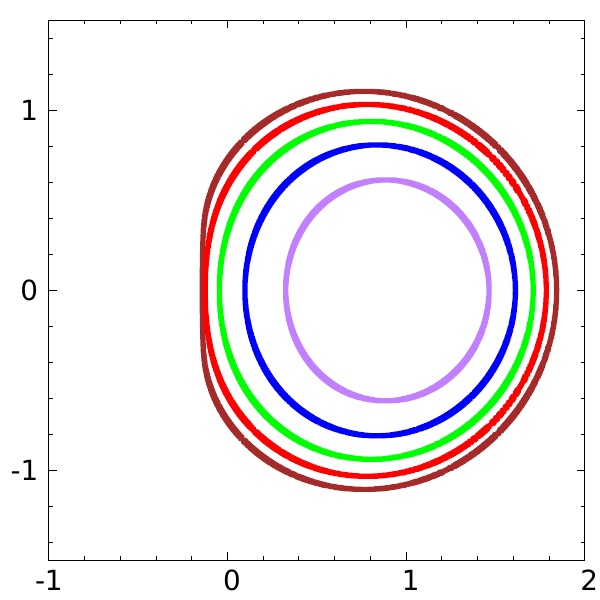}} &
            {\includegraphics[scale=0.30,trim=0 0 0 0]{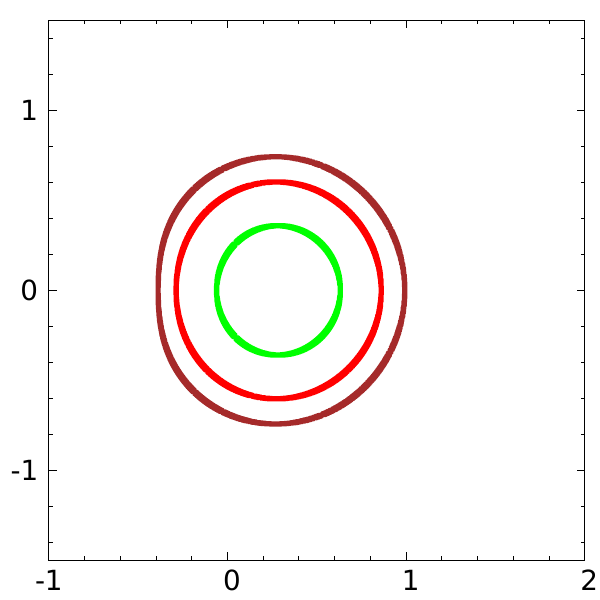}} \\
        \end{tabular}
        \caption{Plasma profile 1{ with $a/a_{max}=0.999$, $r_O=5m$ and $\vartheta_O=\pi/2$}. In each image $\omega_c^2/\omega_\infty^2$ is equal to $0.00$ (brown), $2.25$ (red), $4.50$ (green), $6.75$ (blue) and $8.90$ (purple). In the top panel $Q_{(p)}/Q_{(p)max}=\sqrt{0.25}$; in the bottom $Q_{(p)}/Q_{(p)max}=\sqrt{0.75}$. 
       {Note that, with the exception of the Modified Kerr and Braneworld metrics, there is no associated shadow for certain frequency ratios. This is because a forbidden region emerges in the equatorial plane for some particular frequency ratio $\omega^2_c/\omega^2_\infty$ (see Fig. \ref{fig:SH_pr}). In this region, the propagation condition \eqref{eq:SH_pc} is not satisfied, and as a result, the shadow is no longer visible.
        }
        }
        \label{fig:SH_p2}
    \end{figure*}

\begin{figure*}[htbp]
        \centering
        \begin{tabular}{ccccc}
            Kerr-Newman & Modified Kerr & Kerr-Sen & Braneworld & Dilaton \\
            {\includegraphics[scale=0.30,trim=0 0 0 0]{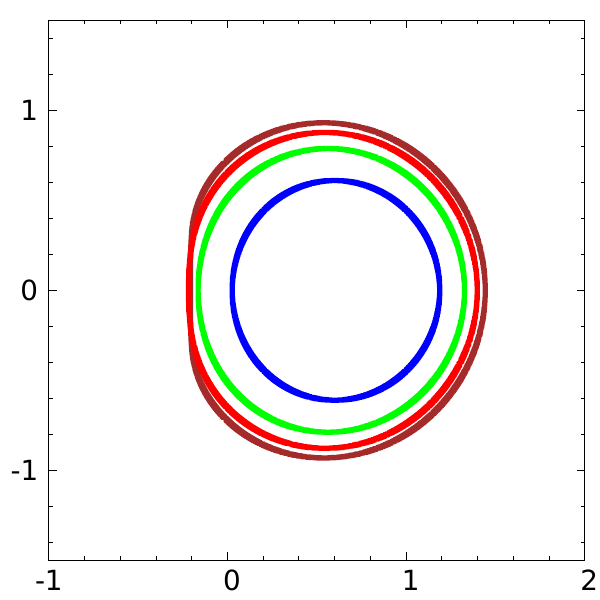}} &
            {\includegraphics[scale=0.30,trim=0 0 0 0]{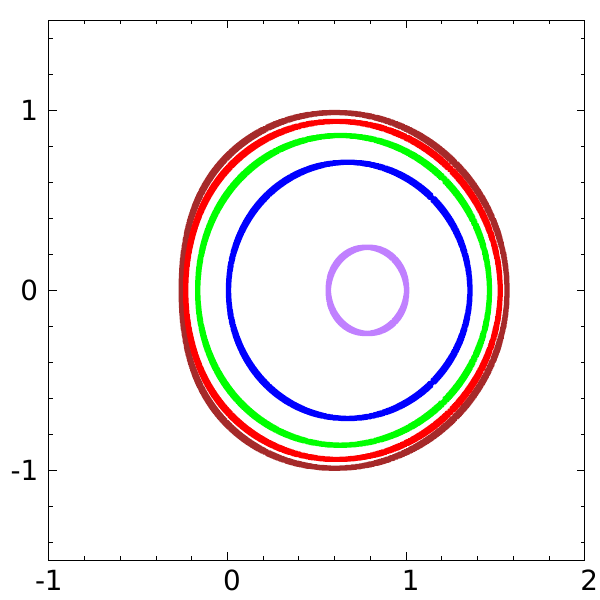}} &
            {\includegraphics[scale=0.30,trim=0 0 0 0]{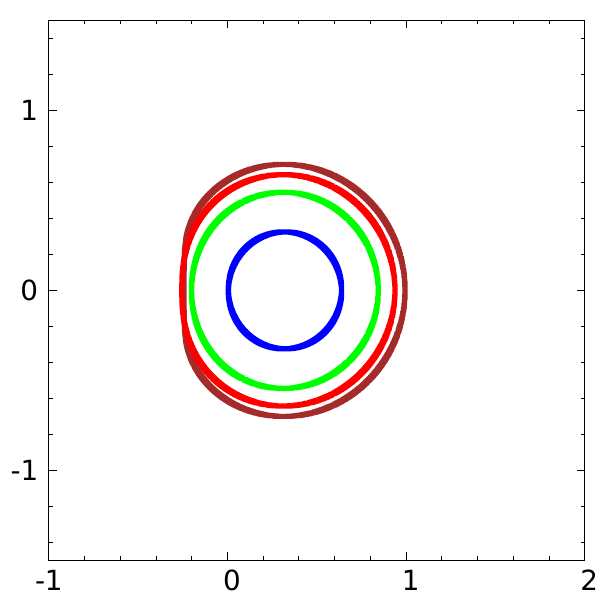}} &
            {\includegraphics[scale=0.30,trim=0 0 0 0]{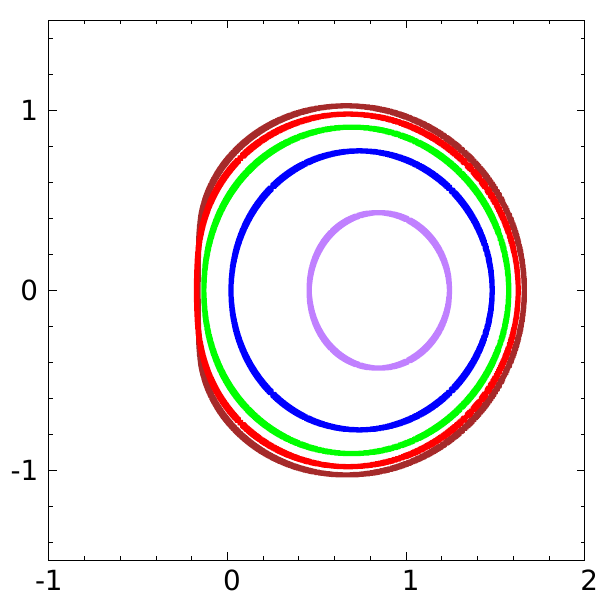}} &
            {\includegraphics[scale=0.30,trim=0 0 0 0]{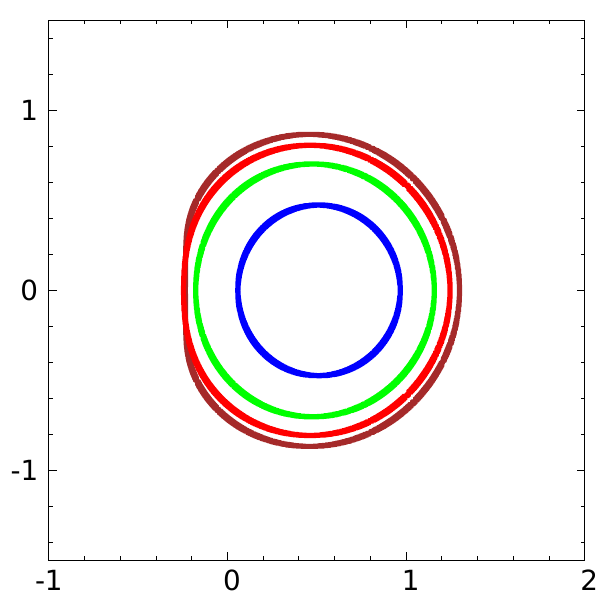}} \\
            {\includegraphics[scale=0.30,trim=0 0 0 0]{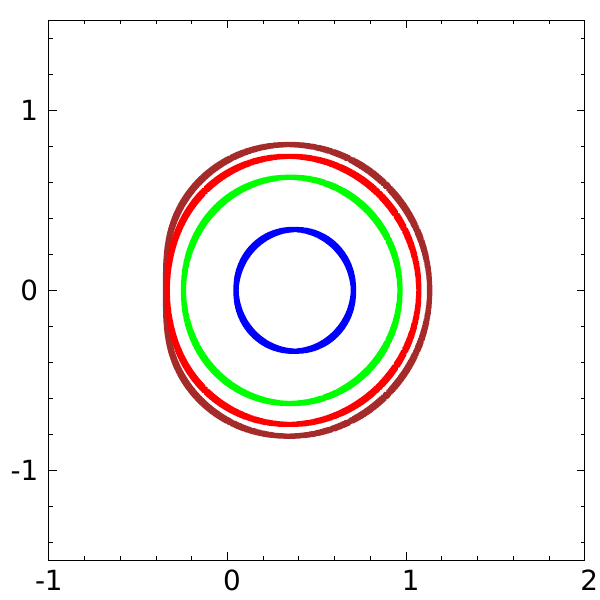}} &
            {\includegraphics[scale=0.30,trim=0 0 0 0]{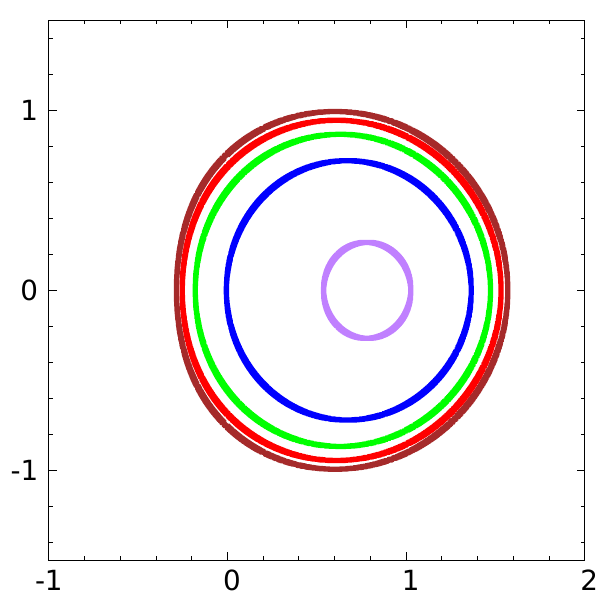}} &
            {\includegraphics[scale=0.30,trim=0 0 0 0]{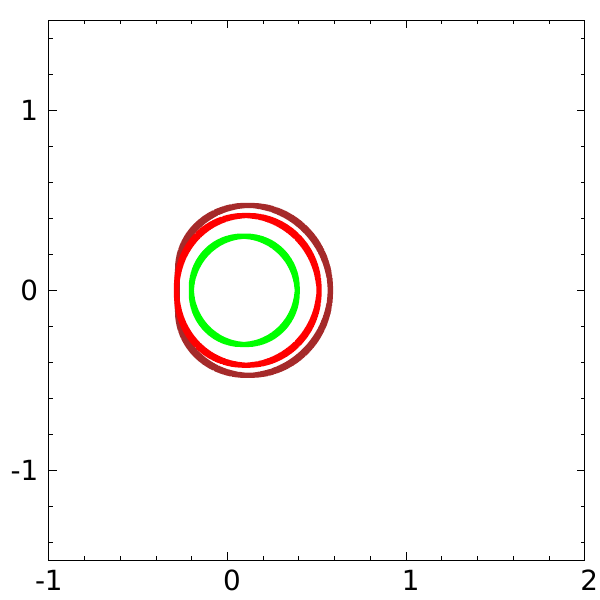}} &
            {\includegraphics[scale=0.30,trim=0 0 0 0]{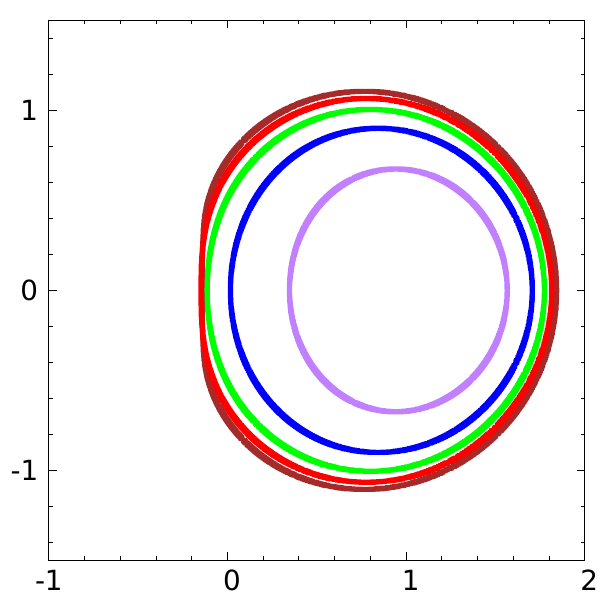}} &
            {\includegraphics[scale=0.30,trim=0 0 0 0]{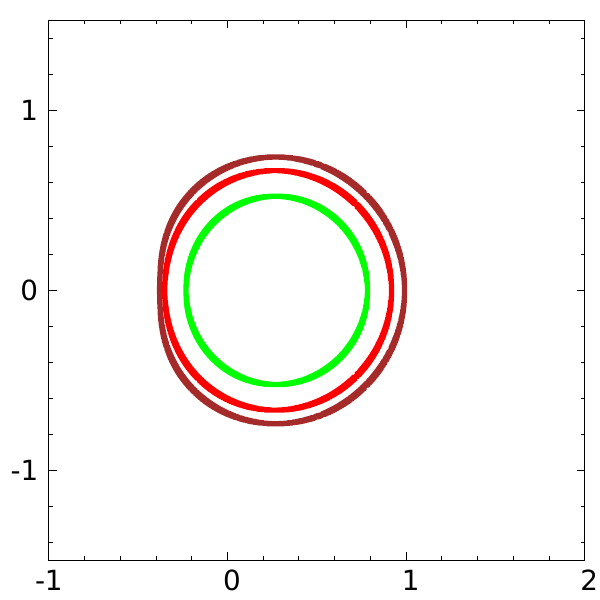}} \\
        \end{tabular}
        \caption{Plasma profile 2{ with $a/a_{max}=0.999$, $r_O=5m$ and $\vartheta_O=\pi/2$}. Brown, red, green, blue and purple curves correspond respectively to the frequency ratios $\omega_c^2/\omega_\infty^2=0.00$, $3.75$, $7.50$, $11.25$ and $15.00$. Top $Q_{(p)}/Q_{(p)max}=\sqrt{0.25}$, bottom $Q_{(p)}/Q_{(p)max}=\sqrt{0.75}$.
       {
        As shown in Fig. \eqref{fig:SH_p2}, some frequencies do not have an associated shadow. For this particular profile, the shadow will vanish for some{critical value $\chi_{cri}$ of the ratio $\chi=\omega^2_c/\omega^2_{\infty}$}, where the forbidden region completely surrounds the black hole and closes in on the equatorial plane (see Fig. \ref{fig:SH_pr}).}
        }
        \label{fig:SH_p3}
    \end{figure*}

\begin{figure*}[htbp]
        \centering
        \begin{tabular}{ccc}
            Kerr-Newman & Modified Kerr & Kerr-Sen \\
            {\includegraphics[scale=0.53,trim=0 0 0 0]{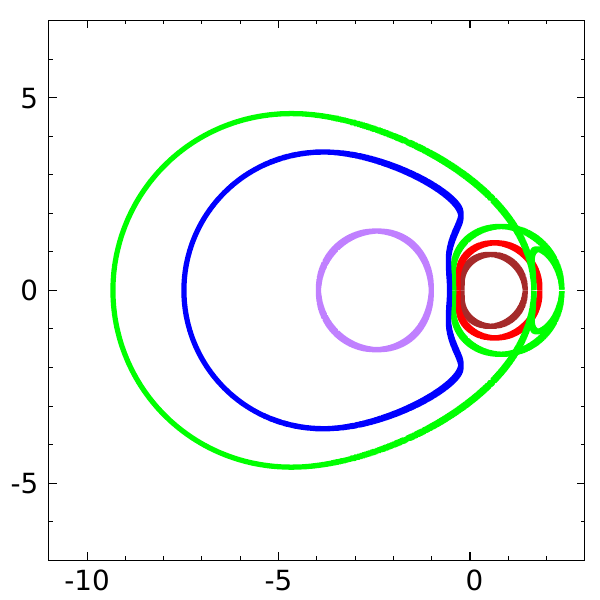}} &
            {\includegraphics[scale=0.53,trim=0 0 0 0]{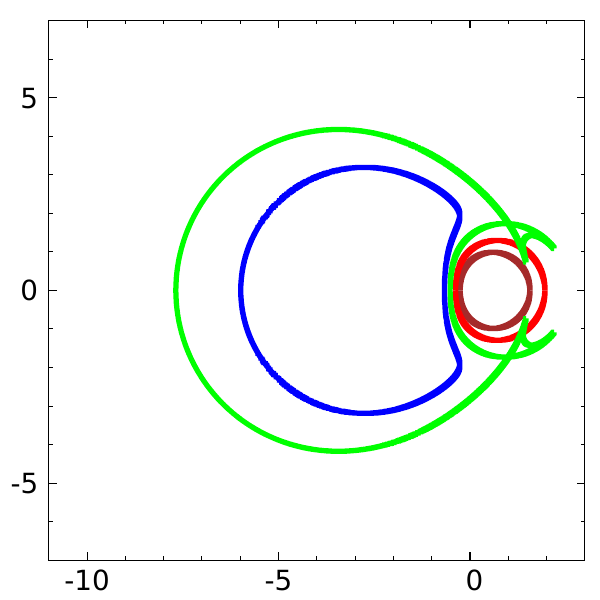}} &
            {\includegraphics[scale=0.53,trim=0 0 0 0]{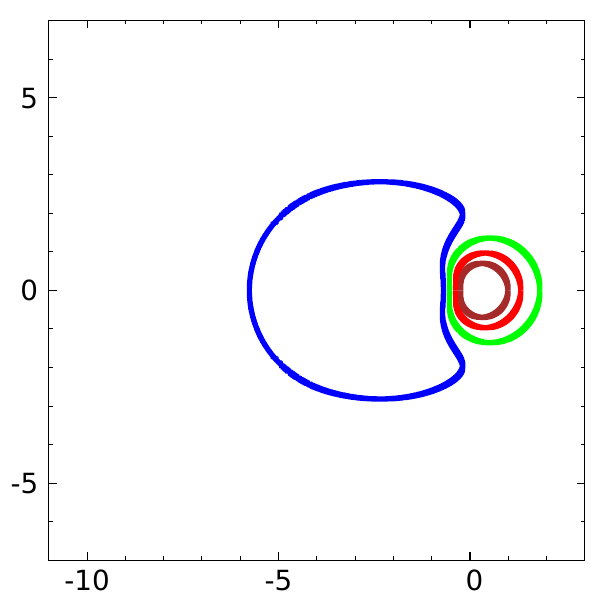}} \\
            {\includegraphics[scale=0.53,trim=0 0 0 0]{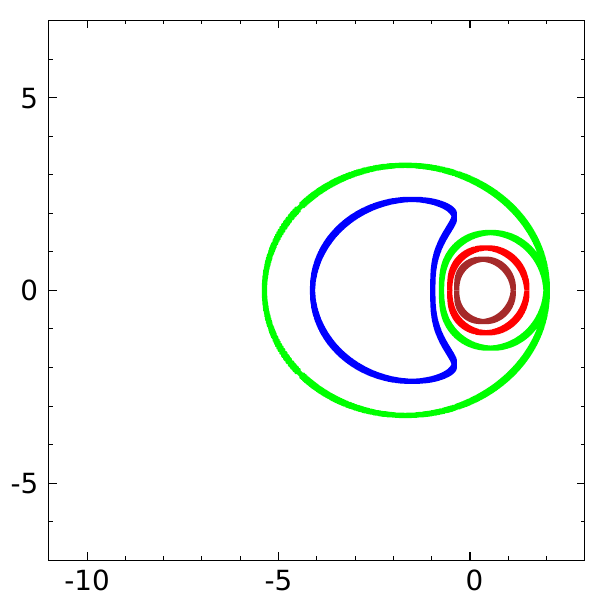}} &
            {\includegraphics[scale=0.53,trim=0 0 0 0]{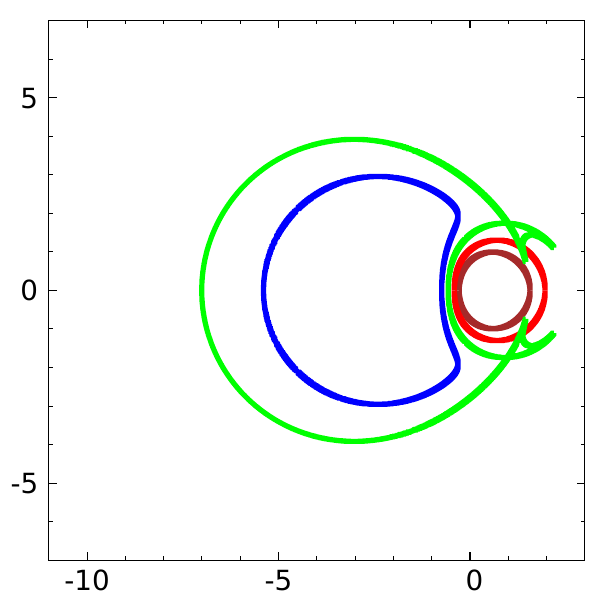}} &
            {\includegraphics[scale=0.53,trim=0 0 0 0]{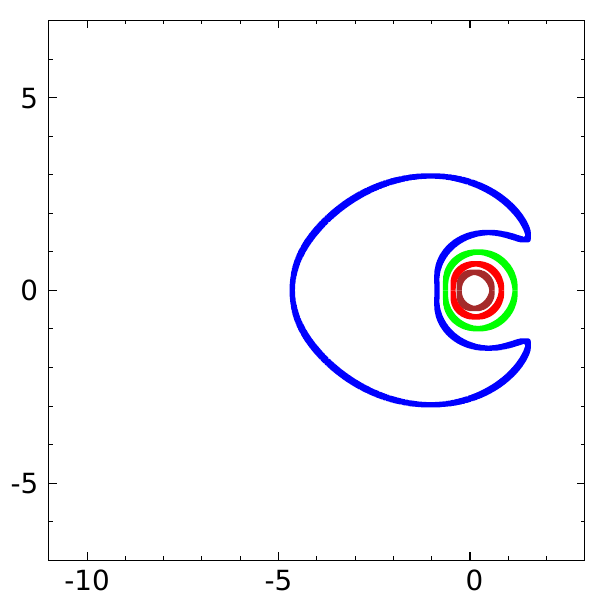}} \\
        \end{tabular}
        \caption{Plasma profile 3{ with $a/a_{max}=0.999$, $r_O=5m$ and $\vartheta_O=\pi/2$}.  In each image $\omega_c^2/\omega_\infty^2$ is equal to $0.000$ (Brown), $0.800$ (red), $1.085$ (green), $1.200$ (blue) and $1.345$ (purple). Top $Q_{(p)}/Q_{(p)max}=\sqrt{0.25}$, bottom $Q_{(p)}/Q_{(p)max}=\sqrt{0.75}$. {
        The brown and red curves always enclose the shadow, leaving outside the illuminated sky. 
        The blue and purple curves always enclose the illuminated sky, leaving outside the shadow. 
        In the Kerr-Newman bottom shadow, the green curve consists of two circles touching at their right end. The inner circle encloses the shadow while the outer circle encloses illuminated sky (except for the inner circle), leaving outside the shadow. 
        For the Kerr-Newman top shadow, the contact point between the green circles shifts, creating three new regions, a lenticular region on the right edge and two triangular regions above and below it. The lenticular region encloses shadow while the triangular ones encloses light. 
        In the Modified Kerr shadows, the right extreme of the green curves splits. Now the inner region of the rightmost circle and the outer region of the leftmost circle are connected through the lenticular region, all containing shadow.
        Finally, for the Kerr-Sen shadows, the green curves enclose the shadows.
        The non plotted curves correspond to $\omega^2_c/\omega^2_\infty > \chi_{cri}$, such that the forbidden region surrounding the black hole now contains the observer (Fig. \ref{fig:SH_pr}) and, consequently, the shadow is not visible. 
        }
        }
        \label{fig:SH_p4}
    \end{figure*}

\begin{figure*}[htbp]
        \centering
        \begin{tabular}{ccccc}
            Kerr-Newman & Modified Kerr & Kerr-Sen & Braneworld & Dilaton \\
            {\includegraphics[scale=0.30,trim=0 0 0 0]{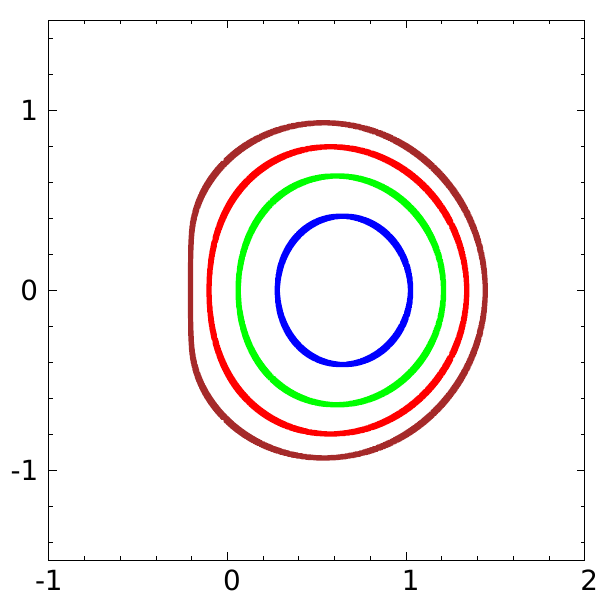}} &
            {\includegraphics[scale=0.30,trim=0 0 0 0]{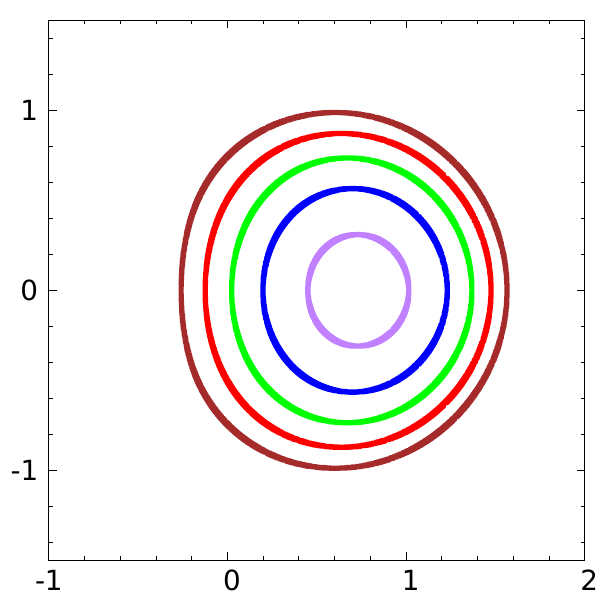}} &
            {\includegraphics[scale=0.30,trim=0 0 0 0]{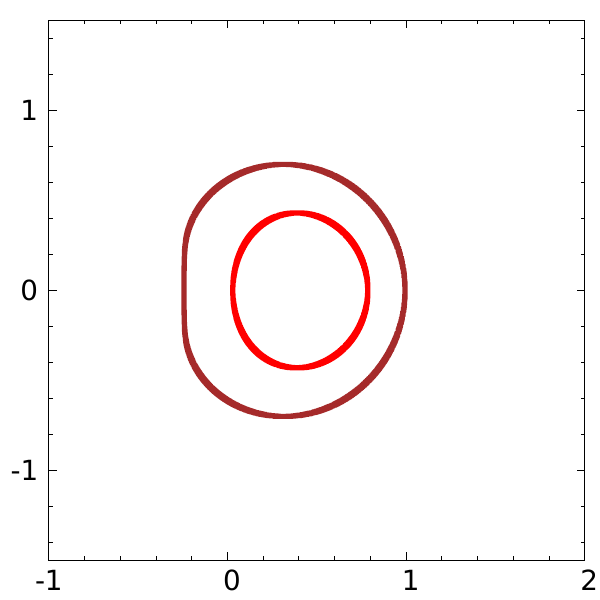}} &
            {\includegraphics[scale=0.30,trim=0 0 0 0]{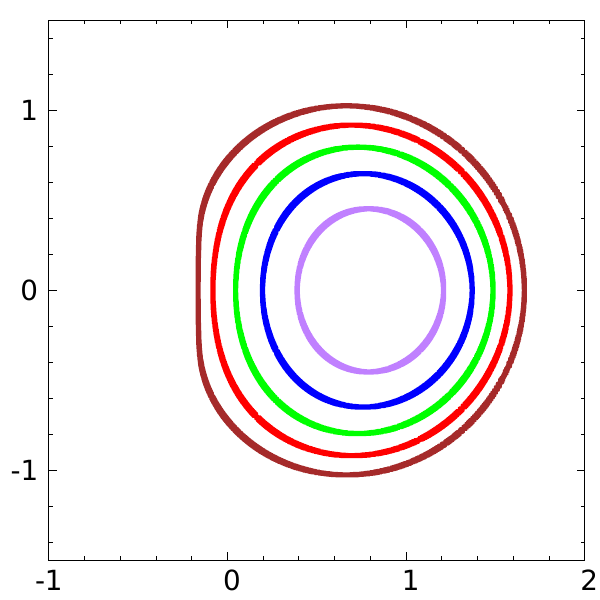}} &
            {\includegraphics[scale=0.30,trim=0 0 0 0]{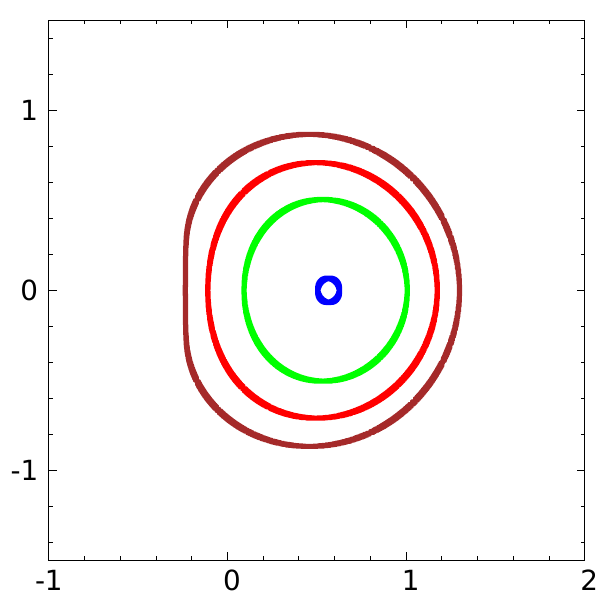}} \\
            {\includegraphics[scale=0.30,trim=0 0 0 0]{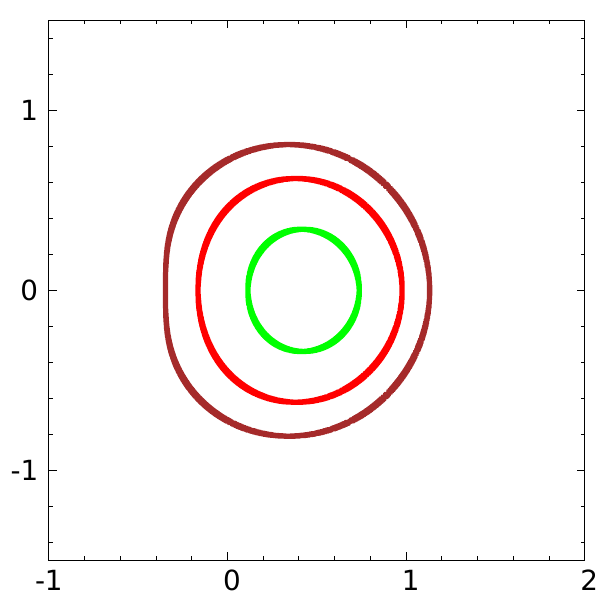}} &
            {\includegraphics[scale=0.30,trim=0 0 0 0]{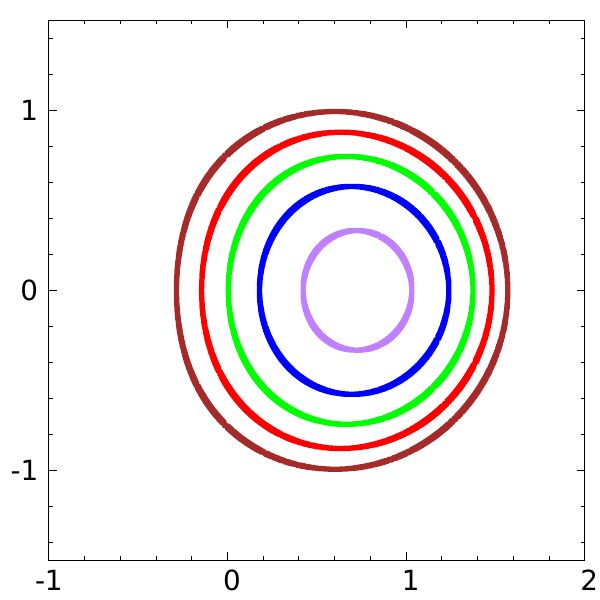}} &
            {\includegraphics[scale=0.30,trim=0 0 0 0]{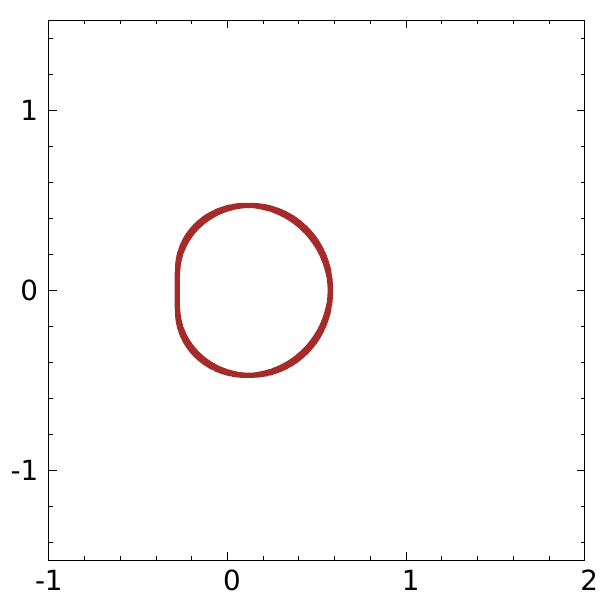}} &
            {\includegraphics[scale=0.30,trim=0 0 0 0]{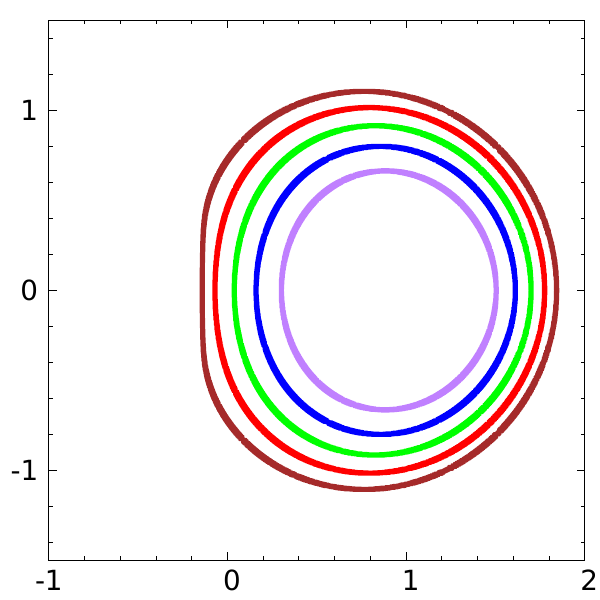}} &
            {\includegraphics[scale=0.30,trim=0 0 0 0]{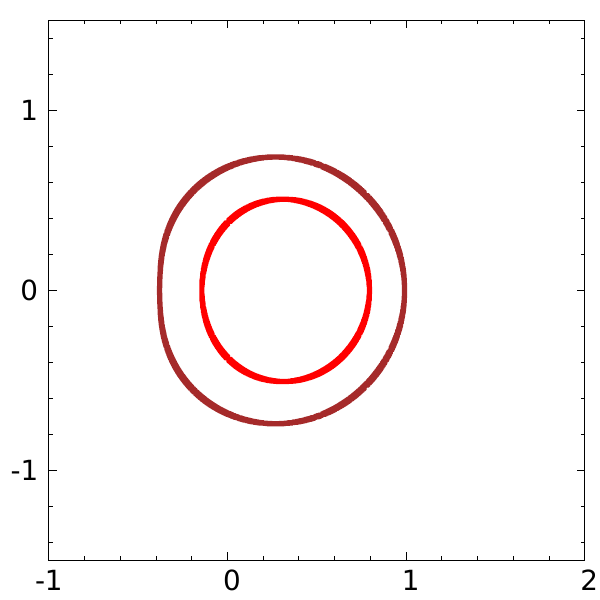}} \\
        \end{tabular}
        \caption{Plasma profile 4{ with $a/a_{max}=0.999$, $r_O=5m$ and $\vartheta_O=\pi/2$}. In each image, brown, red, green, blue and purple curves correspond respectively to the frequency ratios $\omega_c^2/\omega_\infty^2=0.0$, $17.5$, $35.0$, $52.5$ and $70.0$. Top $Q_{(p)}/Q_{(p)max}=\sqrt{0.25}$, bottom $Q_{(p)}/Q_{(p)max}=\sqrt{0.75}$.
       {
        The non plotted curves correspond to $\omega^2_c/\omega^2_\infty > \chi_{cri}$ as explained in Fig.\eqref{fig:SH_p3}.
        }
        }
        \label{fig:SH_p5}
    \end{figure*}

\begin{figure*}[htbp]
        \centering
        \begin{tabular}{ccccc}
            Kerr-Newman & Modified Kerr & Kerr-Sen & Braneworld & Dilaton \\
            {\includegraphics[scale=0.30,trim=0 0 0 0]{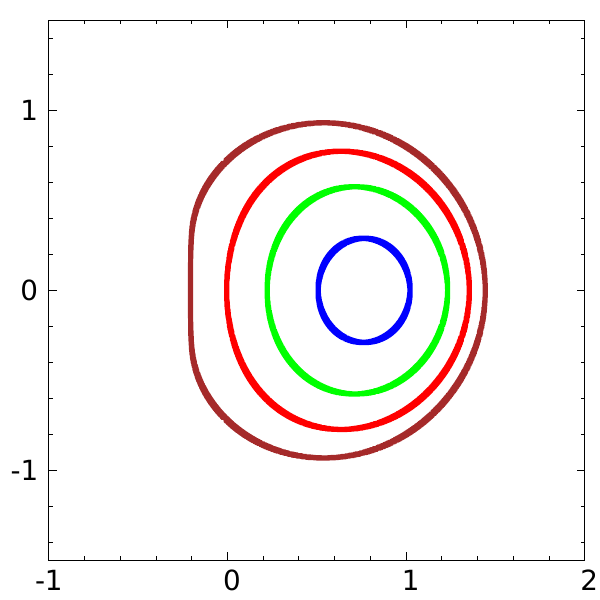}} &
            {\includegraphics[scale=0.30,trim=0 0 0 0]{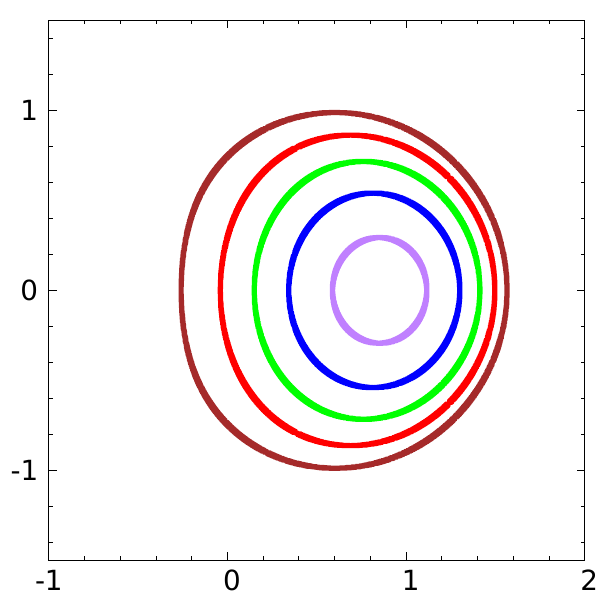}} &
            {\includegraphics[scale=0.30,trim=0 0 0 0]{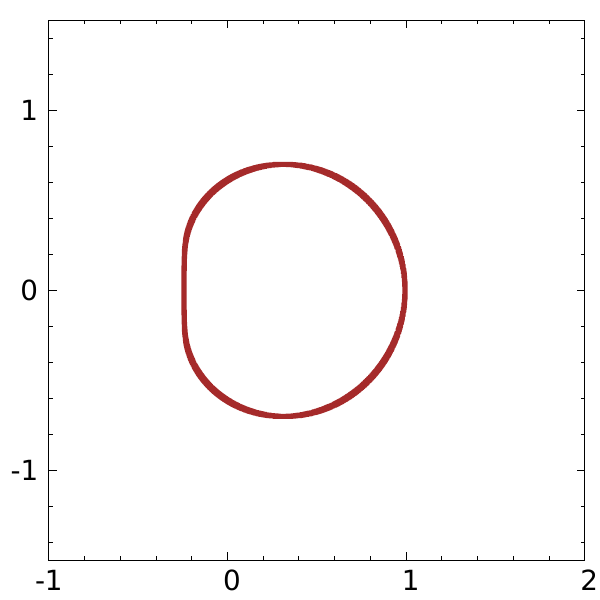}} &
            {\includegraphics[scale=0.30,trim=0 0 0 0]{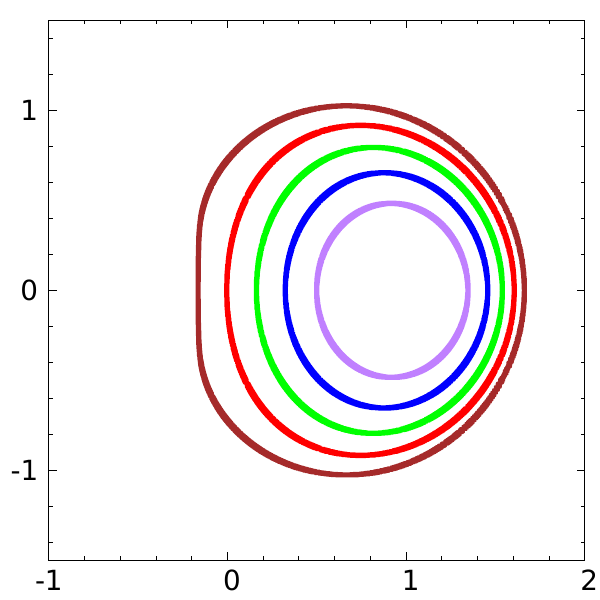}} &
            {\includegraphics[scale=0.30,trim=0 0 0 0]{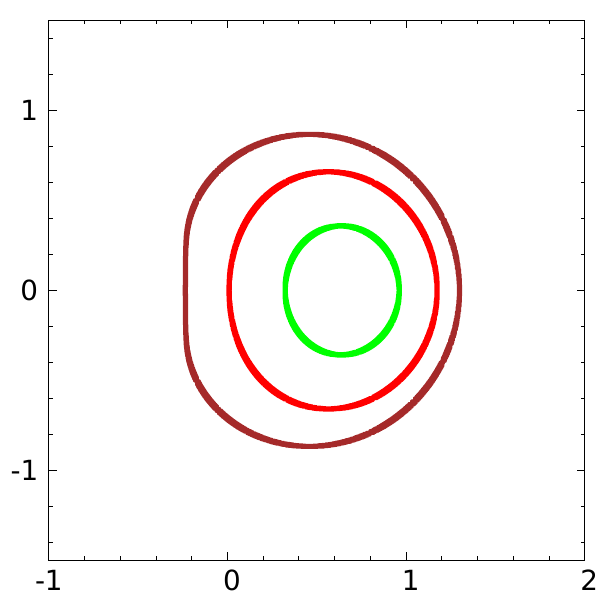}} \\
            {\includegraphics[scale=0.30,trim=0 0 0 0]{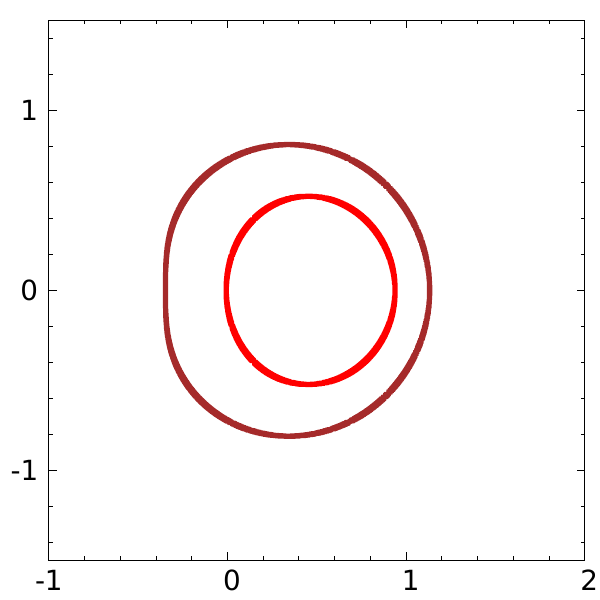}} &
            {\includegraphics[scale=0.30,trim=0 0 0 0]{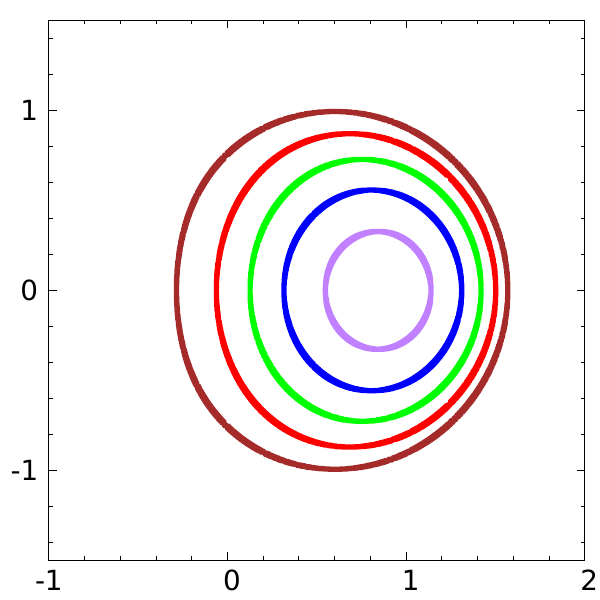}} &
            {\includegraphics[scale=0.30,trim=0 0 0 0]{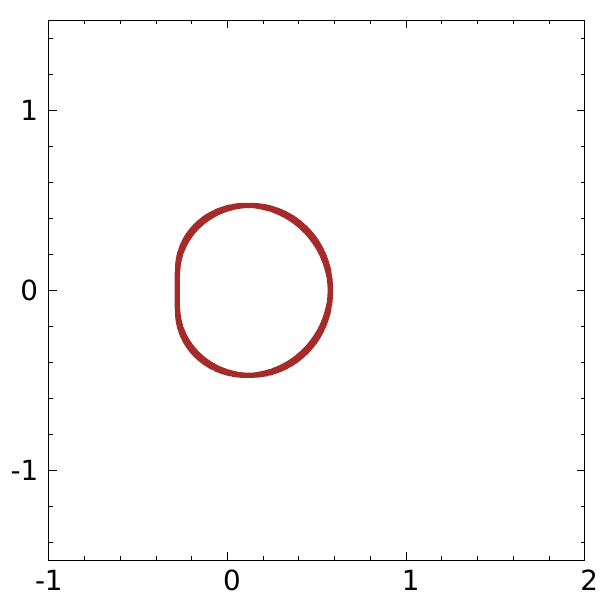}} &
            {\includegraphics[scale=0.30,trim=0 0 0 0]{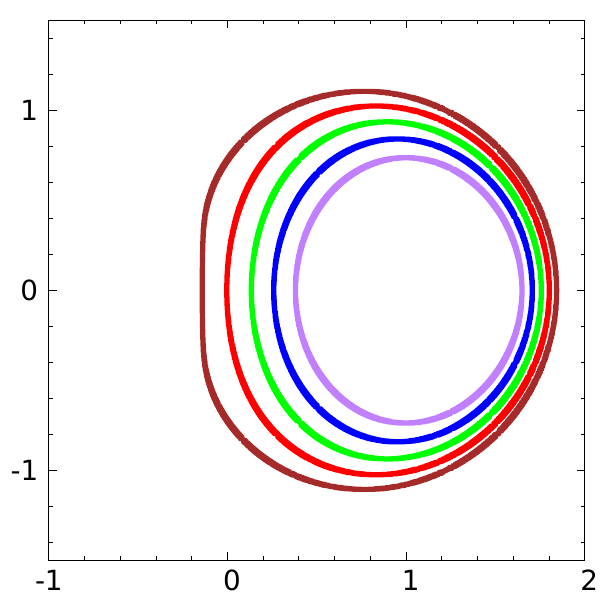}} &
            {\includegraphics[scale=0.30,trim=0 0 0 0]{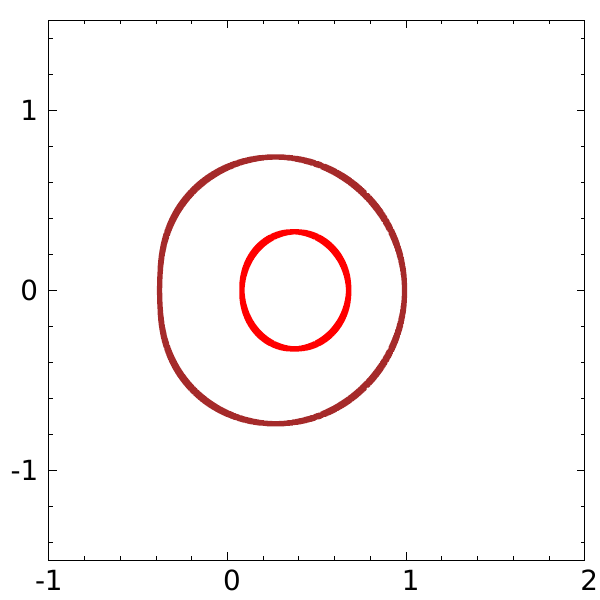}} \\
        \end{tabular}
        \caption{Plasma profile 5{ with $a/a_{max}=0.999$, $r_O=5m$ and $\vartheta_O=\pi/2$}. Brown, red, green, blue and purple curves correspond respectively frequency ratios $\omega_c^2/\omega_\infty^2=0.00$, $32.5$, $65.0$, $97.5$ and $130.0$. Top $Q_{(p)}/Q_{(p)max}=\sqrt{0.25}$, bottom $Q_{(p)}/Q_{(p)max}=\sqrt{0.75}$.
        The non plotted curves correspond to $\omega^2_c/\omega^2_\infty > \chi_{cri}$, as explained in Fig.\eqref{fig:SH_p3}.
        }
        \label{fig:SH_p6}
    \end{figure*}

    The shape of these profiles can be seen depicted in Fig.\eqref{fig:SH_p}, where the plasma frequency as a function of $r$ and $\vartheta$ is shown for profiles of type 1, 2, 4 and 5.
   {In addition, in Fig. \eqref{fig:SH_pr} we present some photon regions obtained for these plasma profiles in the different metrics. The morphology of the photon regions (whether it arises from the poles or from the equator) depends exclusively on the plasma profile. Metrics and other parameters can affect the shape of these regions. In general, when the frequency ratio $\chi=\omega_c^2/\omega_{\infty}^2\ll1$, there is no forbidden region and the photon region resembles the top left figure of Fig. \eqref{fig:SH_pr} for all metrics and plasma profiles. As $\omega_c^2/\omega_{\infty}^2$ reaches a critical value (dependent of the metric and plasma profile), a forbidden region appears and becomes bigger as this ratio increases. }

    For Figs. \eqref{fig:SH_p2} to \eqref{fig:SH_p4} we take $a/a_{max}=0.999$, $r_O=5m$ and $\vartheta_O=\pi/2$. In each figure, the top row takes $Q_{(p)}/Q_{(p)max}=\sqrt{0.25}$ while the bottom row takes $Q_{(p)}/Q_{(p)max}=\sqrt{0.75}$.

    {    
    In Fig.\eqref{fig:SH_p2} we see that for some plasma frequencies the shadow is no longer visible, resulting in a fully bright sky. 
    Through the study of the photon region in Fig. \eqref{fig:SH_pr} it can be seen that there is a certain critical frequency ratio $\chi_{cri}=\omega^2_c/\omega^2_{\infty}$ above which 
    a forbidden region emerges, where the propagation condition (Eq. \eqref{eq:SH_pc}) is no longer satisfied. A forbidden region appears, for profile 1, around the equatorial plane and splits the photon region into two unconnected regions. 
    As the plasma frequency increases, the forbidden region becomes larger, until it completely surrounds the black hole. 
    Low energy photons cannot penetrate the forbidden region and are deflected. 
    For this reason, observers close to the equatorial plane ($\vartheta_O=\pi/2$) stop seeing the shadow first. 
    The resulting sky will be completely bright for observers beyond the forbidden region and completely dark for observers between the forbidden region and the black hole.  
    }  
    
    {
    In Fig.\eqref{fig:SH_p3} we notice that, similar to profile 1, there are some plasma frequencies for which the black hole does not produce a shadow. 
    By studying the photon regions (some of them are shown in Fig. \eqref{fig:SH_pr}) for different frequency ratios $\chi=\omega^2_c/\omega^2_\infty$, it can be checked that there exist a{threshold}  value of this ratio  at which a forbidden region  emerges around the poles of the rotation axis.
    Thus, observers near from the equatorial plane will still observe a shadow, which will vanish as they approach the rotation axis. 
   {As the ratio $\chi$ increases, the forbidden region grows until it completely surrounds the black hole when a critical value $\chi_{cri}$ is reached.}
    Then, the shadow disappears for all observers, resulting in a fully illuminated sky.
    The study of the photon region in Fig. \eqref{fig:SH_pr} shows that the same is true for plasma profiles 4 and 5, since these are similar to profile 2.    
    }
    
   {In general,{$\chi_{cri}$} depends on both the metric model and the $Q_{(p)}$ parameter. As we increase $Q_{(p)}$ the value of $\chi_{cri}$ decreases in Kerr-Newman, Kerr-Sen and Dilaton, while it increases in modified Kerr and Braneworld. 
    In addition,{as the ratio $\chi=\omega^2_c/\omega^2_{\infty}$ goes from $0$ to $\chi_{cri}$} the shadow area decreases until vanishing, loosing its characteristic $D$ shape while tending to a smaller and smaller circle.
    {
    As the plasma frequency grows, the plasma effects becomes more relevant, producing an opposed effect to the gravitational one \cite{BG_2023}. This is because for over-density plasma distributions the plasma acts as a repulsive media instead as the  attractive one of the gravitational field. Therefore the shadow becomes smaller for higher plasma frequencies.
    }
    It is noteworthy that, when shrinking, the shadow does not maintain its relative position with respect to the center of coordinates, but tends to be centered in the geometric center of the shadow corresponding to $\omega_c=0$.}
    Moreover, in Kerr-Newman, Kerr-Sen and Dilaton the value of $\chi_{cri}$ is smaller the higher the plasma density gradient is, i.e., the faster $\omega_p$ decays with $r$. At the same time, in modified Kerr and Braneworld the relationship is inverse.
    
    In Fig.\eqref{fig:SH_p4}, we observe a peculiarity with profile 3. 
   
    {
    Contrary to what we observed for other profiles, the presence of plasma in Profile 3 leads to a magnification in the observed size of the shadow. As the ratio $\chi=\omega^2_c/\omega^2_\infty$ increases, the contour curve of the shadow becomes larger. Beyond a certain threshold of this ratio, the right edge of the curve shifts to the left of the original shadow, resulting in a structure known as "fishtails" (described in \cite{PyT_2017}). The contour curve now encloses the illuminated sky, which is surrounded by the shadow. If the ratio $\chi$ keeps increasing, the fishtail shrinks, decreasing the bright fraction of the sky. When $\chi_{cri}$ is reached, the fishtail vanishes, and the shadow extends over the entire sky.
    These curious effects are fully explained for the Kerr metric in \cite{PyT_2017} through the analysis of the associated photon regions, which have a similar qualitative appearance for all the metrics considered here. In Fig. \eqref{fig:SH_pr}, we observe the appearance of a stable photon region in addition to the unstable photon region. Here, photons from the observer can remain indefinitely without ending up at either the event horizon or infinity.
    Furthermore, the forbidden region now lies beyond the photon region and extends to infinity. Therefore, the shadow is visible only to observers located between the forbidden region and the horizon. Any observer residing inside the forbidden region will see a completely dark sky.}

\section{Aberration in plasma environments}
\subsection{General setting}\label{c3s4}

    The expressions derived in the previous section are valid for a standard observer $\mathcal{O}$ located at position $(r_O,\vartheta_O)$ with 4-velocity $e_0$. However, as explained in \cite{Grenzebach_2015}, if there exists a non-zero $3$-velocity between an observer $\mathcal{O}'$ and $\mathcal{O}$, we would have to deal with relativistic aberration effects.  
    
    The shape of the shadow depends on the observer's 
    state of motion. Therefore, we will have to modify 
    the chosen tetrad if another observer located at 
    $(r_O,\vartheta_O)$ moves with $3-$velocity 
    $\Vec{v}= (v_1,v_2,v_3)$, being $|v|<c=1$, relative 
    to $\mathcal{O}$. {{Here 
    $(v_1,v_2,v_3)$ are the components of the 3-vector 
    in the basis of the spatial vectors 
    $\{e_1,e_2,e_3\}$ as defined in \eqref{eq:SH_e}.}} 
    The $4-$velocity of the moving observer will give us 
    an associated tetrad, which is expressed in terms of 
    the standard tetrad as
    \begin{equation}
        \label{eq:Ab_e}
        \begin{split}
        \Tilde{e}_0 &=\frac{e_0+v_1e_1+v_2e_2+v_3e_3}{\sqrt{1-v^2}}, \\
        \Tilde{e}_1 &=\frac{(1-v_2^2)e_1+v_1(e_0+v_2e_2)}{\sqrt{1-v_2^2}\sqrt{1-v_1^2-v_2^2}}, \\
        \Tilde{e}_2 &=\frac{e_2+v_2e_0}{\sqrt{1-v_2^2}}, \\
        \Tilde{e}_3 &=\frac{(1-v_1^2-v_2^2)e_3+v_3(e_0+v_1e_1+v_2e_2)}{\sqrt{1-v^2}\sqrt{1-v_1^2-v_2^2}}.
        \end{split}
    \end{equation}
    As before, the space vector $\Tilde{e}_3$ corresponds to the incoming direction toward the black hole, while $\Tilde{e}_1$ and $\Tilde{e}_2$ indicate the vertical and horizontal directions respectively in the Cartesian plane. 

    As in Sec.\eqref{c3s3}, for any photon with trajectory $\lambda(s)=(t(s),r(r),\vartheta(s),\varphi(s))$ the tangent vector at the observer's position can be written in two different ways, either using the Boyer-Lindquist like coordinate basis or the tetrad introduced above, resulting in
    \begin{equation}
        \label{eq:Ab_l}
        \begin{split}
        \dot{\lambda} &=\dot{t}\partial_t+\dot{r}\partial_r+\dot{\vartheta}\partial_\vartheta+\dot{\varphi}\partial_\varphi, \\
        \dot{\lambda} &=-\alpha \Tilde{e}_0 + \beta\left(\sin\Theta\cos\Phi \Tilde{e}_1+\sin\Theta\sin\Phi \Tilde{e}_2+\cos\Theta \Tilde{e}_3\right), 
        \end{split}
    \end{equation}
    where $\Theta$ and $\Phi$ are the celestial coordinates of the observer, and now, the factors $\alpha$ and $\beta$ must now be calculated taking into account the $4-$velocity of the observer and the plasma presence,
    \begin{equation}
        \begin{split}
        \alpha &=g(\dot{\lambda},\Tilde{e}_0)=k_0^tp_t+k_0^rp_r+k_0^\vartheta p_\vartheta + k_0^\varphi p_\varphi, \\
        \beta  &=\sqrt{(k_0^tp_t+k_0^rp_r+k_0^\vartheta p_\vartheta + k_0^\varphi p_\varphi)^2-\omega_p^2}.
        \end{split}
        \label{eq:Ab_a1}
    \end{equation}
    (recall that both expressions must be evaluated at the observer's coordinates $(r_O,\vartheta_O)$) where the $k_i^\mu$ elements come from the generic expression
    \begin{equation}
        \Tilde{e}_i=k_i^\mu\partial_\mu,
        \label{eq:Ab_ekd}
    \end{equation}
    and are shown in the following matrix
    \begin{widetext} 
    \begin{equation}
        k_i^\mu=
        \left(
        \begin{matrix}
        \frac{e_0^t+v_2e_2^t}{\sqrt{1-v^2}}                            & \frac{v_3e_3^r}{\sqrt{1-v^2}}           & \frac{v_1e_1^\vartheta}{\sqrt{1-v^2}}                        & \frac{e_0^\varphi+v_2e_2^\varphi}{\sqrt{1-v^2}} \\
        \frac{v_1(e_0^t+v_2e_2^t)}{\sqrt{1-v_2^2}\sqrt{1-v_1^2-v_2^2}} & 0                                       & \sqrt{\frac{1-v_2^2}{1-v_1^2-v_2^2}}e_1^\vartheta            & \frac{v_1(e_0^\varphi+v_2e_2^\varphi)}{\sqrt{1-v_2^2}\sqrt{1-v_1^2-v_2^2}} \\
        \frac{v_2e_0^t+e_2^t}{\sqrt{1-v_2^2}}                          & 0                                       & 0                                                            & \frac{v_2e_0^\varphi+e_2^\varphi}{\sqrt{1-v_2^2}}\\
        \frac{v_3(e_0^t+v_2e_2^t)}{\sqrt{1-v^2}\sqrt{1-v_1^2-v_2^2}}   & \sqrt{\frac{1-v_1^2-v_2^2}{1-v^2}}e_3^r & \frac{v_3v_1e_1^\vartheta}{\sqrt{1-v^2}\sqrt{1-v_1^2-v_2^2}} & \frac{v_3(e_0^\varphi+v_2e_2^\varphi)}{\sqrt{1-v^2}\sqrt{1-v_1^2-v_2^2}} \\
        \end{matrix}
        \right).
        \label{eq:Ab_k}
    \end{equation}
    \end{widetext} 
    
    Comparing the expressions in \eqref{eq:Ab_l}, introducing into these Eq. \eqref{eq:Ab_ekd} and grouping the terms with common factor $\partial_\mu$ ($\mu=t,r,\vartheta,\varphi$), we obtain the following set of coupled equations
    \begin{equation}
        \label{eq:Ab_dot}
        \begin{split}
        \dot{t} =&-\alpha k_0^t+\beta k_1^t\sin\Theta\cos\Phi+\beta k_2^t\sin\Theta\sin\Phi+\beta k_3^t\cos\Theta, \\
        \dot{r} =&-\alpha k_0^r + \beta k_3^r \cos\Theta, \\
        \dot{\vartheta} =&-\alpha k_0^\vartheta + \beta k_1^\vartheta\sin\Theta\cos\Phi + \beta k_3^\vartheta\cos\Theta, \\
        \dot{\varphi} =&-\alpha k_0^\varphi + \beta k_1^\varphi\sin\Theta\cos\Phi + \beta k_2^\varphi\sin\Theta\sin\Phi \\&
        + \beta k_3^\varphi\cos\Theta,
        \end{split}
    \end{equation}
    from which we can obtain the following expressions for the celestial coordinates of the observer
    \begin{widetext} 
    \begin{equation}
        \label{eq:Ab_STP}
        \begin{split}
        \cos\Theta &=\frac{\dot{r}+\alpha k_0^r}{\beta k_3^r}, \\
        \sin\Phi &=
        \frac{\dot{\varphi}+\alpha k_0^\varphi-
        \left(\dot{\vartheta}+\alpha k_0^\vartheta\right)\frac{k_1^\varphi}{k_1^\vartheta}
        +\beta\frac{\dot{r}+\alpha k_0^r}{\beta k_3^r}\left(k_3^\vartheta\frac{k_1^\varphi}{k_1^\vartheta}-k_3^\varphi\right)}
        {\beta\sqrt{1-\left(\frac{\dot{r}+\alpha k_0^r}{\beta k_3^r}\right)^2}k_2^\varphi},
        \end{split}
    \end{equation}
    \end{widetext} 
    where $\dot{r}$, $\dot{\vartheta}$ and $\dot{\varphi}$ must be substituted for their expressions in the equations of motion (Eq. \eqref{eq:SH_dot}). In doing so, we must evaluate all the functions shown in Eq. \eqref{eq:SH_dot} at the observer's coordinates $(r_O,\vartheta_O)$, except the explicit expressions for $K$ and $p_\varphi$ which must be evaluated at $(r_p,\vartheta_O)$.
    It can be verified that, if $\vec{v}=\vec{0}$ is taken, the expressions in Eq. \eqref{eq:Ab_STP} reduce to those found in Eq. \eqref{eq:SH_STF}. Finally, the stereographic coordinates $X(r_p)$ and $Y(r_p)$ will continue to be expressed in the same way as in the previous section (Eq. \eqref{eq:SH_XY}).

    In summary, in order to construct the shadow of a black hole including aberration effects one should use the  following step-by-step procedure:
    
    \begin{enumerate}
    
        \item 
        Choose a metric. Some examples are shown in the Table \ref{tab:Metrics} .
        
        \item 
        Choose the spin parameter $a$ and the rest of the characteristic parameters of the metric, verifying that they satisfy the corresponding constraints ($\Delta(r_h)=0$, $r_h\in \mathbb{R}_{>0}$). 
        
        \item 
        Select the position of the standard observer $(r_O,\vartheta_O)$ (with associated tetrad as given by eq.\eqref{eq:Ab_e}). 
        
        \item 
        Select the $3$-velocity of the observer $\vec{v}$ with respect to the standard observer.
        
        \item
        Calculate the coefficients $e_i^\mu$ from Eq. \eqref{eq:Ab_e} evaluated at the observer's position $(r_O,\vartheta_O)$.
        
        \item 
        Calculate from these the coefficients $k_i^\mu$ as expressed in Eq. \eqref{eq:Ab_k}.
        
        \item 
        Select the functions $f_r(r)$ and $f_\vartheta(\vartheta)$, giving the distribution of the plasma $\omega_p(r,\vartheta)$ around the black hole.
        
        \item 
        Choose the photon frequency at infinity $\omega_{\infty}$ or at the observer $\omega_{obs}$ (related by Eq. \eqref{eq:SH_wo}) and verify that Eq. \eqref{eq:SH_pc} is satisfied during the whole trajectory. In the equations we will use, this motion counter will be present only in the quotient $\omega_p(x)/\omega_{\infty/obs}$, so we recommend expressing both frequencies as a function of the same magnitude $\omega_c$.
        
        \item 
        Write the equations of motion in terms of $K$ and $p_\varphi$ as shown in Eq. \eqref{eq:SH_dot}.
        
        \item 
        Write the celestial coordinates $\sin\Theta$ and $\sin\Phi$ as shown in Eq. \eqref{eq:Ab_STP}, replacing the equations of motion by the expressions obtained in the previous step.
        
        \item 
        Substitute into the expressions for $\sin\Theta$ and $\sin\Phi$ the expressions $K(r_p)$ and $p_\varphi(r_p)$ according to Eqs. \eqref{eq:SH_Kr} and \eqref{eq:SH_pfr}, evaluating at $r=r_p$, which belongs to the interval of radial coordinates for which unstable spherical orbits exist. This gives us $\sin\Theta$ and $\sin\Phi$ as functions of $r_p$, so we have parameterized the shadow contour in the observer's sky. 
        
        \item 
        Solve the equation $\sin^2[\Phi(r_p)]=1$ for $r_p$, obtaining the boundary values $r_{p,min}$ and $r_{p,max}$. Note that this equation can have several real solutions, so you must determine, depending on the characteristics of the problem, which boundary is relevant for shadow formation. 
        
        \item  
        Compute $\sin[\Theta(r_p)]$ and $\sin[\Phi(r_p)]$ by evaluating $r_p$ on the interval $[r_{p,min},r_{p,max}]$.
        
        \item 
        Calculate the dimensionless Cartesian coordinates $X(r_p)$ and $Y(r_p)$ of the shadow boundary curve according to Eq. \eqref{eq:SH_XY}.
        
    \end{enumerate}

\subsection{Black holes shadows and relativistic aberrations}
\label{c3s4ss2}

\begin{figure*}[htbp]
        \centering
        \begin{tabular}{ccccc}
             & $v=+0.9c$ & $v=+0.1c$ & $v=-0.1c$ & $v=-0.9c$ \\
            \rotatebox{90}{~~~~~~~~Kerr-Newman} & 
            {\includegraphics[scale=0.36,trim=0 0 0 0]{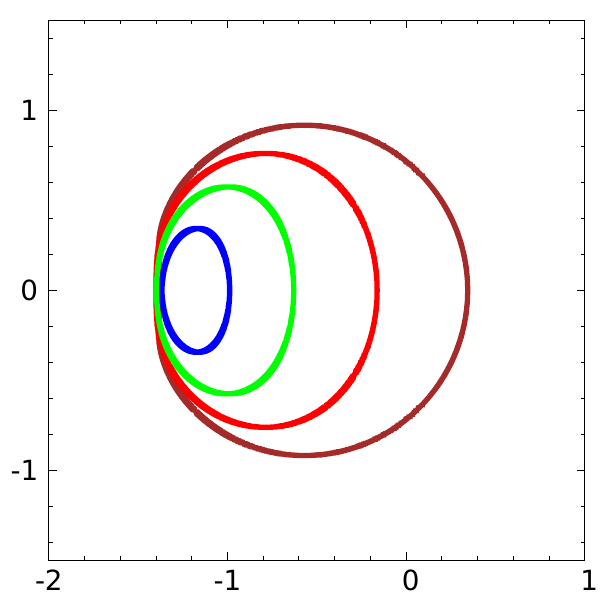}} &
            {\includegraphics[scale=0.36,trim=0 0 0 0]{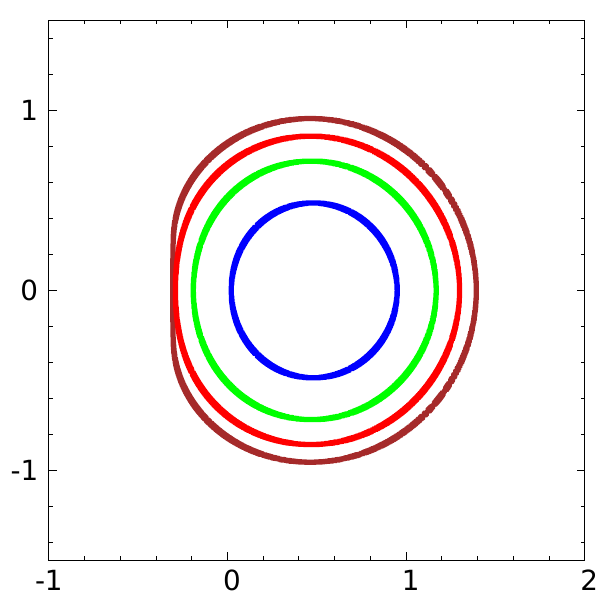}} &
            {\includegraphics[scale=0.36,trim=0 0 0 0]{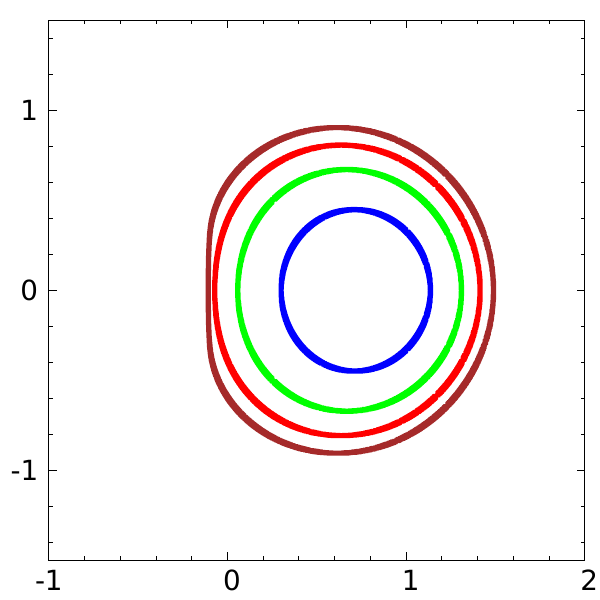}} &
            {\includegraphics[scale=0.39,trim=0 0 0 0]{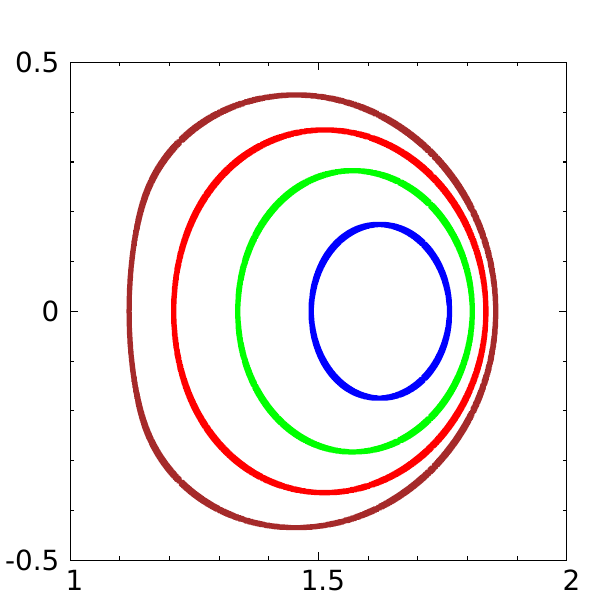}} \\
            \rotatebox{90}{~~~~~~~Modified Kerr} & 
            {\includegraphics[scale=0.36,trim=0 0 0 0]{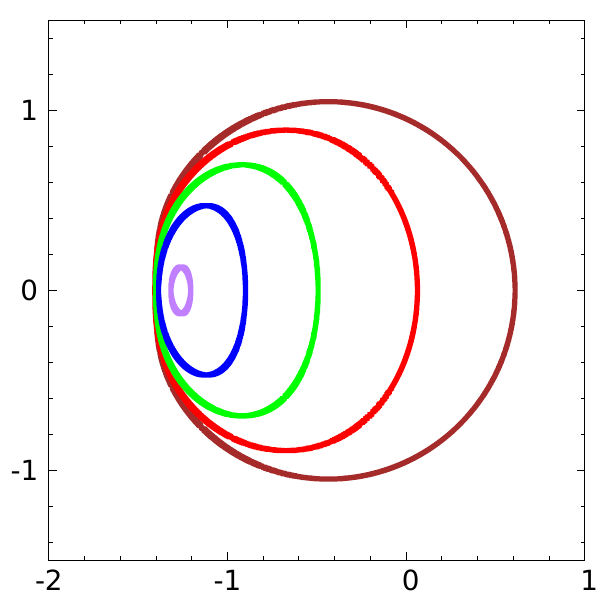}} &
            {\includegraphics[scale=0.36,trim=0 0 0 0]{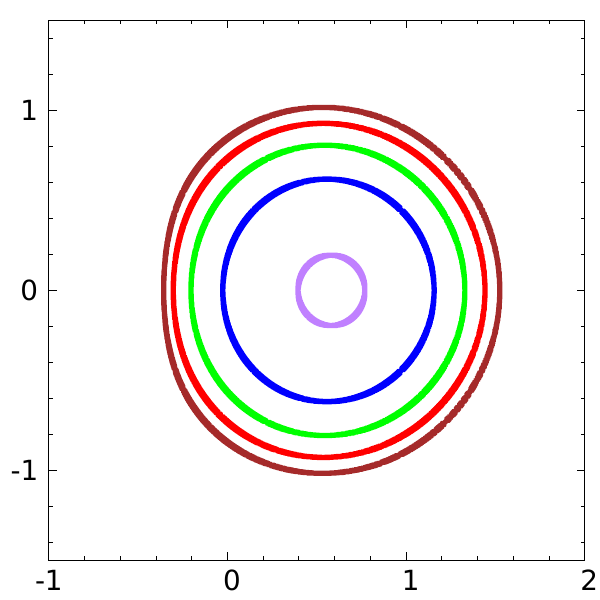}} &
            {\includegraphics[scale=0.36,trim=0 0 0 0]{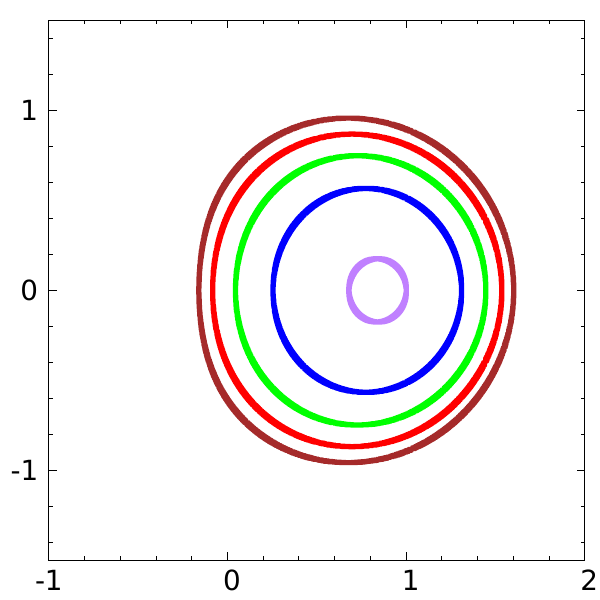}} &
            {\includegraphics[scale=0.39,trim=0 0 0 0]{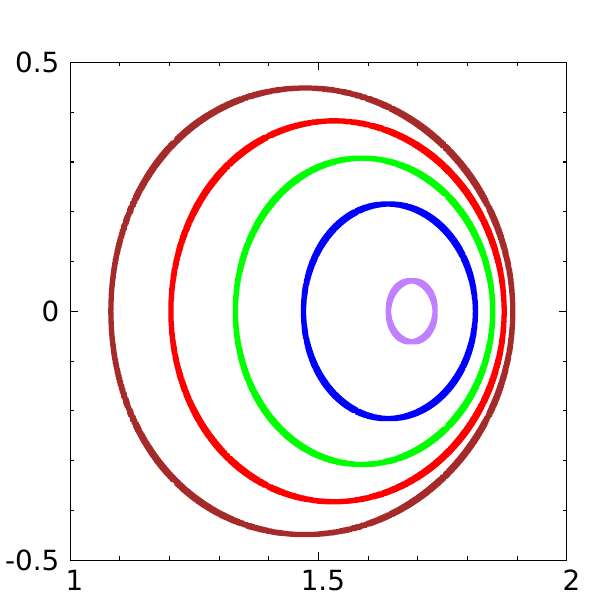}} \\
            \rotatebox{90}{~~~~~~~~~~~Kerr-Sen} & 
            {\includegraphics[scale=0.36,trim=0 0 0 0]{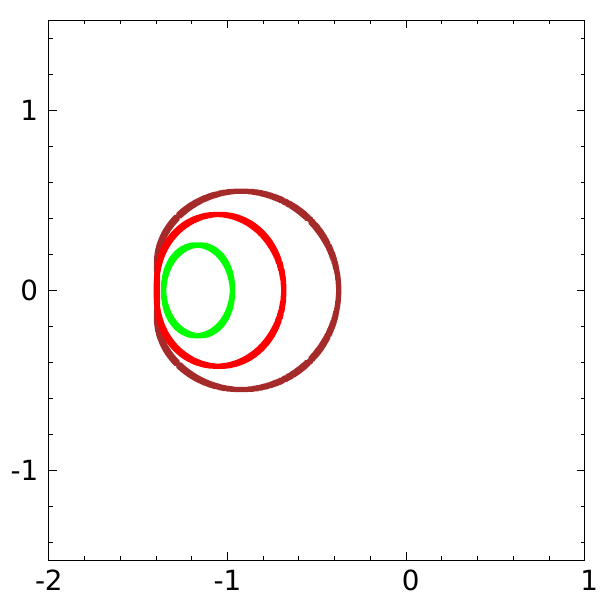}} &
            {\includegraphics[scale=0.36,trim=0 0 0 0]{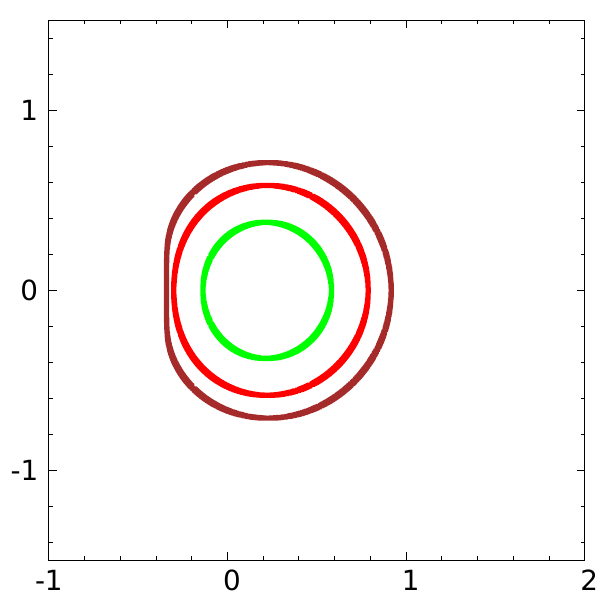}} &
            {\includegraphics[scale=0.36,trim=0 0 0 0]{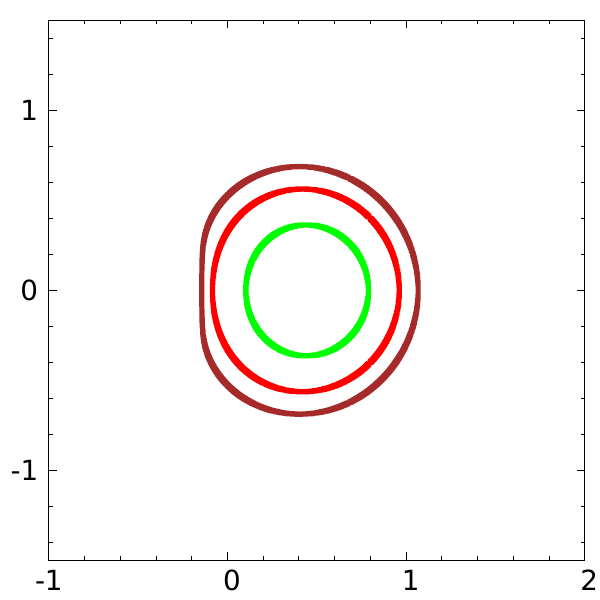}} &
            {\includegraphics[scale=0.39,trim=0 0 0 0]{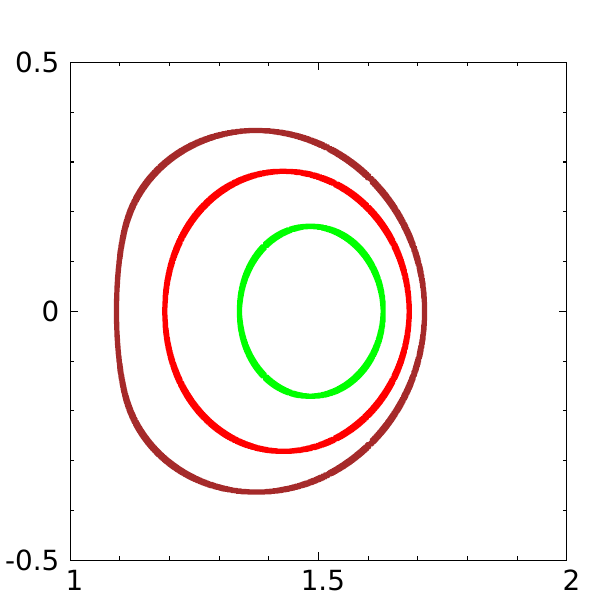}} \\
            \rotatebox{90}{~~~~~~~~~~Braneworld} & 
            {\includegraphics[scale=0.36,trim=0 0 0 0]{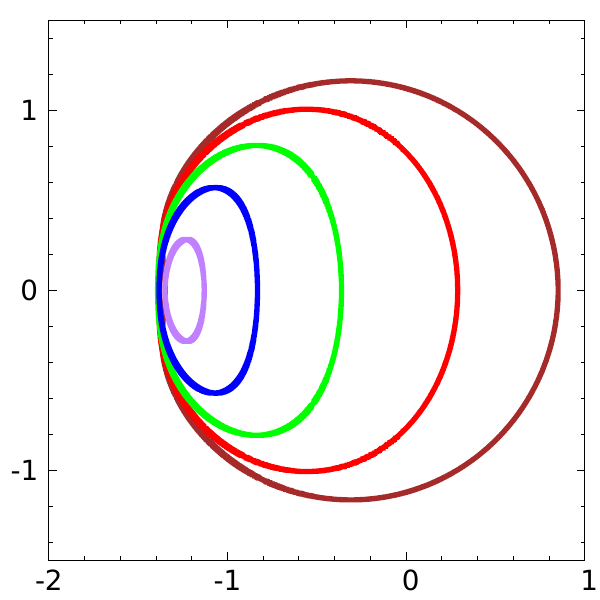}} &
            {\includegraphics[scale=0.36,trim=0 0 0 0]{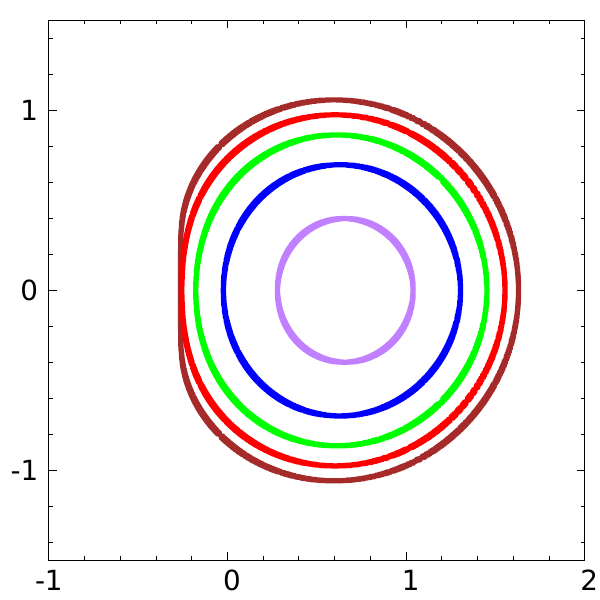}} &
            {\includegraphics[scale=0.36,trim=0 0 0 0]{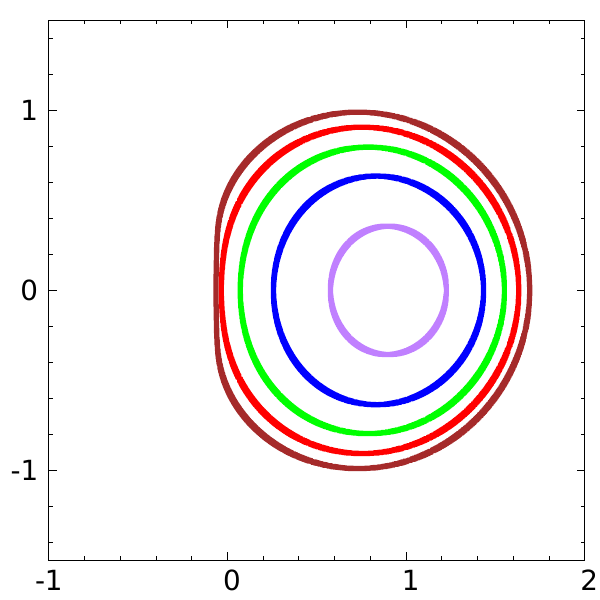}} &
            {\includegraphics[scale=0.39,trim=0 0 0 0]{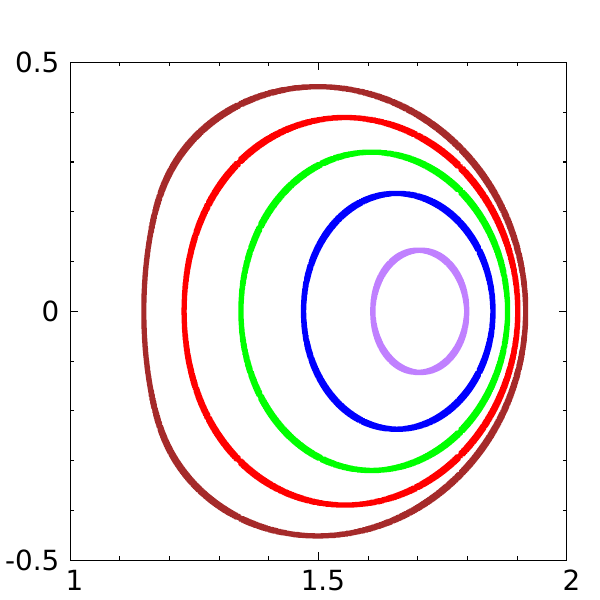}} \\
            \rotatebox{90}{~~~~~~~~~~~~Dilaton} &
            {\includegraphics[scale=0.36,trim=0 0 0 0]{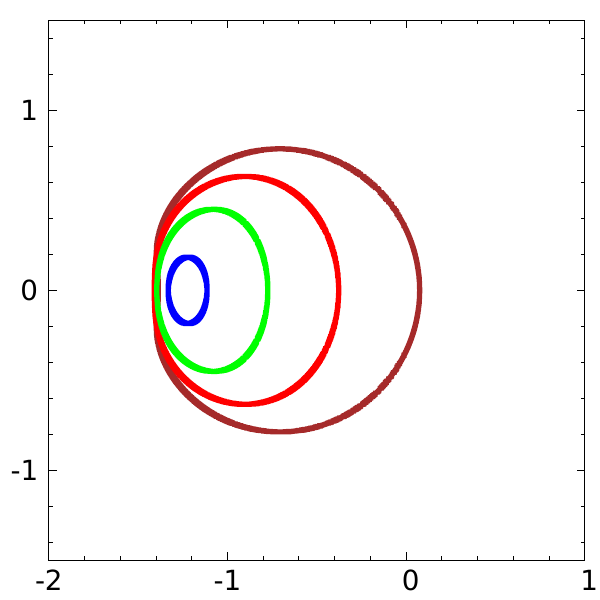}} &
            {\includegraphics[scale=0.36,trim=0 0 0 0]{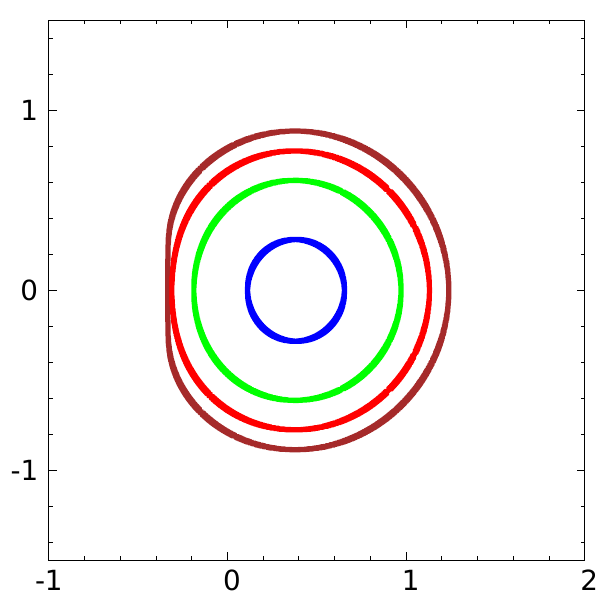}} &
            {\includegraphics[scale=0.36,trim=0 0 0 0]{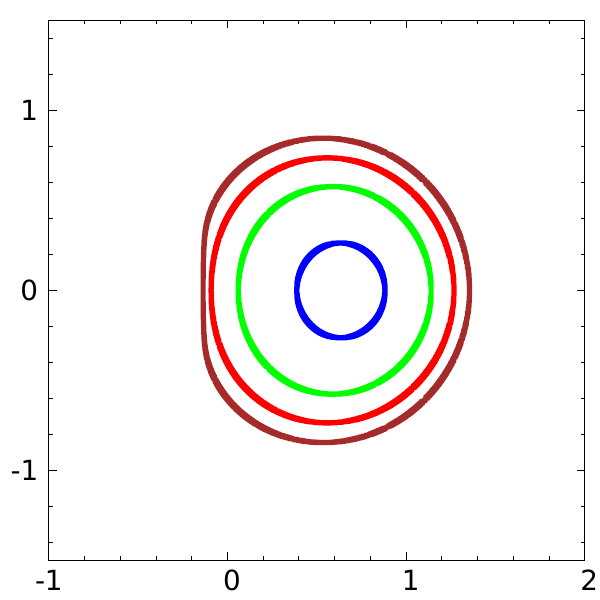}} &
            {\includegraphics[scale=0.39,trim=0 0 0 0]{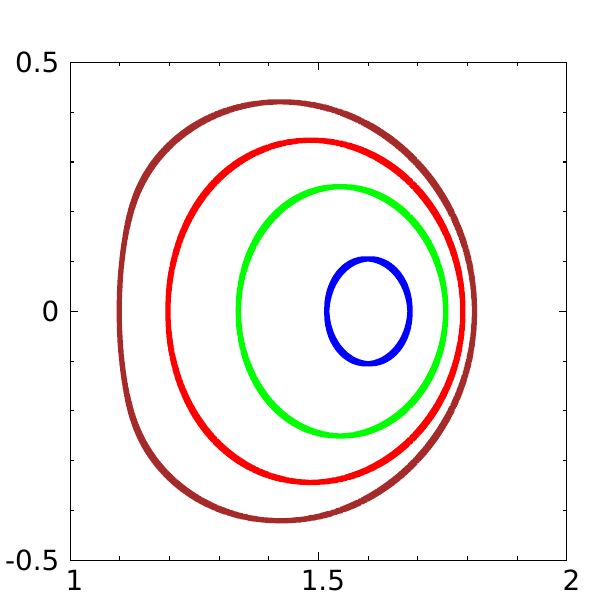}} \\
        \end{tabular}
        \caption{Relativistic aberration in plasma profile 1{ with $a/a_{max}=0.999$, $r_O=5m$ and $\vartheta_O=\pi/2$}. Brown, red, green, blue and purple curves correspond respectively to the frequency ratios $\omega_c^2/\omega_\infty^2=0.00$, $2.25$, $4.50$, $6.75$ and $8.90$. 
       {As explained in Fig.\eqref{fig:SH_p2},
        for some particular {$\omega^2_c/\omega^2_\infty = \chi_{cri}$}, a forbidden region emerges and consequently, the shadow is no longer visible.
        Note that for the plots shown in the right column (associated with a relative velocity of $v=-0.9$), in order to facilitate the visualization of the different shadows, we have modified the scale of the axes.
        }
        }
        \label{fig:Ab_p2}
    \end{figure*}

\begin{figure*}[htbp]
        \centering
        \begin{tabular}{ccccc}
             & $v=+0.9$ & $v=+0.1$ & $v=-0.1$ & $v=-0.9$ \\
            \rotatebox{90}{~~~~~~~~Kerr-Newman} & 
            {\includegraphics[scale=0.36,trim=0 0 0 0]{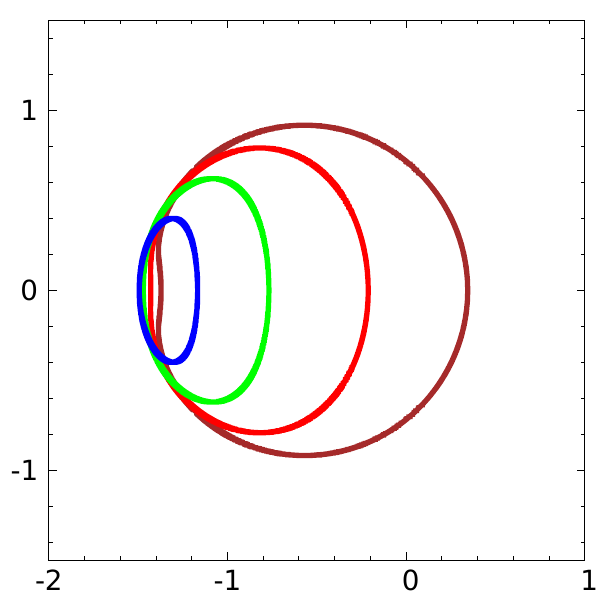}} &
            {\includegraphics[scale=0.36,trim=0 0 0 0]{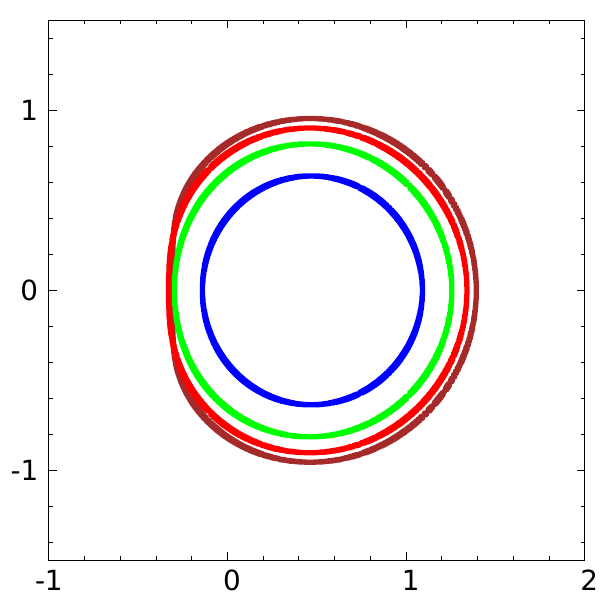}} &
            {\includegraphics[scale=0.36,trim=0 0 0 0]{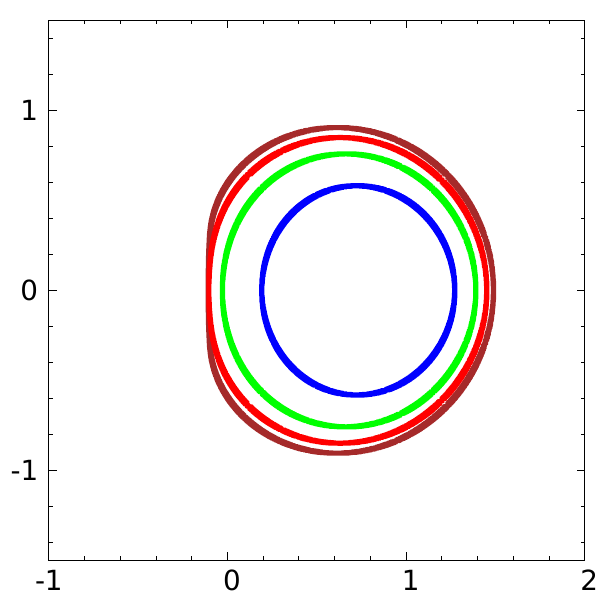}} &
            {\includegraphics[scale=0.39,trim=0 0 0 0]{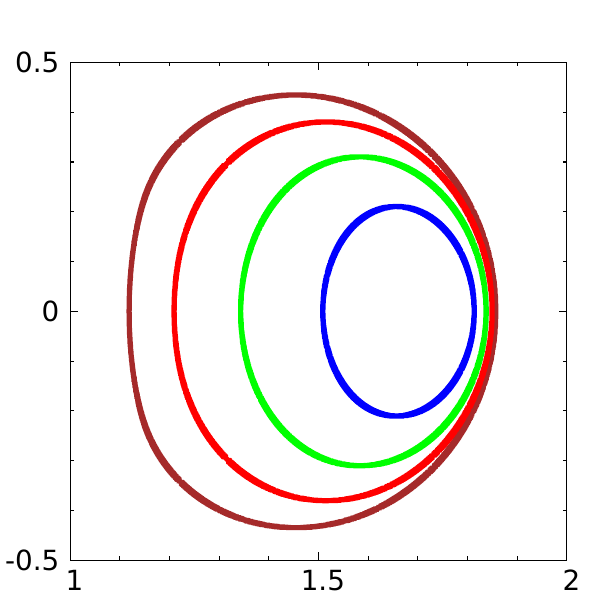}} \\
            \rotatebox{90}{~~~~~~~Modified Kerr} & 
            {\includegraphics[scale=0.36,trim=0 0 0 0]{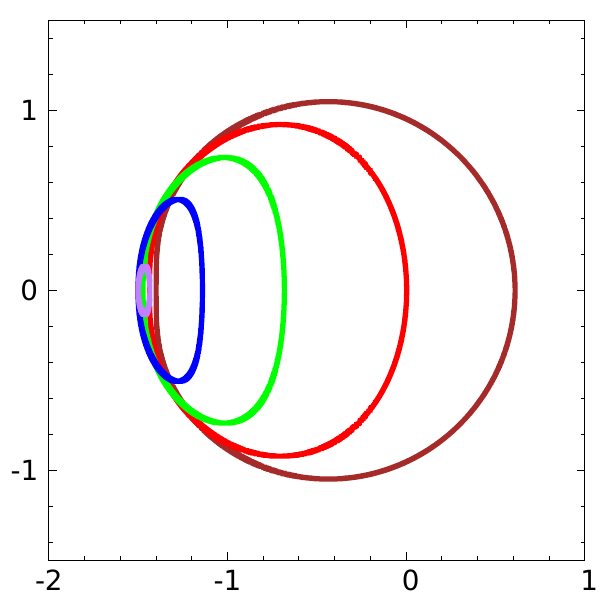}} &
            {\includegraphics[scale=0.36,trim=0 0 0 0]{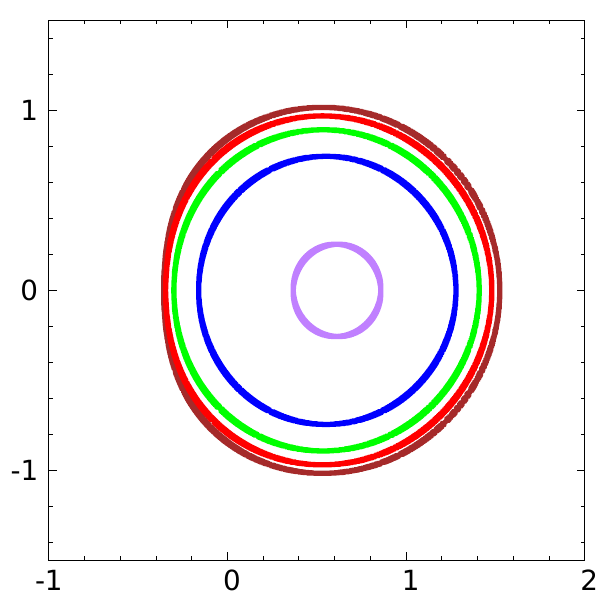}} &
            {\includegraphics[scale=0.36,trim=0 0 0 0]{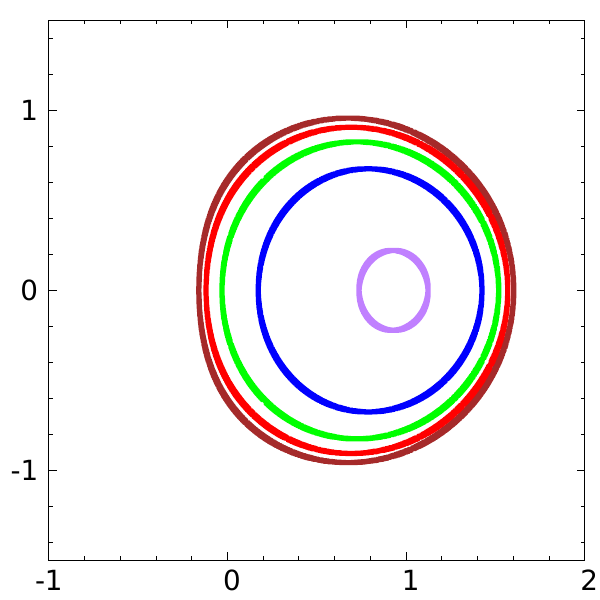}} &
            {\includegraphics[scale=0.39,trim=0 0 0 0]{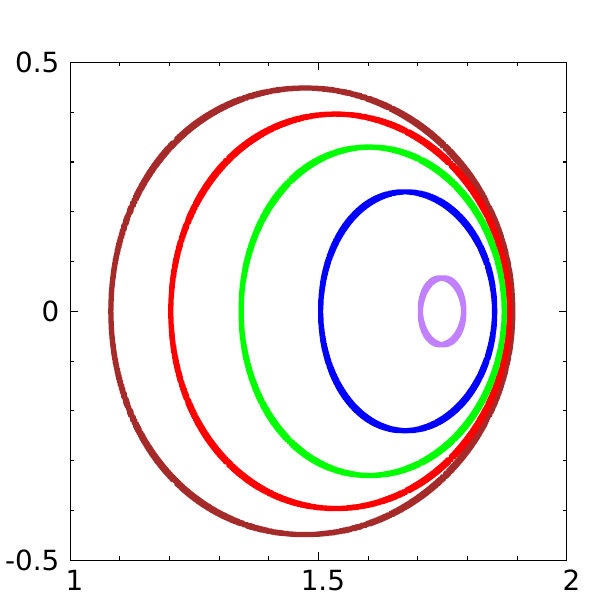}} \\
            \rotatebox{90}{~~~~~~~~~~~Kerr-Sen} & 
            {\includegraphics[scale=0.36,trim=0 0 0 0]{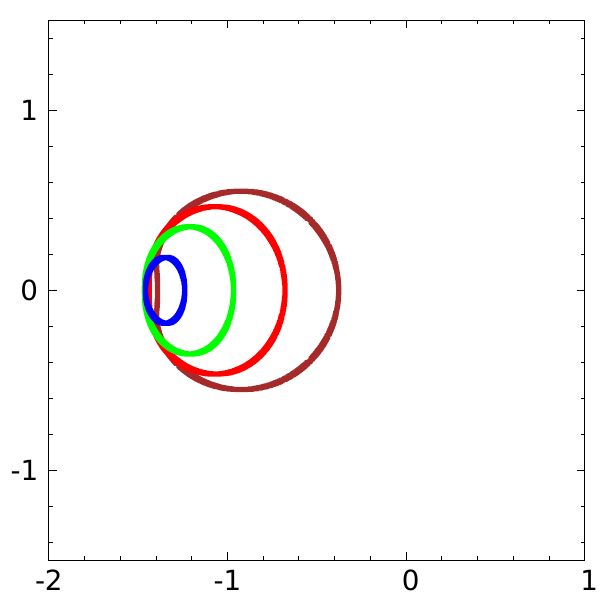}} &
            {\includegraphics[scale=0.36,trim=0 0 0 0]{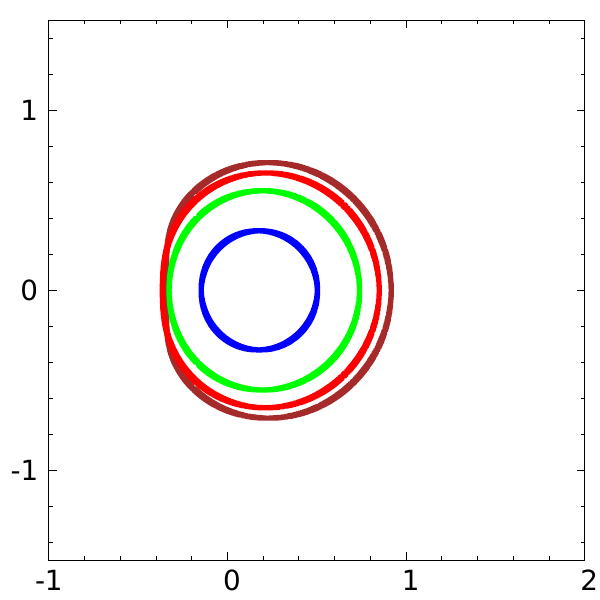}} &
            {\includegraphics[scale=0.36,trim=0 0 0 0]{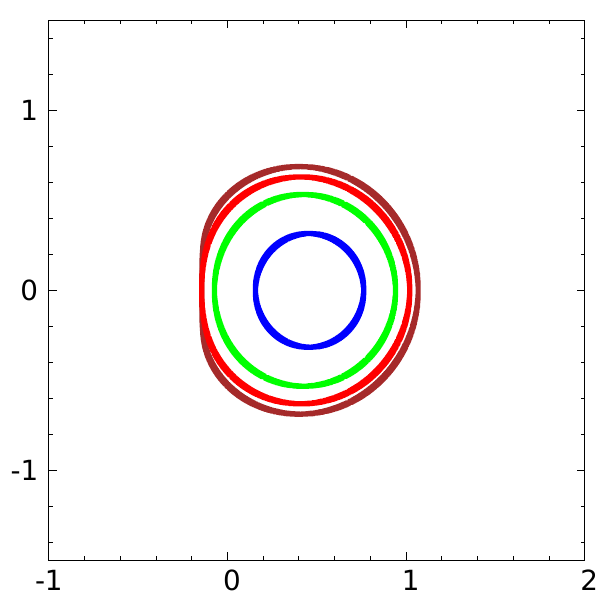}} &
            {\includegraphics[scale=0.39,trim=0 0 0 0]{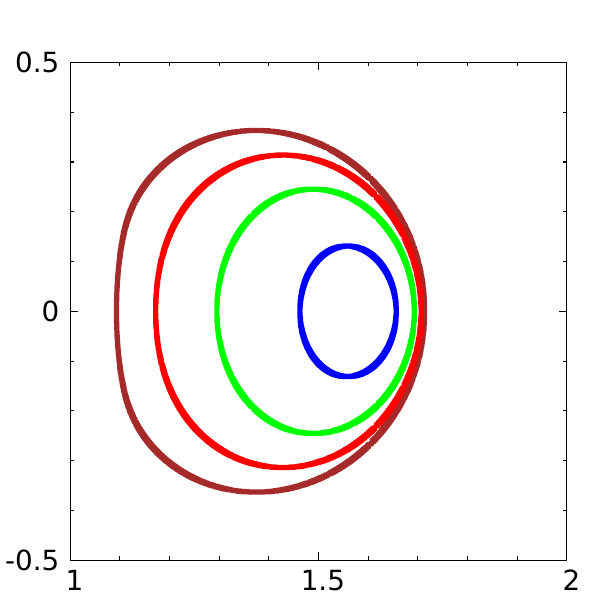}} \\
            \rotatebox{90}{~~~~~~~~~~Braneworld} & 
            {\includegraphics[scale=0.36,trim=0 0 0 0]{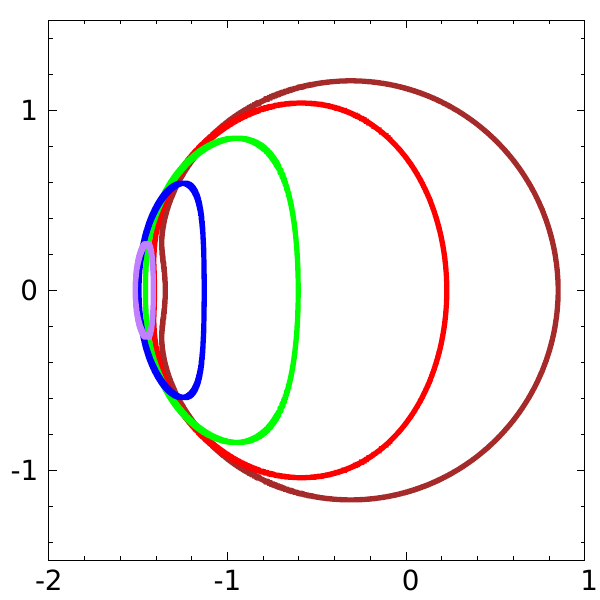}} &
            {\includegraphics[scale=0.36,trim=0 0 0 0]{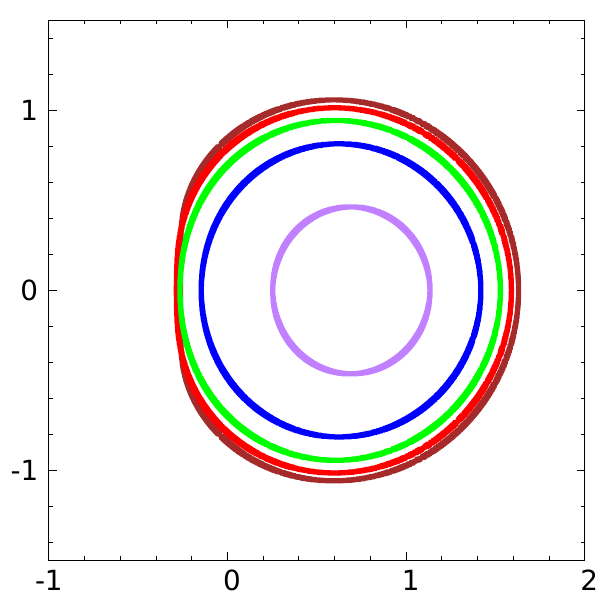}} &
            {\includegraphics[scale=0.36,trim=0 0 0 0]{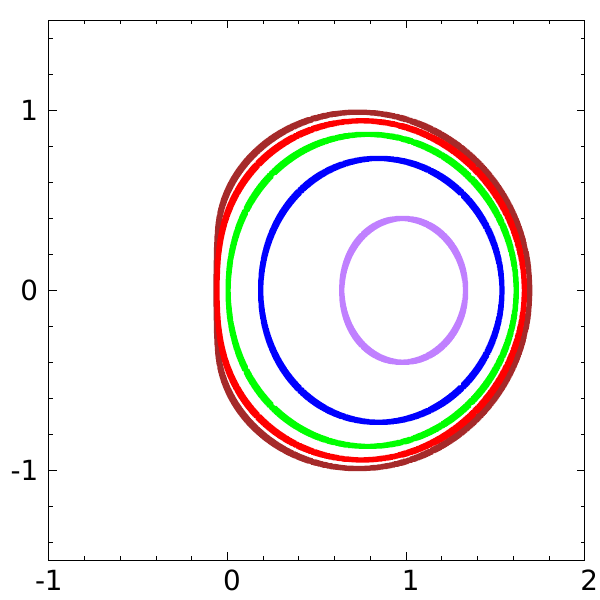}} &
            {\includegraphics[scale=0.39,trim=0 0 0 0]{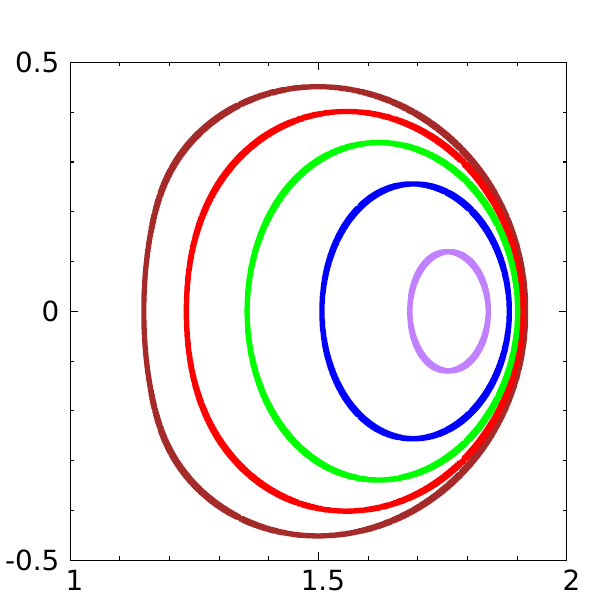}} \\
            \rotatebox{90}{~~~~~~~~~~~~Dilaton} &
            {\includegraphics[scale=0.36,trim=0 0 0 0]{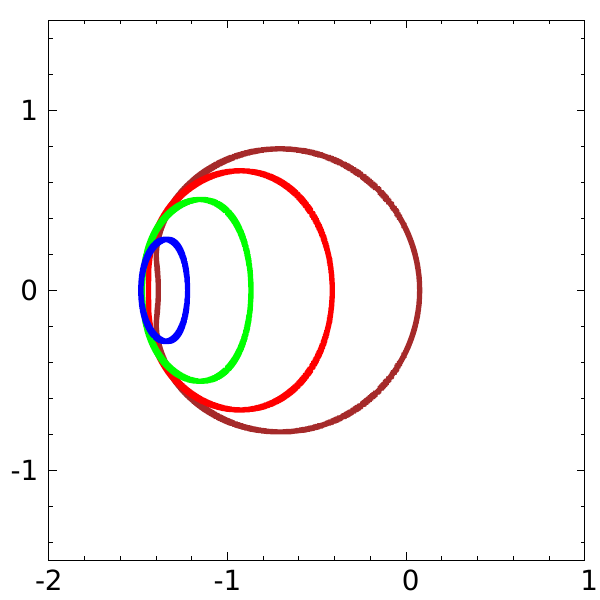}} &
            {\includegraphics[scale=0.36,trim=0 0 0 0]{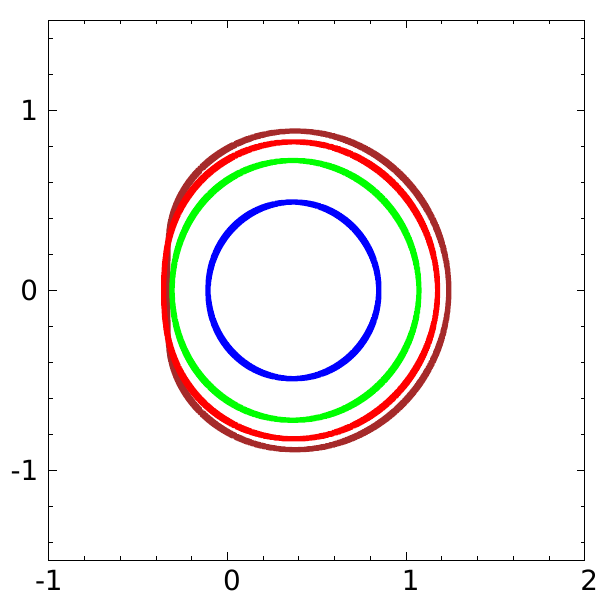}} &
            {\includegraphics[scale=0.36,trim=0 0 0 0]{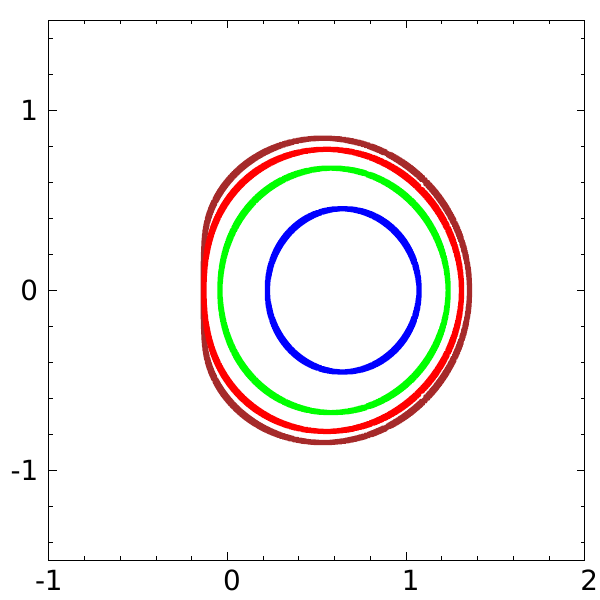}} &
            {\includegraphics[scale=0.39,trim=0 0 0 0]{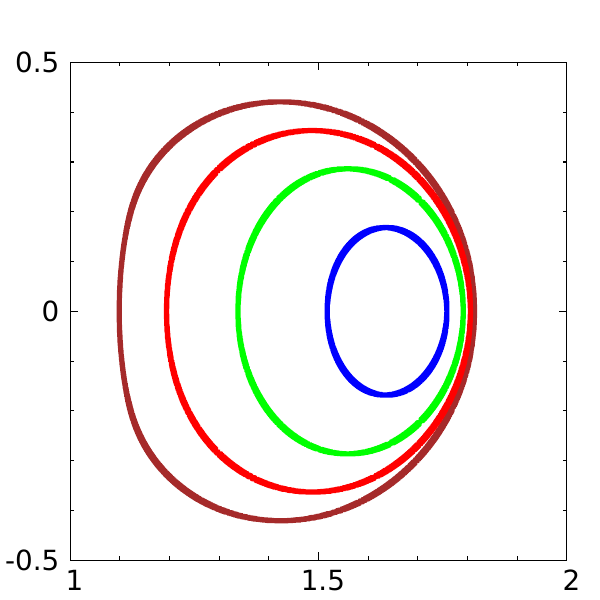}} \\
        \end{tabular}
        \caption{Relativistic aberration in plasma profile 2{ with $a/a_{max}=0.999$, $r_O=5m$ and $\vartheta_O=\pi/2$}. Brown, red, green, blue and purple curves correspond respectively to the frequency ratios $\omega_c^2/\omega_\infty^2=0.00$, $3.75$, $7.50$, $11.25$ and $15.00$.
       {As explained in Fig.\eqref{fig:SH_p3},
        for some particular {$\omega^2_c/\omega^2_\infty = \chi_{cri}$}, the forbidden region reaches the equatorial plane and consequently, the shadow is no longer visible.
        As in Fig.\eqref{fig:Ab_p2} we have modified the scale of the axes in the plots of the last column.
        }
        }
        \label{fig:Ab_p3}
    \end{figure*}

\begin{figure*}[htbp]
        \centering
        \begin{tabular}{ccccc}
             & $v=+0.9$ & $v=+0.1$ & $v=-0.1$ & $v=-0.9$ \\
            \rotatebox{90}{~~~~~~~~Kerr-Newman} & 
            {\includegraphics[scale=0.36,trim=0 0 0 0]{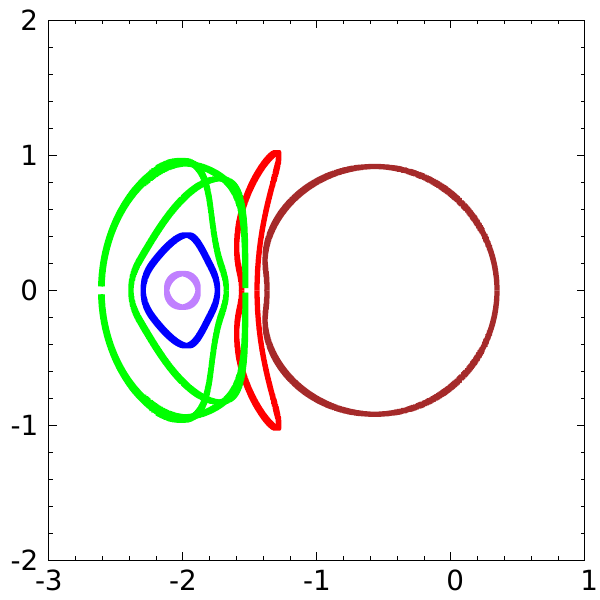}} &
            {\includegraphics[scale=0.36,trim=0 0 0 0]{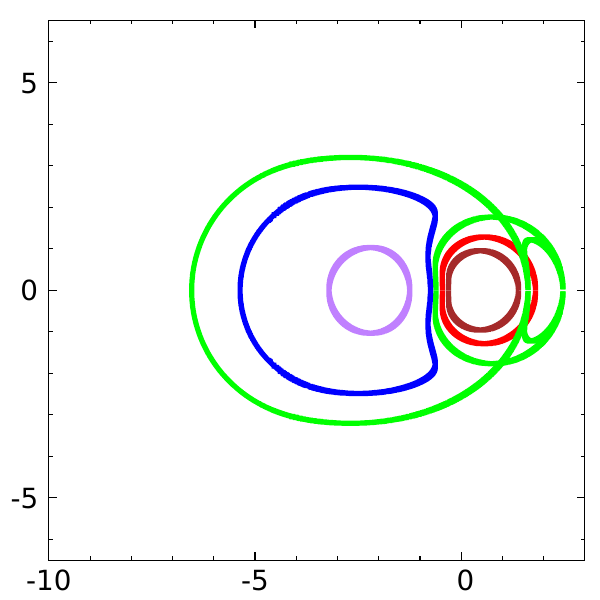}} &
            {\includegraphics[scale=0.36,trim=0 0 0 0]{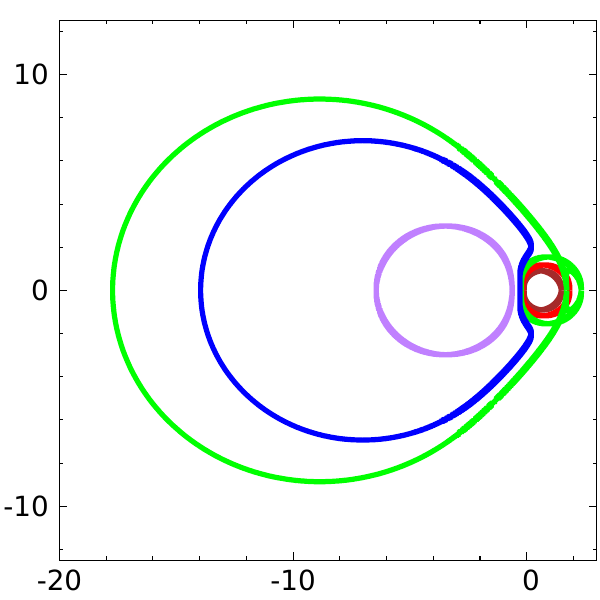}} &
            {\includegraphics[scale=0.36,trim=0 0 0 0]{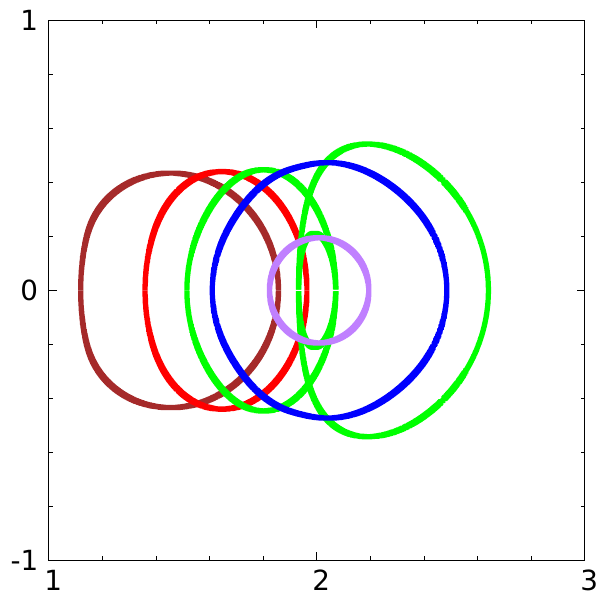}} \\
            \rotatebox{90}{~~~~~~~Modified Kerr} & 
            {\includegraphics[scale=0.36,trim=0 0 0 0]{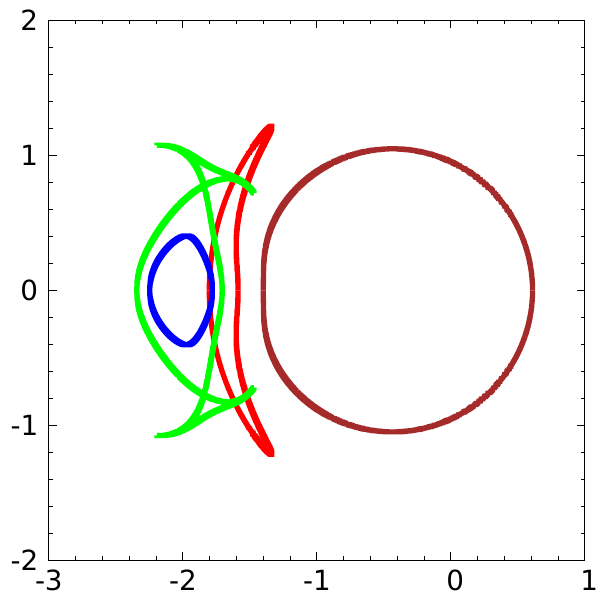}} &
            {\includegraphics[scale=0.36,trim=0 0 0 0]{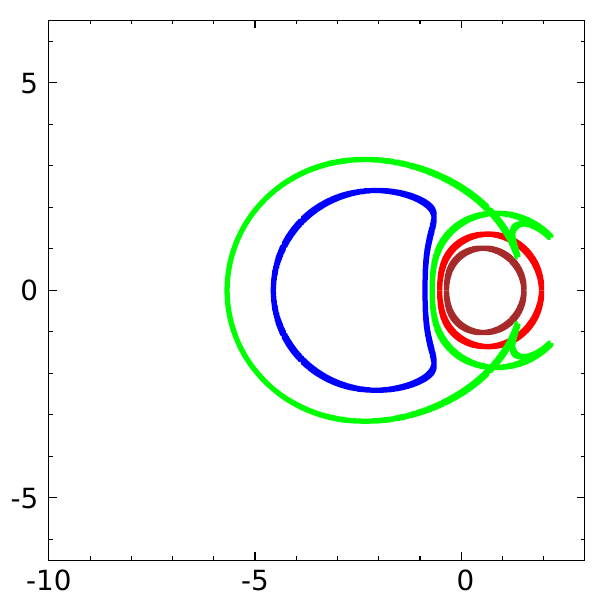}} &
            {\includegraphics[scale=0.36,trim=0 0 0 0]{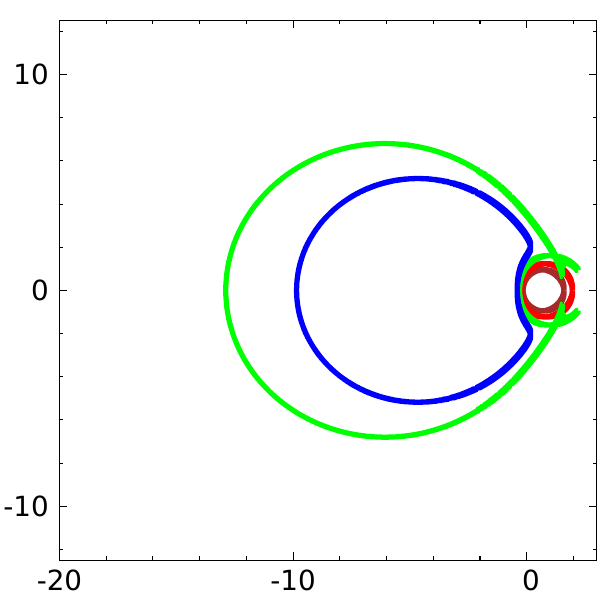}} &
            {\includegraphics[scale=0.36,trim=0 0 0 0]{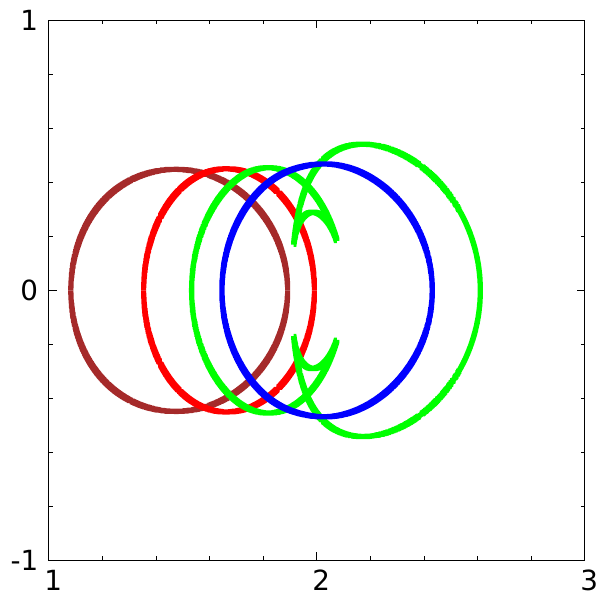}} \\
            \rotatebox{90}{~~~~~~~~~~~Kerr-Sen} & 
            {\includegraphics[scale=0.36,trim=0 0 0 0]{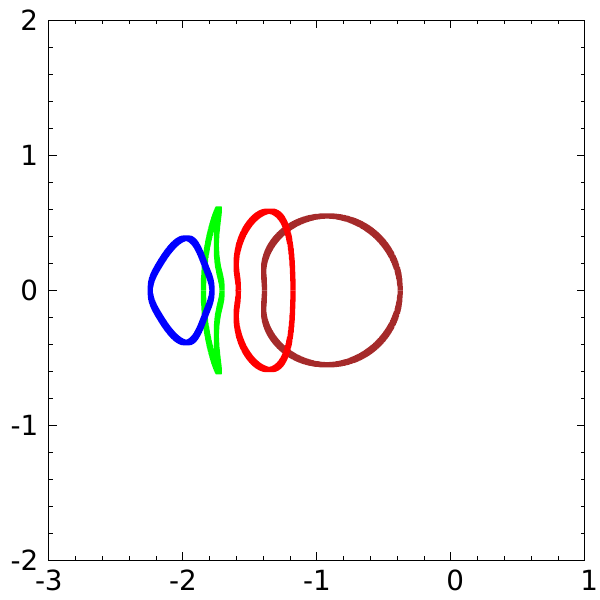}} &
            {\includegraphics[scale=0.36,trim=0 0 0 0]{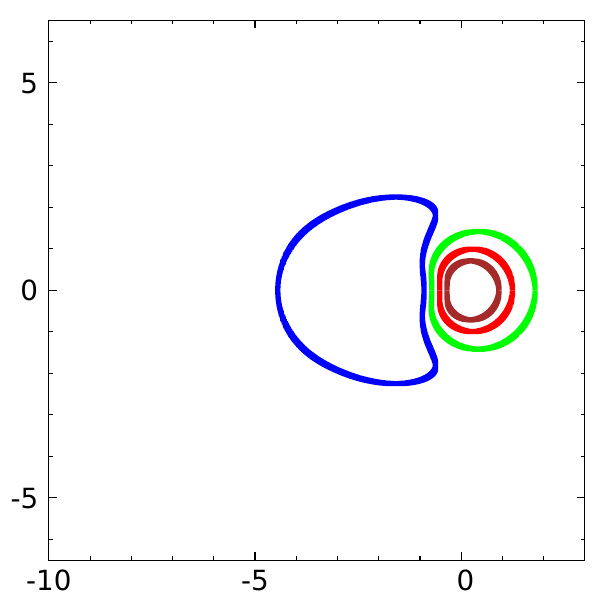}} &
            {\includegraphics[scale=0.36,trim=0 0 0 0]{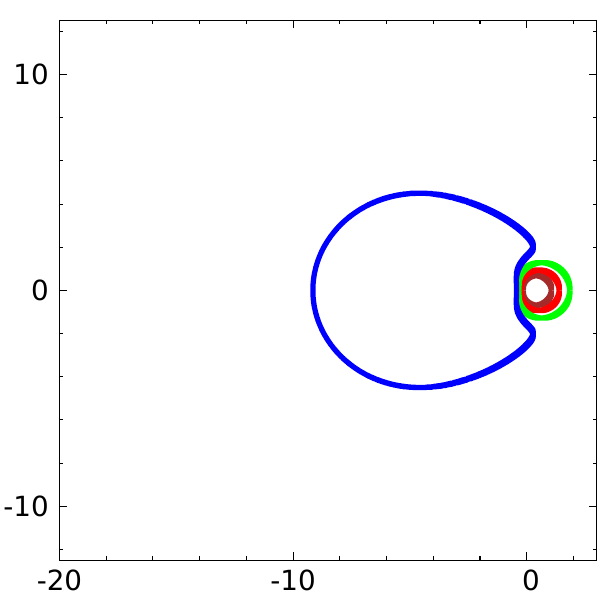}} &
            {\includegraphics[scale=0.36,trim=0 0 0 0]{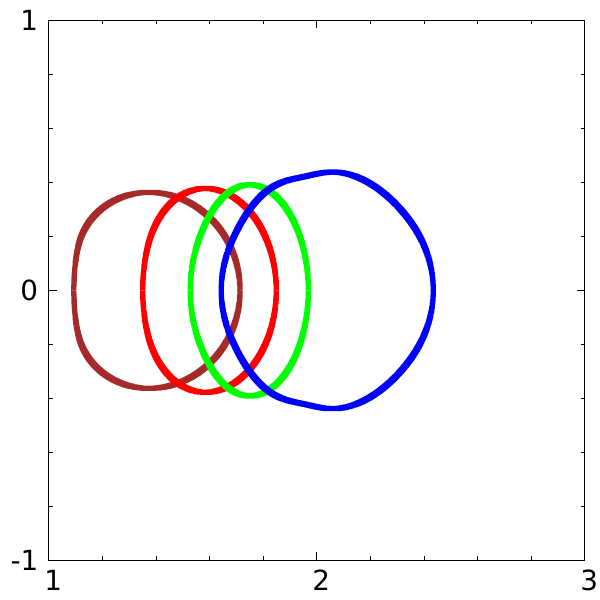}} \\
        \end{tabular}
        \caption{Relativistic aberration in the plasma profile 3{with $a/a_{max}=0.999$, $r_O=5m$, and $\vartheta_O=\pi/2$}. The brown, red, green, blue, and purple curves correspond, respectively, to the frequency ratios $\omega_c^2/\omega_\infty^2=0.000$, $0.800$, $1.085$, $1.200$, and $1.345$.
       {For $v=0.9$, $0.1$, and $-0.1$, the brown and red curves always enclose the shadow, leaving the illuminated sky outside. Meanwhile, the blue and purple curves always enclose the illuminated sky, leaving the shadow outside. For $v=-0.9$, all curves enclose the shadow. The brightness and shadow locations for the green curve for $v=0.9$, $0.1$, and $-0.1$ are similar to the ones explained in Fig. \eqref{fig:SH_p4} (the leftmost circle and triangular regions enclose light, while the rightmost circle and lenticular region enclose shadow). The Kerr-Newman $v=0.9$ green curve is an extreme variant of the Modified Kerr $v=0.9$ one, where the extremes of the triangular regions above and below touch each other, such that the innermost region encloses light, the left and right regions enclose shadow, the triangular regions enclose light, and the outside is shadow. For $v=-0.9$, all green curves enclose the shadow, except for the triangular regions that enclose light. The non-plotted curves correspond to {$\omega^2_c/\omega^2_\infty > \chi_{cri}$}, such that the forbidden region surrounding the black hole now contains the observer (Fig. \ref{fig:SH_pr}), and consequently, the shadow is not visible.
        Note that for the plots shown in each column (associated with different relative velocities), in order to facilitate the visualization of the different shadows, we have modified the scale of the axes.
        }
        }
        \label{fig:Ab_p4}
    \end{figure*}

\begin{figure*}[htbp]
        \centering
        \begin{tabular}{ccccc}
             & $v=+0.9$ & $v=+0.1$ & $v=-0.1$ & $v=-0.9$ \\
            \rotatebox{90}{~~~~~~~~Kerr-Newman} & 
            {\includegraphics[scale=0.36,trim=0 0 0 0]{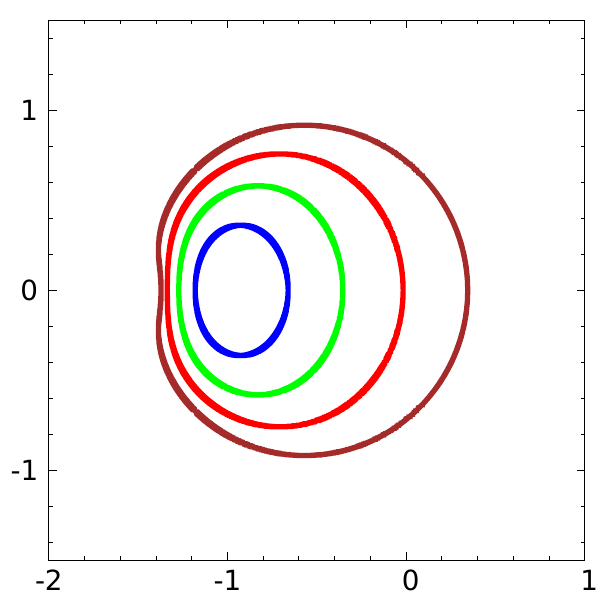}} &
            {\includegraphics[scale=0.36,trim=0 0 0 0]{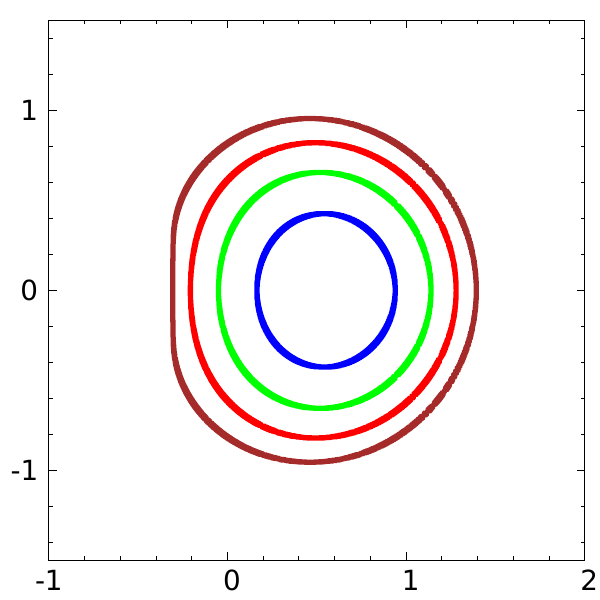}} &
            {\includegraphics[scale=0.36,trim=0 0 0 0]{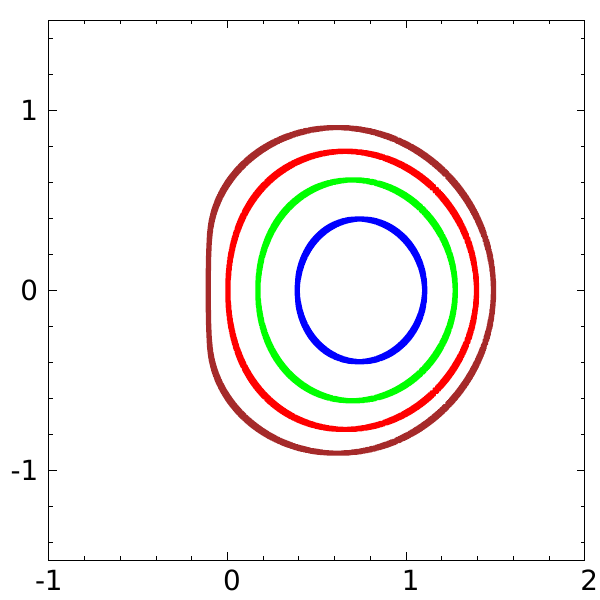}} &
            {\includegraphics[scale=0.39,trim=0 0 0 0]{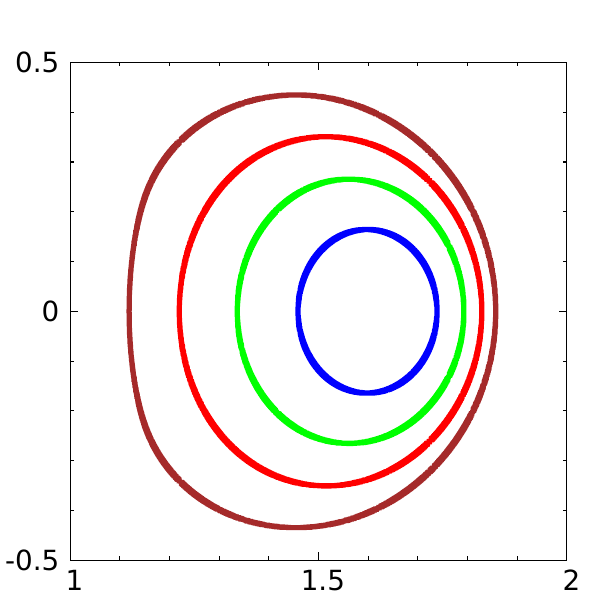}} \\
            \rotatebox{90}{~~~~~~~Modified Kerr} & 
            {\includegraphics[scale=0.36,trim=0 0 0 0]{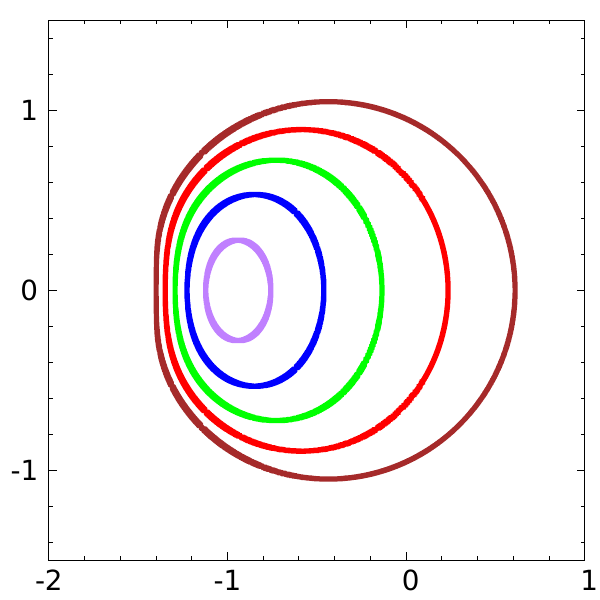}} &
            {\includegraphics[scale=0.36,trim=0 0 0 0]{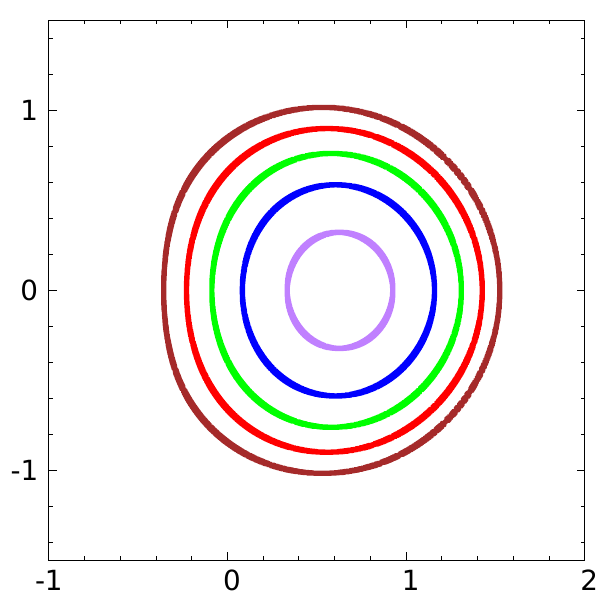}} &
            {\includegraphics[scale=0.36,trim=0 0 0 0]{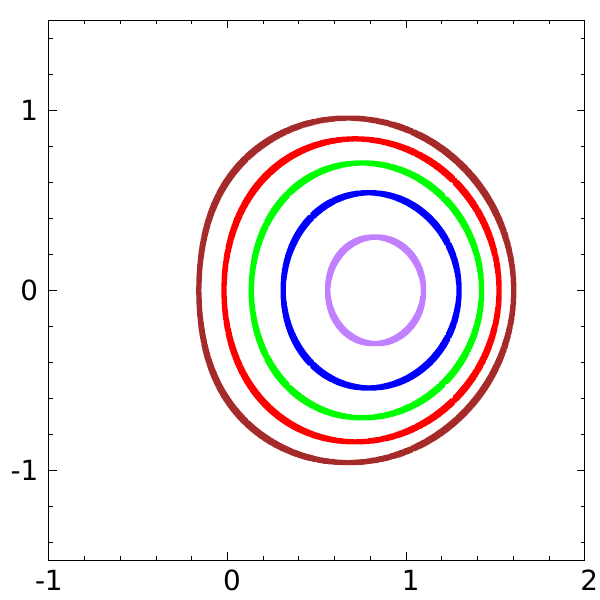}} &
            {\includegraphics[scale=0.39,trim=0 0 0 0]{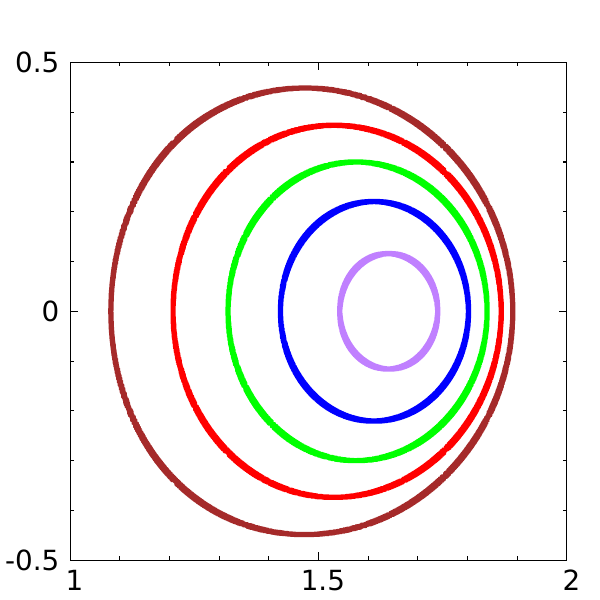}} \\
            \rotatebox{90}{~~~~~~~~~~~Kerr-Sen} & 
            {\includegraphics[scale=0.36,trim=0 0 0 0]{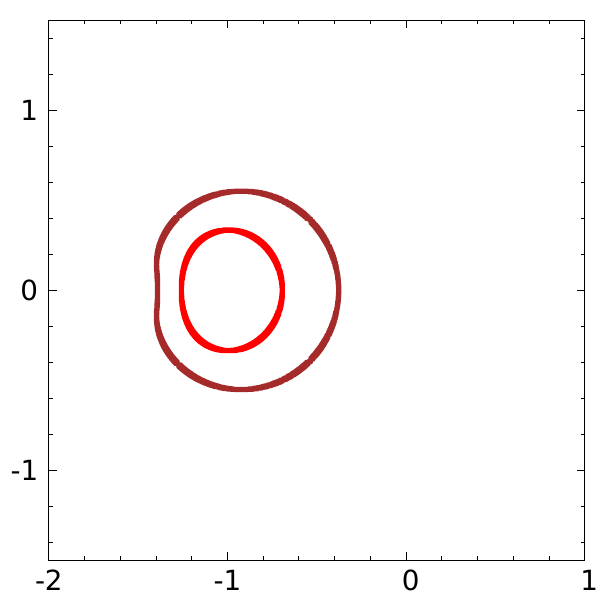}} &
            {\includegraphics[scale=0.36,trim=0 0 0 0]{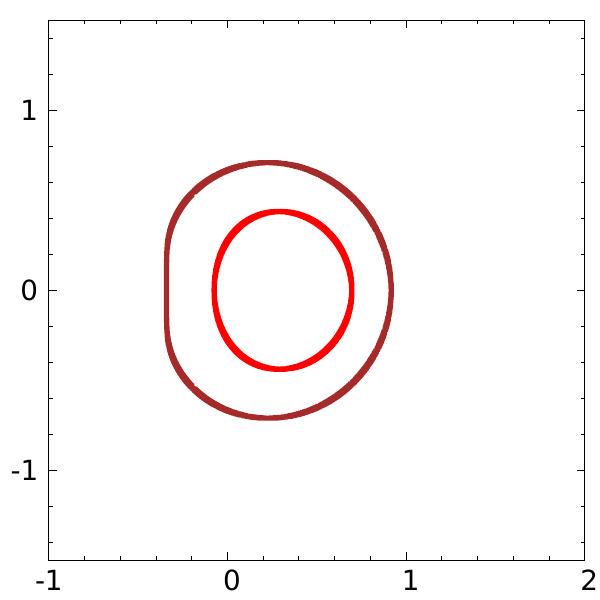}} &
            {\includegraphics[scale=0.36,trim=0 0 0 0]{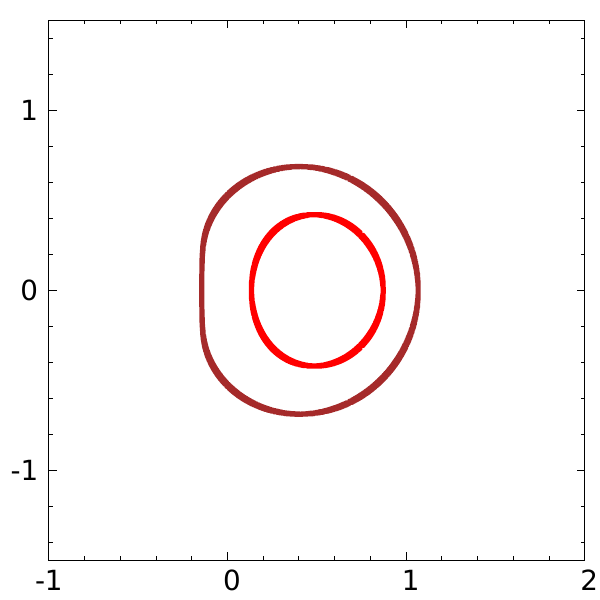}} &
            {\includegraphics[scale=0.39,trim=0 0 0 0]{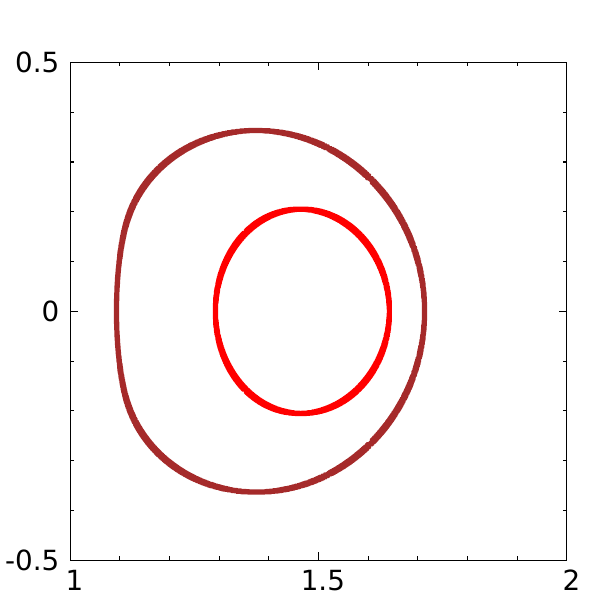}} \\
            \rotatebox{90}{~~~~~~~~~~Braneworld} & 
            {\includegraphics[scale=0.36,trim=0 0 0 0]{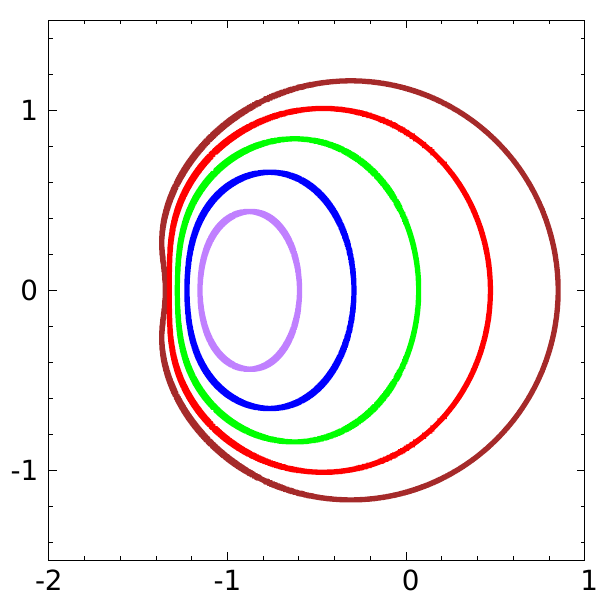}} &
            {\includegraphics[scale=0.36,trim=0 0 0 0]{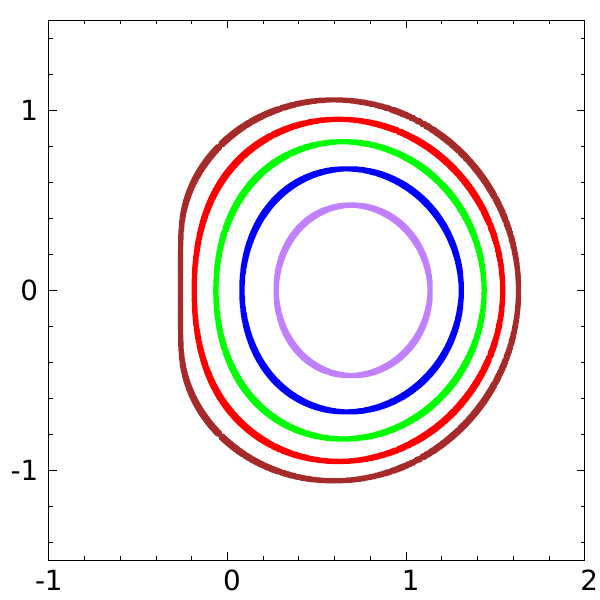}} &
            {\includegraphics[scale=0.36,trim=0 0 0 0]{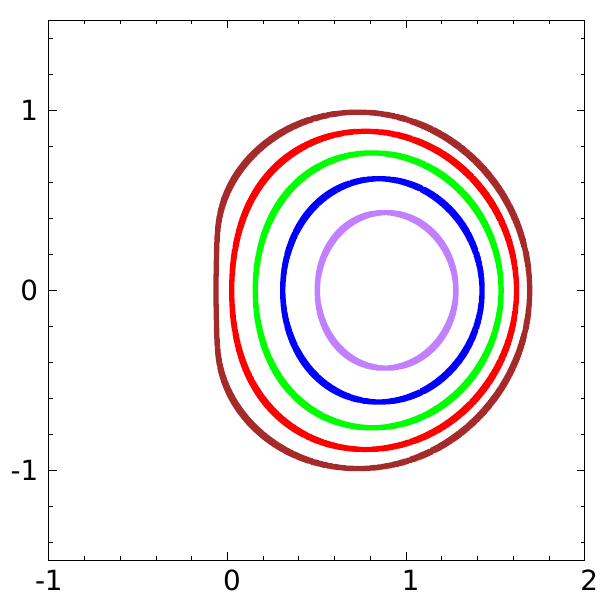}} &
            {\includegraphics[scale=0.39,trim=0 0 0 0]{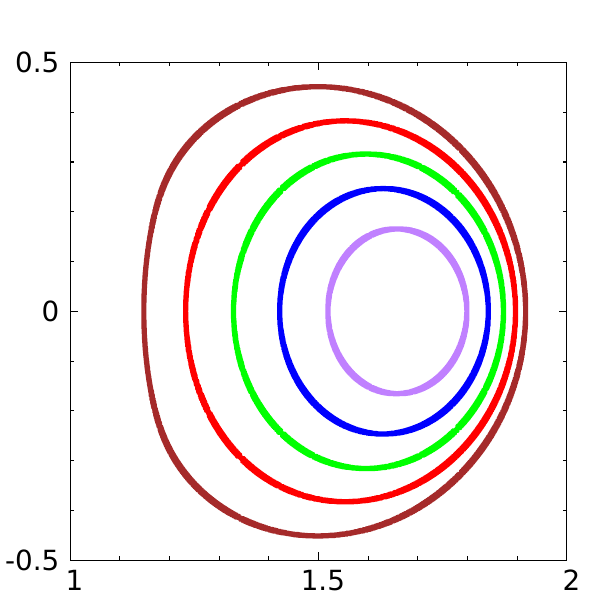}} \\
            \rotatebox{90}{~~~~~~~~~~~~Dilaton} &
            {\includegraphics[scale=0.36,trim=0 0 0 0]{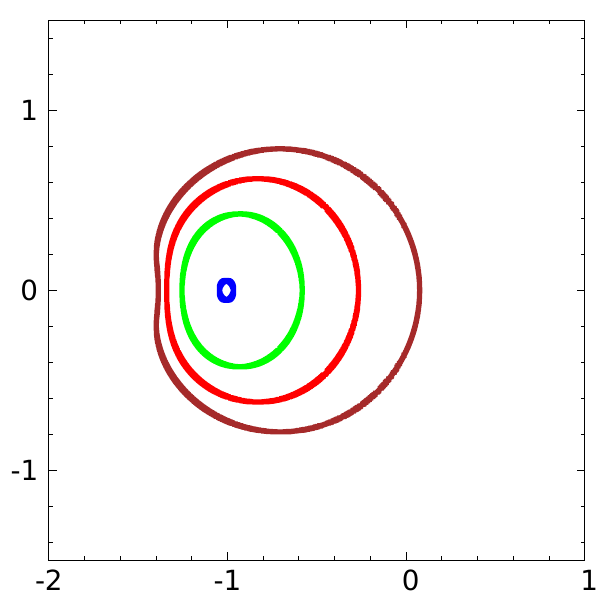}} &
            {\includegraphics[scale=0.36,trim=0 0 0 0]{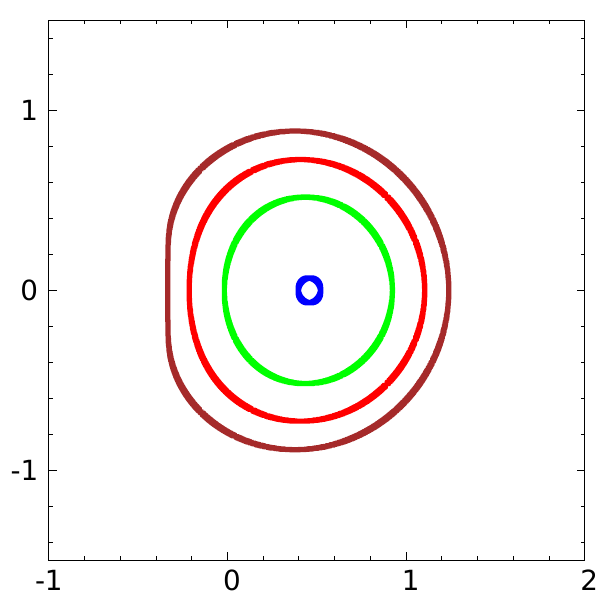}} &
            {\includegraphics[scale=0.36,trim=0 0 0 0]{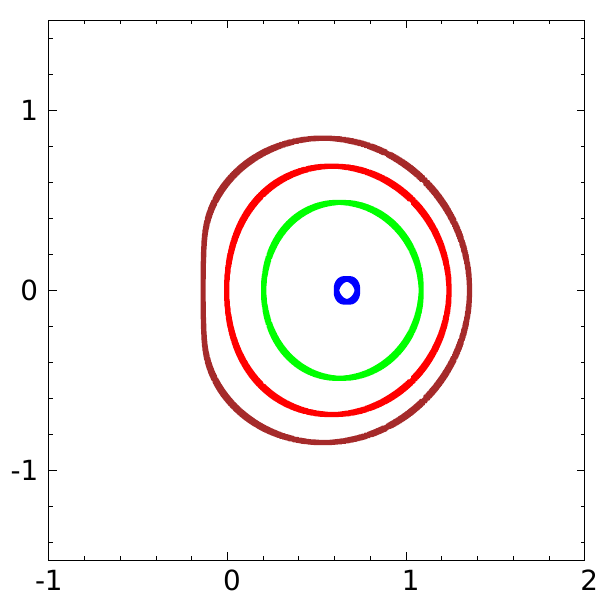}} &
            {\includegraphics[scale=0.39,trim=0 0 0 0]{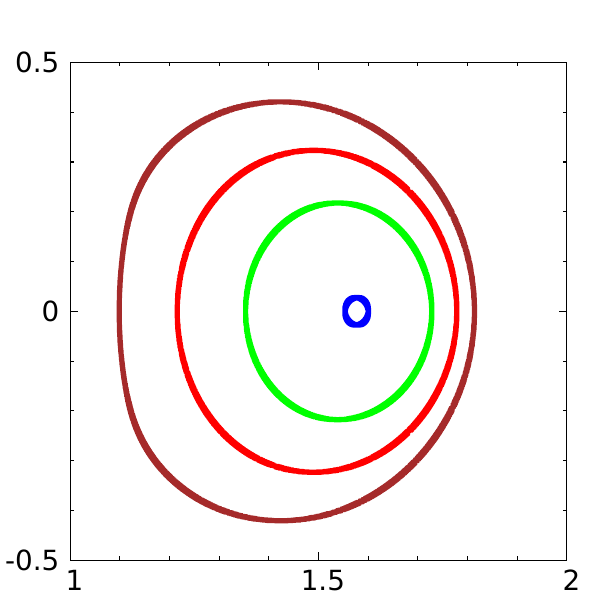}} \\
        \end{tabular}
        \caption{Relativistic aberration in the plasma profile 4{ with $a/a_{max}=0.999$, $r_O=5m$ and $\vartheta_O=\pi/2$}. Brown, red, green, blue and purple curves correspond respectively to the frequency ratios $\omega_c^2/\omega_\infty^2=0.0$, $17.5$, $35.0$, $52.5$ and $70.0$.
       {As explained in Fig.\eqref{fig:SH_p3},
        for some particular {$\omega^2_c/\omega^2_\infty = \chi_{cri}$}, the forbidden region reaches the equatorial plane and consequently, the shadow is no longer visible. As in Fig.\eqref{fig:Ab_p2} we have modified the scale of the axes in the plots of the last column.
        }
        }
        \label{fig:Ab_p5}
    \end{figure*}

\begin{figure*}[htbp]
        \centering
        \begin{tabular}{ccccc}
             & $v=+0.9$ & $v=+0.1$ & $v=-0.1$ & $v=-0.9$ \\
            \rotatebox{90}{~~~~~~~~Kerr-Newman} & 
            {\includegraphics[scale=0.36,trim=0 0 0 0]{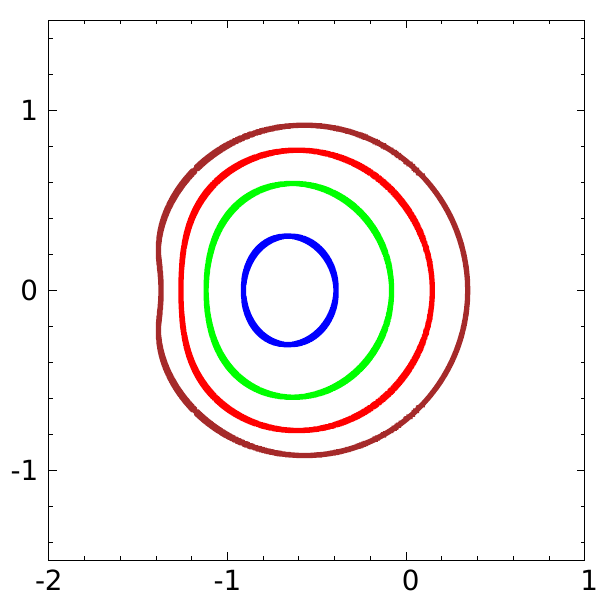}} &
            {\includegraphics[scale=0.36,trim=0 0 0 0]{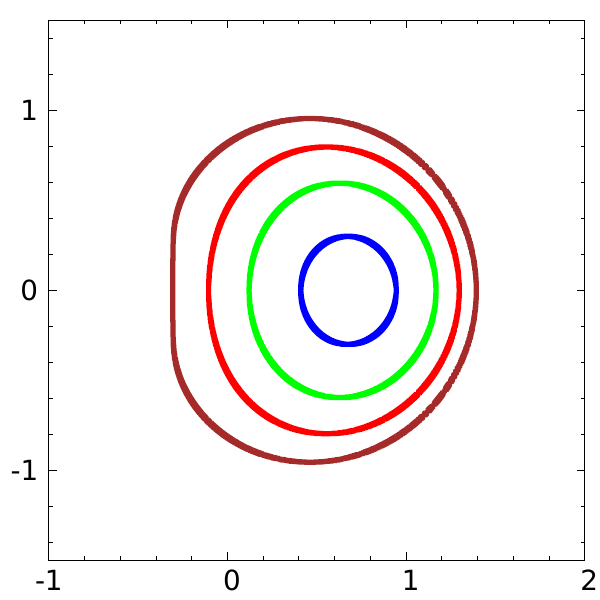}} &
            {\includegraphics[scale=0.36,trim=0 0 0 0]{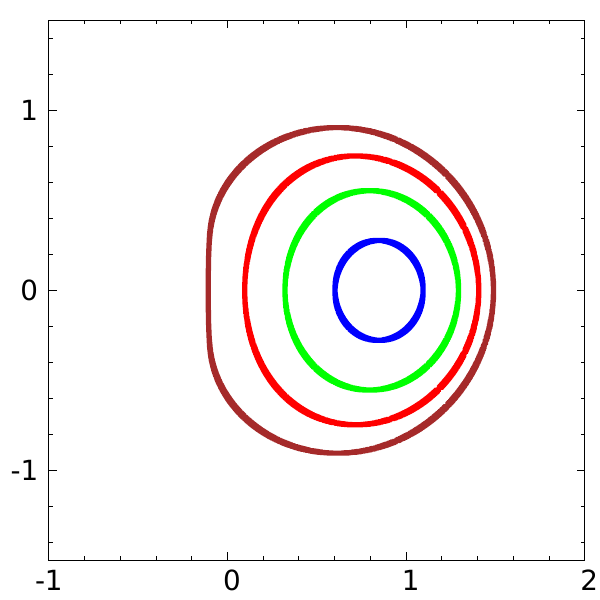}} &
            {\includegraphics[scale=0.39,trim=0 0 0 0]{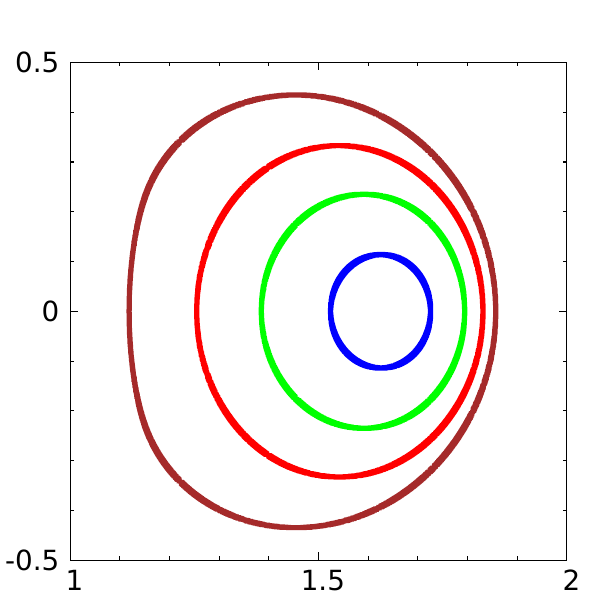}} \\
            \rotatebox{90}{~~~~~~~Modified Kerr} & 
            {\includegraphics[scale=0.36,trim=0 0 0 0]{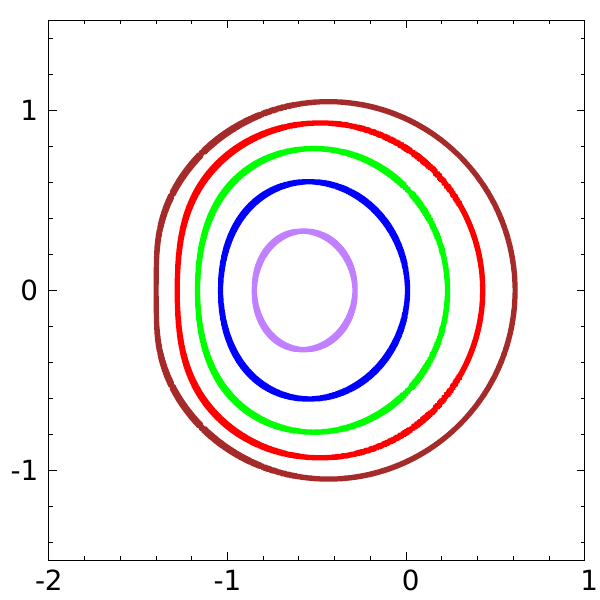}} &
            {\includegraphics[scale=0.36,trim=0 0 0 0]{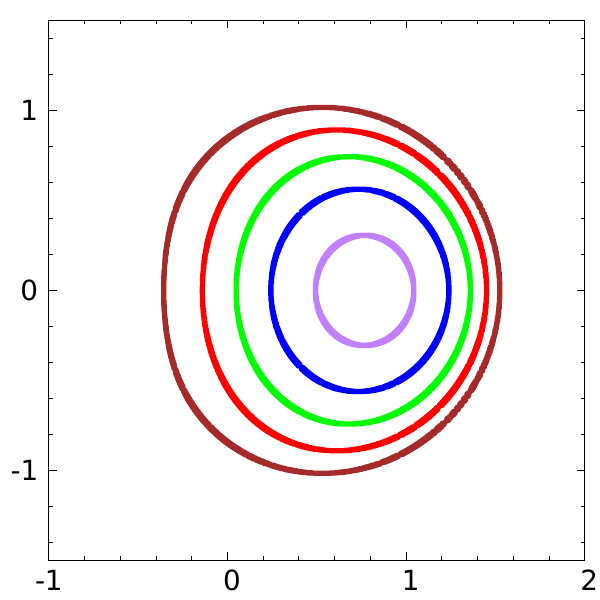}} &
            {\includegraphics[scale=0.36,trim=0 0 0 0]{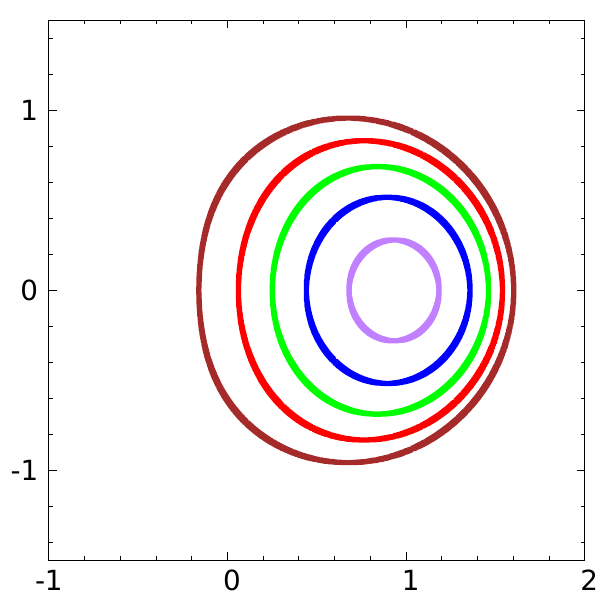}} &
            {\includegraphics[scale=0.39,trim=0 0 0 0]{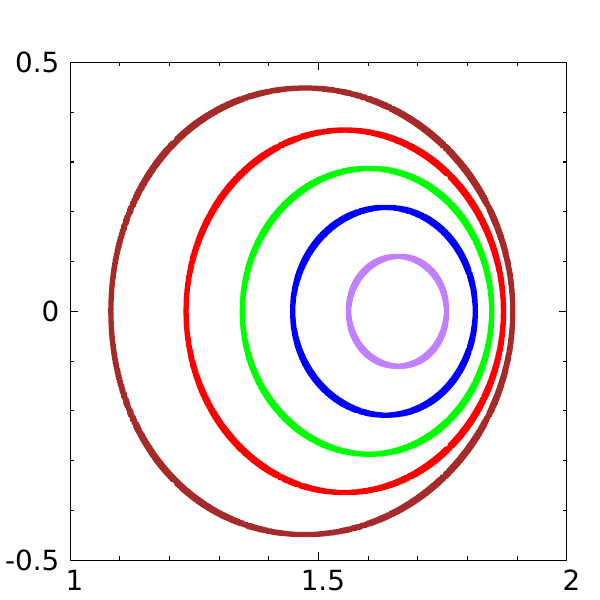}} \\
            \rotatebox{90}{~~~~~~~~~~~Kerr-Sen} & 
            {\includegraphics[scale=0.36,trim=0 0 0 0]{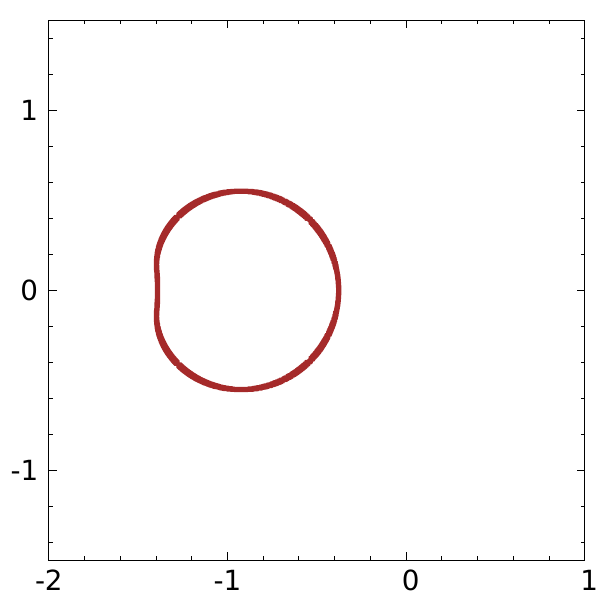}} &
            {\includegraphics[scale=0.36,trim=0 0 0 0]{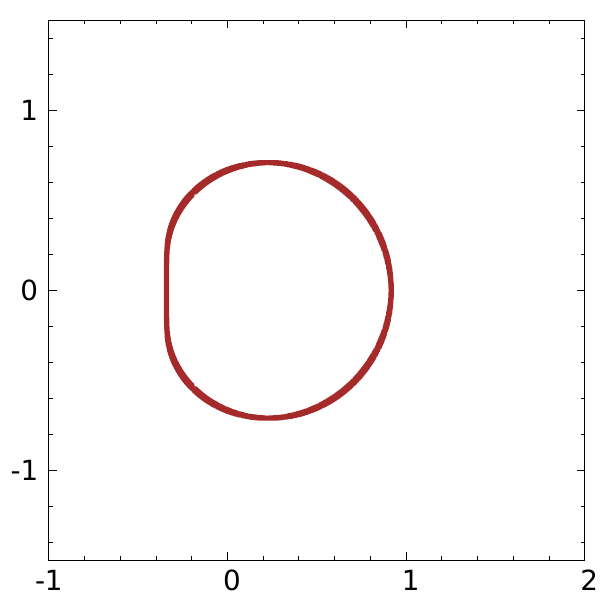}} &
            {\includegraphics[scale=0.36,trim=0 0 0 0]{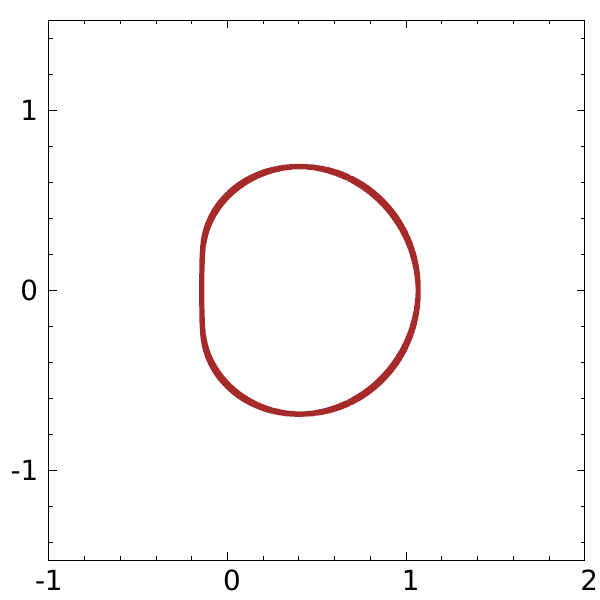}} &
            {\includegraphics[scale=0.39,trim=0 0 0 0]{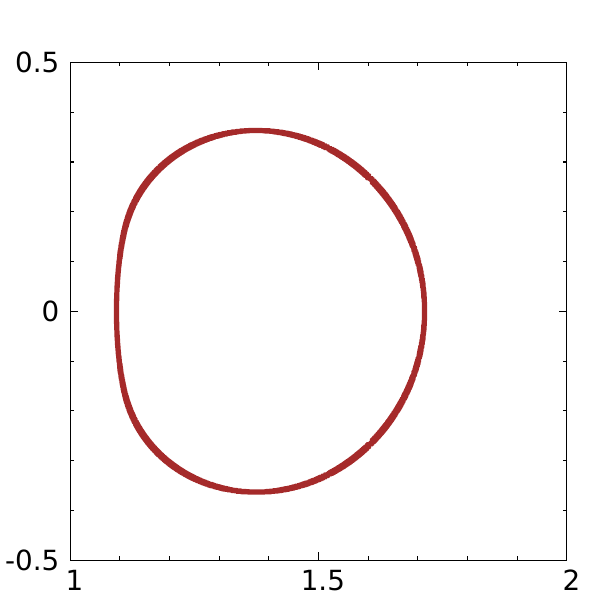}} \\
            \rotatebox{90}{~~~~~~~~~~Braneworld} & 
            {\includegraphics[scale=0.36,trim=0 0 0 0]{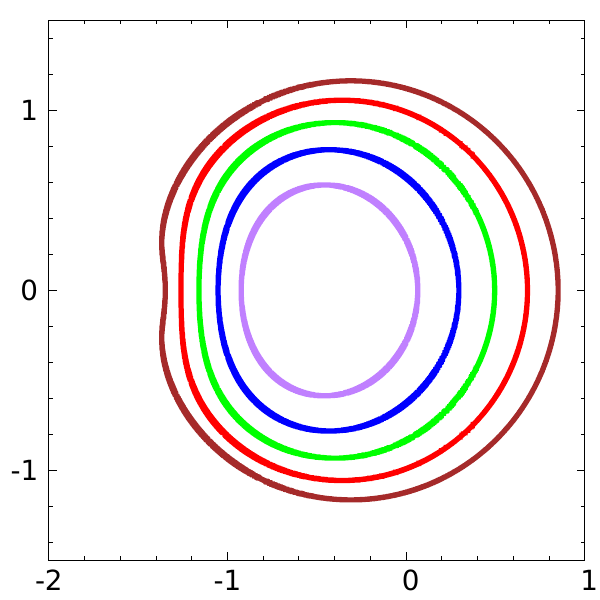}} &
            {\includegraphics[scale=0.36,trim=0 0 0 0]{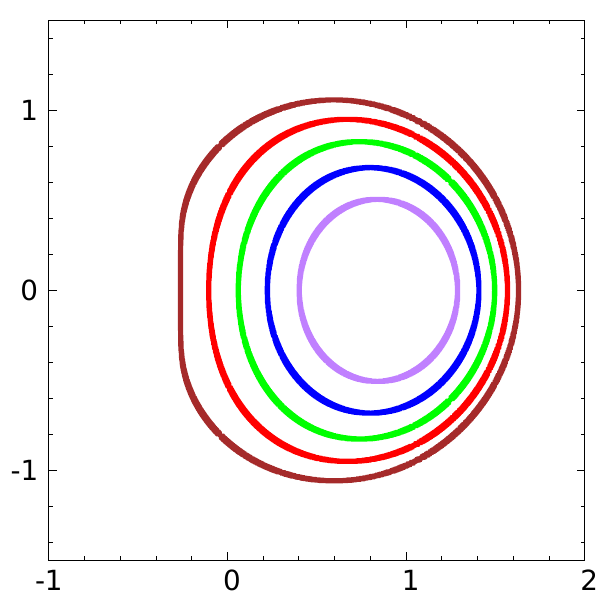}} &
            {\includegraphics[scale=0.36,trim=0 0 0 0]{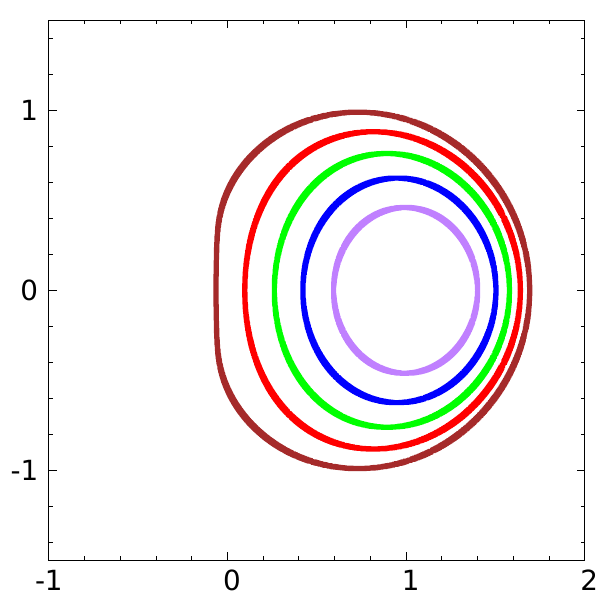}} &
            {\includegraphics[scale=0.39,trim=0 0 0 0]{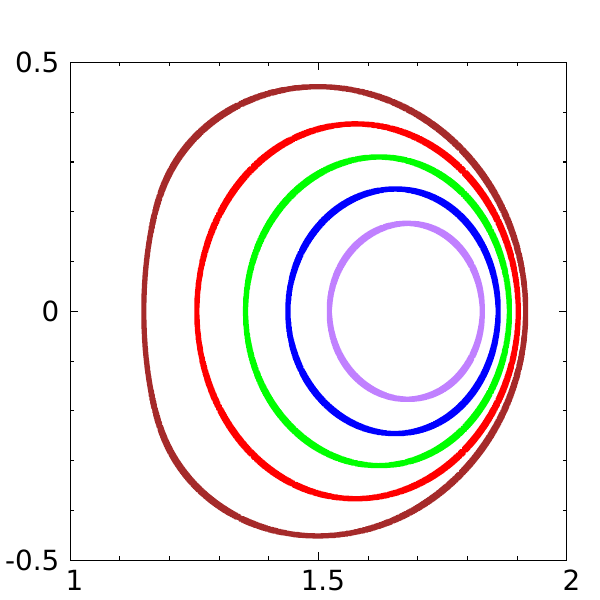}} \\
            \rotatebox{90}{~~~~~~~~~~~~Dilaton} &
            {\includegraphics[scale=0.36,trim=0 0 0 0]{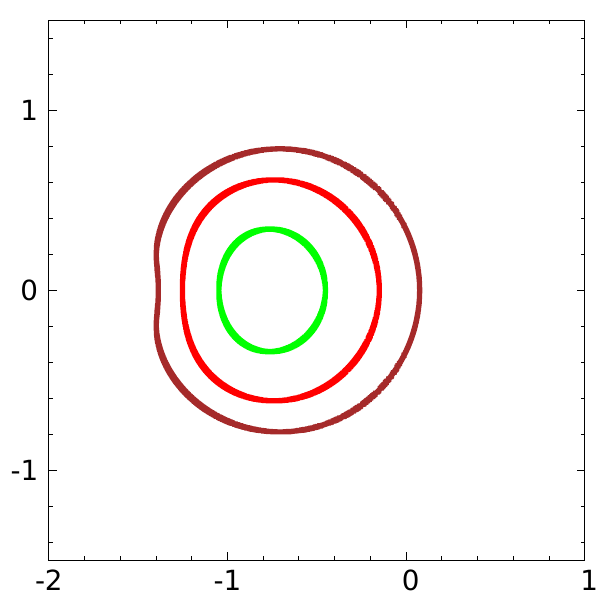}} &
            {\includegraphics[scale=0.36,trim=0 0 0 0]{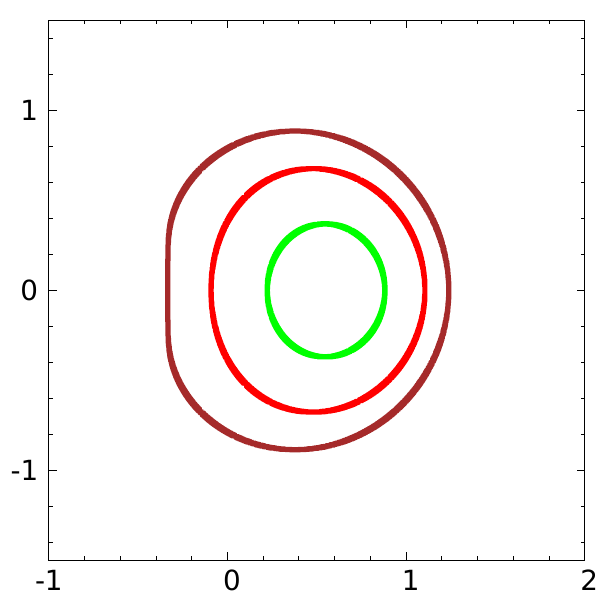}} &
            {\includegraphics[scale=0.36,trim=0 0 0 0]{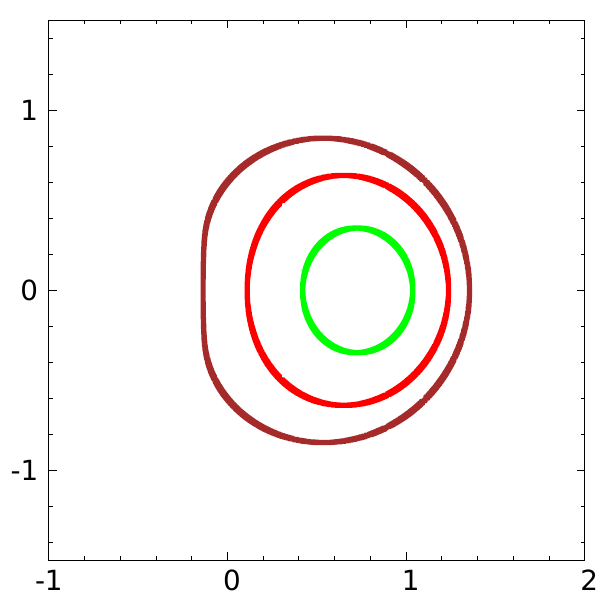}} &
            {\includegraphics[scale=0.39,trim=0 0 0 0]{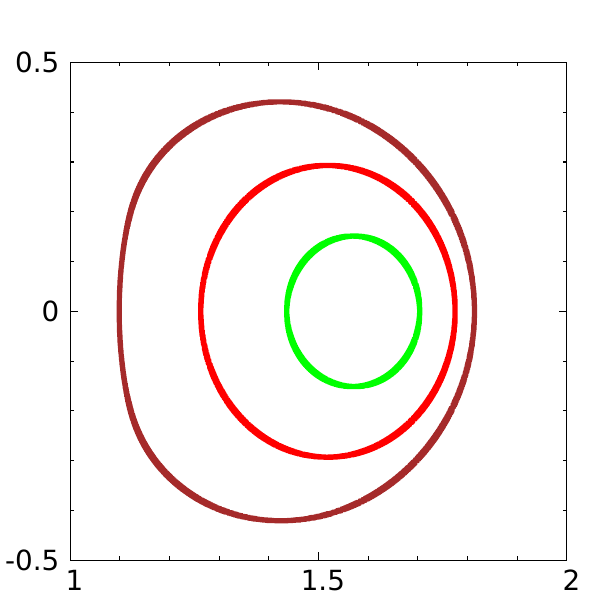}} \\
        \end{tabular}
        \caption{Relativistic aberration in the plasma profile 5{ with $a/a_{max}=0.999$, $r_O=5m$ and $\vartheta_O=\pi/2$}. Brown, red, green, blue and purple curves correspond respectively to the frequency ratios $\omega_c^2/\omega_\infty^2=0.00$, $32.5$, $65.0$, $97.5$ and $130.0$.
       {As explained in Fig.\eqref{fig:SH_p3},
        for some particular {$\omega^2_c/\omega^2_\infty=\chi_{cri}$}, the forbidden region reaches the equatorial plane and consequently, the shadow is no longer visible. As in Fig.\eqref{fig:Ab_p2} we have modified the scale of the axes in the plots of the last column.
        }
        }
        \label{fig:Ab_p6}
    \end{figure*}

    Here, we will only consider motion of observers $\mathcal{O'}$ in the equatorial plane ($\vartheta_O=\pi/2)$) with velocities in the direction of the vector $e_2$ {(i.e. in the $\varphi$ direction)} relative to the standard observers.
    The parameters $a/a_{max}=0.999$, $r_O=5m$ and $Q_{(p)}/Q_{(p)max}=\sqrt{0.25}$ are chosen.
    Thus, the $3$-velocity of the observer will be given by {$\vec{v}= v e_2$}, with $v<0$ indicating that the displacement is {in the sense of rotation of the black hole and $v>0$ indicating a motion in the opposite direction relative to the standard observer}.
    The resulting shadows for different values of the observer's velocity $v$ are shown in Figs. \eqref{fig:Ab_p2} to \eqref{fig:Ab_p6}, considering the different plasma profiles introduced above, and are visualized via stereographic projection from the celestial sphere to a plane. Each figure shows the resulting shadows for different values of the frequency ratio $\chi=\omega_c^2/\omega_\infty^2$.


    In the aforementioned figures  we can see that the shadows are shifted towards the apex, i.e. in the direction of the observer's motion, as is to be expected in an aberration phenomenon. At the same time, the size and shape of the shadow are affected. These effects increase the higher the relative velocity, and can be explained if we relate the direction of the observer's motion to the spin of the black hole and to the equatorial plane as the plane of symmetry, as stated in \cite{Grenzebach_2015}.
    For relatively low velocities ($|v|\leq0.1$), the 
    aberration effects do not affect too drastically 
    the shape or size of the observed shadow. The shift 
    towards the apex is small but becomes more 
    noticeable for larger ratios $\chi$. In 
    general, for $v=+0.1$ the shadow is a little larger 
    than for $v=-0.1$. The shape of 
    the shadow is similarly affected. {The asymmetry observed in the second and third columns of Figs. \eqref{fig:Ab_p2}, \eqref{fig:Ab_p3}, \eqref{fig:Ab_p5}, and \eqref{fig:Ab_p6} can be explained as follows. If the black hole were not rotating, the standard observers would be static, denoted as $\mathcal{O''}$. In such a scenario, an observer $\mathcal{O'}$ in relative motion to $\mathcal{O''}$ with velocity $\vec{v}=ve_2$ would perceive a reduced-sized image. However, the size of the image would not depend on the sense of motion (i.e., the sign of $v$), but only on its magnitude (refer to Fig. (3) of \cite{Grenzebach_2015}). However, due to the presence of angular momentum, our standard observers $\mathcal{O}$ are themselves rotating around the black hole. Consequently, in relation to distant static observers $\mathcal{O''}$ (with associated tetrad derived from Eqs. \eqref{eq:SH_e} in the limit of $r_O\to\infty$), an observer $\mathcal{O'}$ with a relative velocity $\vec{v}=-|v|e_2$ compared to $\mathcal{O}$ will experience a higher speed when orbiting the black hole than an observer moving with respect to the standard observer with $\vec{v}=|v|e_2$. This difference in motion leads to the observed asymmetry in the aberration effects.}

   {
    However, as shown in Fig.\eqref{fig:Ab_p4}, for the case of uniform plasma (Profile 3), for an observer moving in the sense or rotation of the black hole ($v<0$), even a slight variation in the observer's velocity can significantly transform the observed shadow, leading to a small shadow in an illuminated sky to become a spot of light in a dark sky when approaching a critical velocity $v_c<0$.
    To explain this phenomenon, we need to bear in mind that we are projecting the celestial sphere onto a plane using stereographic projection. By analyzing the shadows for increasing values of $|v|$ (not shown), it is evident that as $v$ approaches a critical value $v_c<0$, the left end of the contour curve becomes increasingly larger. For $|v|>|v_c|$, the curve passes behind the observer (on the celestial sphere), and in the stereographic projection, it is plotted on the right side. As a result, for low velocities of the observer, the blue and violet curves enclose the illuminated sky, whereas for a specific critical value, they enclose the shadow.} 
    
    {
    Note also that for high velocities ($|v|\approx0.9$) the modification in the shape of the shadows is increased. For low ratios $\chi$ the shadow tends to be magnified while for high $\chi$ tends to be demagnified, but this effect is different depending on both the metric and the plasma profile.  
    For $v=0.9$ the left side of the shadow (which corresponds to the straight side of the $D$) shrinks. Thus, for low {$\chi$} the shadow becomes almost circular, with a slight indentation on the left side. As {$\chi$} increases, the shadow shrinks adopting a lenticular shape. The apparent size of the shadows will depend strongly on the spacetime model and the observed frequency. For $v=-0.9$, the straight side of the $D$ grows in proportion, resulting in a more flattened shadow at lower 
    {$\chi$}. As {$\chi$} grows, contrary to the previous case, the shadow becomes more circular in shape as it shrinks, losing the $D$ shape. In this case, the image is clearly demagnified, regardless of the plasma profile from which it originates.
    In all cases, the rings structure formed by the shadows corresponding to different frequencies is crowded in the direction of motion.    
    In the case of uniform plasma, the fishtail structure mentioned above is transformed, giving rise to others somewhat more exotic. The analysis of these structures is not the objective of this work, but we include them as a curiosity. 
    }

\section{Final remarks}
\label{c3s5}

    Using the KSZ metric family, we employed various spacetime models to investigate the propagation of light in a non-magnetized, pressureless plasma - a dispersive medium with a frequency-dependent refractive index -. The Hamiltonian formalism was employed to describe the photon dynamics within the plasma environment.
    This class of black holes includes several well-known exact analytic black hole solutions and many other black hole metrics obtained by deformations of the Kerr metric or stipulated by some cosmological or braneworld scenarios. Moreover, as shown in \cite{KSZ_2018}, these can serve to effectively approximate more complex metrics that do not allow separation of variables. 

        It is important to clarify that the class of black holes considered here (KSZ) is not the most general among those that allow the separation of variables in the HJ equation. We use them since our intention was to describe in a practical way black holes with the same symmetries of the Kerr model, in asymptotically flat and axially symmetric spacetimes presenting a event horizon and admitting a generalized Carter constant. This allowed us a simple analytical treatment by being able to completely separate the null geodesic equations.

    We derived general analytical formulas  to find the contour curve of the shadow cast by the black hole on the observer's sky in terms of angular celestial coordinates, for a wide range of metrics. The obtained expressions are valid for any photon frequency at infinity $\omega_\infty$, any value of the spin parameter $a$, any position of the observer within the external communication domain, any plasma distribution satisfying the separability condition, and any relative velocity between the observer and the black hole. These formulas served as the zeroth-order approximation to numerically study more realistic situations, and a step-by-step procedure for shadow construction with the inclusion of aberration effects was provided. We also elaborated different configurations considering specific plasma distributions in different alternatives to Kerr geometries. 

     It was observed that different spacetime models affected the shape and size of the shadow differently, and the characteristic parameter $Q_{(p)}$ played a crucial role in each of them. In the Kerr-Newman, Kerr-Sen, and Dilaton metrics, the effects arising from angular momentum were reduced when considering high values of $Q_{(p)}$, where $a$ became negligible. On the other hand, in modified Kerr and Braneworld, the effects associated with the angular momentum were conserved even for values close to $Q_{(p)max}$, increasing in Braneworld.

     {Of course, if the plasma frequency is small compared to the photon frequency, the shadow is not very different from the pure gravity case.}  On the contrary, if the plasma frequency is close to the photon frequency, the properties of the shadow change drastically depending on the plasma distribution. We note that there is a certain plasma frequency ratio $\chi_{cri}$ above which the shadow is no longer visible. This depends on both the metric model and the $Q_{(p)}$ parameter. On the other hand, when considering a{uniform} plasma the shadows adopt extremely exotic shapes called ``fishtails''.
     Here, different bright and dark regions appear, generally enclosing an illuminated spotlight in a shadowed sky. Consequently, the size and shape of the shadow were modified in the presence of a plasma environment around a black hole, depending heavily on the ratio between the plasma frequency and the photon frequency.
 
    {As one of the main results of this article, we  have investigated the influence of aberration on the size and shape of the black hole shadow, which is shifted towards the apex. The degree of shift increases for higher plasma frequencies. The magnification or demagnification of the shadow, as well as its deformation due to aberration, depend strongly on the frequency ratio $\omega^2_c/\omega^2_\infty$, the metric, and the plasma distribution. These effects become more pronounced at higher velocities. Notably, when the observer's direction of motion coincides with the rotation direction of the black hole, the effects resulting from the angular momentum, such as the $D$ shape and the increase in size, are diminished due to the coupling of the motions. When a uniform plasma distribution is considered, we observed that the fishtails, first observed in \cite{PyT_2017}, are now deformed into more complex shapes. Additionally, we identified a critical velocity at which the contour curve transitions from enclosing the illuminated sky to enclosing the shadow.}
    
    The processes of absorption or scattering of photons, as well as the gravitational field produced by the plasma environment, were not taken into account. Hence, the presence of the plasma manifested itself only through perturbations in the photon trajectories which resulted in a chromatic description of these phenomena, depending on their frequency.  Moving forward, the next step would be to explore how the luminosity of the accretion disk itself, in the radio-frequency regime, affects the final image of rotating black holes. A model of magnetized plasma, whose dynamics have previously been discussed by Broderick and Blandford \cite{2003BroBla, Broderick:2003fc}, should also be included in this situation. In this scenario, analytical methods must be eschewed in favor of numerical techniques, which can perform ray tracing and implement radiative transfer equations.

   {Finally, it is worth mentioning that Chang {\it et.al.} have recently introduced a new framework for computing the shadows of black holes using astrometric measurements for observers at different states of motion and located at finite distances from the black hole \cite{Chang:2020miq}. Although their findings on the shape of shadows are consistent with other definitions for observers situated far from the black hole, this does not seem to apply to observers located at finite distances. The cause of this difference is still an open problem\cite{Chang:2020lmg,Chang:2021ngy}. It would also be desirable to extend their framework to include the study of shadows in plasma environments and compare the results with those obtained using the Perlick-Tsupko-Grenzebach framework. Work in this area is currently underway. }

\subsubsection*{Acknowledgements}
 We are very grateful to Oleg Tsupko for  illuminating discussions and valuable comments.
We acknowledge financial support from CONICET, SeCyT-UNC. T. M. thanks H. Hosseini for discussion on black holes shadows and acknowledges financial support from the  FONDECYT de iniciaci\'on 2019 (Project No. 11190854) of the Chilean National Agency for the Science and Technology (CONICYT).

%\bibliography{biblio}

\end{document}